\documentclass[fleqn,usenatbib]{mnras}

\usepackage{newtxtext}

\usepackage[T1]{fontenc}

\DeclareRobustCommand{\VAN}[3]{#2}
\let\VANthebibliography\thebibliography
\def\thebibliography{\DeclareRobustCommand{\VAN}[3]{##3}\VANthebibliography}

\usepackage{graphicx}	
\usepackage{amsmath}	
\usepackage{amssymb}	

\title[Angular Momentum and Vortices in FDM]{Angular Momentum and the Absence of Vortices in the Cores of Fuzzy Dark Matter Haloes}

\author[Schobesberger, Rindler-Daller \& Shapiro ]{
Sonja O. Schobesberger$^{1}$\thanks{E-mail: sonja.schobesberger@protonmail.ch}, Tanja Rindler-Daller$^{1}$\thanks{E-mail: tanja.rindler-daller@univie.ac.at} and
Paul R. Shapiro$^{2}$\thanks{E-mail: shapiro@astro.as.utexas.edu}
\\
$^{1}$Institut f\"ur Astrophysik, Universit\"atssternwarte Wien, University of Vienna, T\"urkenschanzstr. 17, A-1180 Vienna, Austria\\
$^{2}$Department of Astronomy, The University of Texas at Austin, 2515 Speedway, Austin, TX 78712, USA}

\date{Accepted XXX. Received YYY; in original form ZZZ}

\pubyear{2020}

\begin{document}
\label{firstpage}
\pagerange{\pageref{firstpage}--\pageref{lastpage}}
\maketitle

\begin{abstract}
Scalar Field Dark Matter (SFDM), comprised of ultralight ($\gtrsim 10^{-22}$ eV) bosons, is distinguished from massive ($\gtrsim$ GeV), collisionless Cold Dark Matter (CDM) by its novel structure-formation dynamics as Bose-Einstein condensate (BEC) and quantum superfluid with wave-like properties, described by the Gross-Pitaevski and Poisson (GPP) equations. In the free-field (“fuzzy”) limit of SFDM (FDM), structure is inhibited below the de Broglie wavelength $\lambda_{\text{deB}}$, but resembles CDM on larger scales. Virialized haloes have “solitonic” cores of radius $\sim \lambda_{\text{deB}}$ that follow the ground-state attractor solution of GPP, surrounded by CDM-like envelopes. As superfluid, SFDM is irrotational (vorticity-free) but can be unstable to vortex formation. We previously showed this \textit{can} happen in halo cores, from angular momentum arising during structure formation, when repulsive self-interaction (SI) is present to support them out to a second length scale $\lambda_{\text{SI}}$ with $\lambda_{\text{SI}} > \lambda_{\text{deB}}$ (the Thomas-Fermi regime), but only if SI is strong enough. This suggested FDM cores ($\emph{without}$ SI) would not form vortices. FDM simulations later found vortices, but only \textit{outside} halo cores, consistent with our previous suggestion based upon TF-regime analysis. We extend that analysis now to FDM, to show explicitly that vortices should \textit{not} arise in solitonic cores from angular momentum, modelling them as either Gaussian spheres or compressible, ($n = 2$)-polytropic, irrotational Riemann-S ellipsoids. We find that, for typical halo spin parameters, angular momentum per particle is below $\hbar$, the minimum required even for one singly-quantized vortex in the centre. Even for larger angular momentum, however, vortex formation is not energetically favoured.  
\end{abstract}




\section{Introduction}\label{sec:intro}
In recent years, ultralight bosonic dark matter has attracted increasing attention in the community as an alternative to the
standard cold dark matter (CDM) paradigm. While CDM is modelled astrophysically as a collisionless "gas" without internal particle self-interactions, 
except for gravity, ultralight bosons can exhibit a richer particle phenomenology, affecting structure formation and galactic dynamics. This way, the search for signature effects on galactic and cosmological scales helps to distinguish ultralight bosons from CDM. 
Moreover, the theoretical predictions of excessive clustering in standard CDM face issues, when confronted with observations. These problems have been known for a while as "cusp-core problem", "missing-satellite-problem", "too-big-too-fail-problem" and possibly further issues, see e.g. \cite{2015PNAS..11212249W}; \cite{2017ARA&A..55..343B}; \cite{2019A&ARv..27....2S}.
On the other hand, ultralight bosons with $m \sim (10^{-25}-10^{-18})$ eV$/c^2$, bring about a substantial cutoff in the small-scale structure formation, potentially alleviating the above problems, since structure formation is inhibited below a length scale which depends upon the boson parameters, either the de Broglie wave length $\lambda_{\text{deB}}$ discussed below, or a second length scale $\lambda_{\text{SI}} > \lambda_{\text{deB}}$, if a strongly repulsive particle self-interaction (\textbf{SI}) is present. 

Ultralight bosons encompass a large family of models.
The common thing they share, however, is that they are described by a scalar field, hence also called "scalar field dark matter" (\textbf{SFDM}) and the further requirement that they shall be "cold" (i.e. non-relativistic) as of some point in their cosmic evolution, in order to be able to make up for the present dark matter (DM) energy density. 
In terms of particle models, 
ultralight bosons have been considered in 
string theories and other extra-dimensional models, see e.g. \citet{1995PhRvL..75.2077F}, \citet{1997PhRvD..56.6391G}, \citet{Svrcek_2006}, \citet{2010PhRvD..81l3530A}, or \cite{2016arXiv160306580F}. 
 
In this work, we will consider the simplest model of SFDM with a Lagrangian that contains kinetic energy and a (rest-)mass term\footnote{The distinction between real and complex scalar fields is not important in our study here.}. Furthermore, we will only
consider the non-relativistic description of such DM particles, which is suitable for the physics of DM haloes. In this context, it has been also known as Bose-Einstein-Condensate (\textbf{BEC}) dark matter (or BEC-DM, or BEC-CDM) in order to emphasize the idea that the ultralight bosons undergo Bose-Einstein condensation, which motivates the use of a scalar field in the first place. SFDM without SI has been also called fuzzy DM, $\psi$DM, wave DM or free SFDM. In this work, we will call it fuzzy DM (\textbf{FDM}), in observance of the early paper
by \citet{2000PhRvL..85.1158H}, wherein that term was coined. 

However, the "fuzziness" of FDM in the sense originally introduced by these authors actually 
requires more than just the lack of SI. It basically demands that the de Broglie wavelength of the bosons, for given mass $m$ and collective velocity $v$,
\begin{equation}
\label{eq:lambda-1}
    \lambda_{\text{deB}} = \frac{h}{mv},
\end{equation}
becomes of the order of galactic scales,
\begin{equation}
\label{eq:lam-as}
    \frac{\lambda_{\text{deB}}}{2 \pi} = \frac{\hbar}{mv} = 1.92 \text{ kpc} \left( \frac{10^{-22} \text{ eV}/c^2}{m}\right) \left( \frac{10 \text{ km s}^{-1}}{v}\right),
\end{equation}
which implies that the corresponding velocity should be of order the virial velocity of the gravitationally bound object. 

In this model, the de Broglie length is of same order of magnitude as the characteristic Jeans length, or the size of the smallest gravitationally bound object expected in BEC-DM. BEC-DM as FDM can cure the small-scale-crisis, once the scale that fits the smallest (sub)halo will host the smallest observed types of galaxies. In other words,  if we believe that the smallest galactic DM structures should not undershoot a size of about 1-2 kpc, then we require a boson mass of order $m \gtrsim 10^{-22}$ eV$/c^2$. As a result, this regime of very small masses has received considerable amount of interest in the recent literature.

The quantum nature of SFDM as a BEC and quantum superfluid distinguishes its dynamics from that of a
collisionless gas like CDM, in more ways than just the suppression of structure below the de Broglie wavelength
of the former. For example, BECs are generally irrotational, i.e. vorticity-free. However, as known from laboratory experiments, quantum vortices can arise, either from
the rotation of a single BEC, or from the merging of multiple BECs which initially had no net angular momentum. Vortices result that represent topological defects, outside of which the rest of
the system remains vorticity-free. Also, vorticity in quantum fluids is discrete and quantized.  Like "tornadoes", in which the density drops to zero towards the centre,
while velocity diverges, such quantum vortices can be a significant departure from the smooth background,
with dynamical and structural consequences.  As a quantum superfluid, SFDM, or BEC-DM,  must also be
irrotational - free of vorticity. However, even with irrotational initial conditions, BEC-DM can also
become unstable to vortex formation, as found for BECs in the lab.
 
This paper will focus on a well-posed problem, namely the question of whether vortices can form within rotating FDM ground-state structures in gravitational equilibrium, more precisely we will study the appearance of a single vortex.

Earlier literature considered the possibility of vortices
(either single or in a lattice) 
in BEC-DM haloes, with and without SI, in a somewhat heuristic way, by estimating a force equilibrium, or using formulae derived in a gravity-free setting, see e.g. \citet{2009arXiv0902.0605B}; \citet{Silverman2002}; \citet{2002CQGra..19L.157Y}; \citet{Zinner2011}.

Our previous papers \citet{2010ASPC..432..244R} and \citet{2012MNRAS.422..135R} (henceforth abbreviated \textbf{RS12})
were the first to ask and answer the question of whether vortex formation can actually happen in realistic BEC-DM haloes.
Our analysis here is inspired by the analytical approach of RS12.
Motivated by the knowledge gained from laboratory experiments of BEC's and the theory behind them, which showed 
that BECs held in a trap, if made to rotate with 
sufficient angular velocity, form quantum vortices from an 
initially vortex-free state, we considered the possibility 
that the angular momentum of haloes associated with
their formation was sufficient, in the setting of this cosmic ``gravitational'' trap, to trigger the same instability 
with respect to forming vortices.  In particular, by modelling
the gravitational equilibrium of BEC-DM haloes (and halo
cores) in the presence of the angular momentum expected 
to arise during large-scale structure formation, as parameterized by the cosmological halo spin-parameter, RS12 determined
whether vortex formation was energetically favoured, and for which parameters of the BEC-DM boson mass and SI coupling strength that would happen, if so. The analysis of RS12 was carried out in the regime of strongly repulsive SI, or the so-called Thomas-Fermi (\textbf{TF}) regime, where gravity is balanced by the SI pressure. 

It was shown that vortex formation is favoured for a large part of the boson parameter space. In fact, the parameter space of favoured vortex formation overlaps the parameter space of the TF regime to a great extent. In addition, approximate halo models for this regime were introduced in RS12, which were shown to be viable models for rotating BEC-DM haloes, having different amount of angular momentum. In particular, rotating haloes (or halo cores) without vortices were shown to be well represented by so-called irrotational Riemann-S ellipsoidal figures, whose equation of state is an ($n=1$)-polytrope, as appropriate for the TF regime.


Now, when RS12 found that vortices were likely to form in
BEC-DM haloes with SI, their analysis showed that, even within
the TF regime, a minimum SI coupling strength was required to make
vortex formation energetically favoured.  It was reasoned, therefore, that,
if SI were absent altogether, as in FDM, vortices would not
form at all, i.e. in halo cores supported against gravity only
by FDM quantum pressure and the amount of rotation expected from
large-scale structure formation. The purpose of the present paper is
to revisit this suggestion by RS12 that no vortices are expected for
FDM, by performing a new, detailed analysis along the lines of that which RS12 performed in the TF regime\footnote{The analysis in RS12 does not cover the case of attractive boson self-interactions.}.  As we will show by this new analysis, the
expectation expressed in RS12 was correct and in accordance
with 3D simulations of FDM halo formation which came later.
In particular, while vorticity has been reported in FDM structure formation
simulations in some locations, it is entirely absent from the halo cores.
Our results help to explain this exclusivity.

When haloes form cosmologically, their size and mass can extend
well beyond that of their cores, to include an envelope region, whose properties on average should resemble CDM, since structure formation in BEC-DM on scales much larger than either the
de Broglie wavelength or the characteristic scale of SI pressure support resembles that of CDM. In fact, we suggested in \citet{2014MPLA...2930002R} that the polytropic, SI pressure support that set the size of halo cores in the TF regime would be supplemented on larger scales by the support of wave motion, generated by the wave nature of BEC-DM and its quantum pressure during virialization of haloes that assemble from infall and mergers, making it possible for haloes to be much larger than their polytropic cores in which only SI dominates. 

Meanwhile, FDM has been studied in more detail, including
simulations of halo formation that report all virialized haloes have solitonic cores of the size of the de Broglie
wavelength (as evaluated inside haloes),
supported against gravity by quantum pressure, see \cite*{2014NatPh..10..496S}, \cite*{2016PhRvD..94d3513S} and \cite{2017MNRAS.471.4559M}.
And all haloes also show a wave-supported envelope
outside this solitonic core, with a profile that resembles
that in CDM haloes, but in which wave motions provide the
random internal motions responsible for virial equilibrium,
instead of random particle orbits, as first described by
\cite{2014MPLA...2930002R}. These same simulations find that vorticity and signs of quantum turbulence\footnote{Vortices are fundamental building blocks of quantum turbulence, because quantum turbulence exemplifies itself in the form of large disordered vortex tangles, where individual vortices interact via vortex reconnections. Moreover, these turbulent flows may also exhibit partial spatial polarization by organizing into vortex bundles enabling large-scale energy flows. The field of quantum turbulence is far from being understood (see e.g. \citet{2016PhR...622....1T} for lab examples). We do not study quantum turbulence in this paper and only note that the vortices found in the halo envelopes of simulations could arise from complicated wave dynamics and interference. They are distinct
from the vortices we study here in at least one respect: they
do not appear to arise in ``self-gravitating'' equilibrium
objects like our solitonic cores, in which the quantum pressure
force is balanced locally by the gravity of the mass associated 
only with the vortex and its near environment.} are generated during structure formation (from vorticity-free initital conditions), but only outside of the solitonic cores. The origin of this vorticity has not been well-studied, but its absence from solitonic cores is consistent with our suggestion in RS12 that vortex formation by instability in the presence
of angular momentum requires a sufficiently strong repulsive SI in order to be energetically favoured. However, the analysis in RS12 was limited to the TF regime, so our purpose here is to extend that analysis to the case of FDM.


The fuzzy regime presents an important challenge for analytical methods, in that all length scales of interest, in particular the characteristic size of the system under question and perturbations in that system like vortices, are of similar order of magnitude. This is in stark contrast to the TF regime studied in RS12, where length scales are separated by many orders of magnitude, allowing to focus on leading-order considerations.
We shall show that for solitonic cores in FDM, where gravity is balanced by quantum pressure and rotation alone, vortex formation cannot be triggered by angular momentum. This is in direct contrast to the TF regime, in which strong SI enables vortex formation.
For vortex formation to be triggered by angular momentum,
the specific angular momentum must first satisfy a necessary
condition that it exceeds the minimum value that gives each
particle an angular momentum of $\hbar$.   If this necessary condition
is satisfied, then it is further required that vortex formation be
energetically favoured, in order to establish that vortex formation
will take place.   In the TF regime, both conditions can be
met for the typical amounts of specific angular momentum for cosmological haloes,
for a large range of particle mass and SI strength.  However, for FDM, we will show here that the necessary (minimum) condition is generally not
met for typical amounts of halo angular momentum.  
We further show that, even for angular momentum large enough to meet the necessary condition,
vortex formation is nevertheless not energetically favoured.   
This is consistent with and can explain the fact that simulations
of structure formation in the FDM model do not find vortices
in the solitonic cores of FDM haloes.  

Our current analysis will share certain assumptions of RS12: we will also confine to a DM-only analysis; we will limit the consideration to rotating, equilibrium systems, i.e. regions whose size is comparable to the characteristic length scale of hydrostatic objects, and whose density profile corresponds to the (approximate) ground state of the underlying equations of motion, the Gross-Pitaevskii-Poisson system of equations.
In FDM, considered in this paper, the characteristic scale is the de Broglie length of bosons, which are distributed according to a numerically calculated solitonic profile, while in the TF regime the characteristic scale is proportional to the radius of an ($n=1$)-polytrope of the underlying density profile. 
The (approximate) ground-state solutions considered in RS12 and here are viable models for the central core region of a big halo, or of the entire region of a small halo which consists only of the core.
The reader shall keep this in mind, when we talk about "haloes" and "halo cores" in our work.  

However, the (approximate) ground-state solutions that we study here also apply to bound systems composed of DM bosons of higher mass, not just ultra-light, e.g. those arising from QCD axions, a fundamental (pseudo-) scalar particle with a mass of about $m \sim 10^{-5}$ eV$/c^2$, introduced to resolve the CP problem of QCD (\citet{1977PhRvL..38.1440P}). It has been suggested as a DM particle in \cite*{PhysRevLett.40.223} and \cite*{PhysRevLett.40.279}. 
The QCD axion and generally other bosons with $m \ggg 10^{-22}$ eV$/c^2$ have de Broglie lengths $\lambda_{\text{deB}} \lll 1$ kpc, i.e. invisible on any galactic scales of interest, hence they would not be examples for FDM in the strong sense, for their wave nature is not potent on galactic scales anymore, implying they would fail to solve the small-scale problems of CDM. Yet, they remain valid models, as long as the nature of DM is unsettled and as long as no other reasons have been found to exclude them definitely. Since we are interested here to study the implications for FDM halo cores, we will not consider the high-mass range of QCD axions in this paper, however.

This paper is organized as follows.  In \autoref{sec:funda}, we present the fundamental equations. In \autoref{sec:fuzzy}, we review the basic concepts of SFDM in the fuzzy limit (FDM) and present models of non-rotating solitonic cores. The latter are the building blocks of our models with rotation, which are introduced in \autoref{sec:virial-halos}. \autoref{sec:energyanalysis} includes the derivation of our main results, namely the analysis of the conditions of vortex formation, where we show that vortices due to angular momentum are not expected in FDM halo cores. Finally, our conclusions and discussions are presented in \autoref{sec:con}. Also, some more technical background is provided in two appendices.

\section{Fundamentals of gravitationally bound BEC-DM structures}\label{sec:funda}

In the non-relativistic limit,
BEC-DM objects under self-gravity are generally described by the Gross-Pitaevskii (GP) equation of motion (\cite{pitaevskii1961vortex}, \cite{1961NCim...20..454G}, \cite{1969PhRv..187.1767R}) for the complex scalar wavefunction $\psi(\vec{r},t)$ of the bosons in the Bose-Einstein condensate (BEC), 
\begin{equation}
\label{eq:GP}
i \hbar \frac{\partial \psi}{\partial t} = -\frac{\hbar^2}{2m}\Delta \psi + (m\Phi + g |\psi|^2)\psi,
\end{equation}
which is coupled to the Poisson equation
\begin{equation}
\label{eq:P}
\Delta \Phi = 4\pi Gm |\psi|^2\ ,
\end{equation}
where $|\psi|^2(\vec{r},t) = n(\vec{r},t)$ corresponds to the number probability density, $m$ denotes the boson mass and $\Phi$ the gravitational potential. The system of equations of motion (\ref{eq:GP}-\ref{eq:P}) is called the Gross-Piaevskii-Poisson (\textbf{GPP}) system.
We assume that all $N$ particles comprising a given object of volume $V$ are in the condensed state described by $\psi$, hence
\begin{equation}
    N = \int_V |\psi|^2 \text{d}V\ .
\end{equation}
The last term in Eq. (\ref{eq:GP}) describes an effective 2-particle SI with a coupling\footnote{The notation $\lambda$ for $g$ is also common, especially in the context of the relativistic version of GPP.} constant $g=4\pi \hbar^2 a_s/m$. Its sign is determined by the sign of the scattering length $a_s$. Within this framework, $m$ and $g$ are the fundamental particle parameters. In this work, we will set $g=0$, but in order to compare our analysis with previous results we include it in the description of the basic equations.
Equ. (\ref{eq:GP}) admits a quantum fluid description upon introducing hydrodynamical variables via the Madelung transformation (\citet{1927ZPhy...40..322M}):
\begin{equation}
\label{eq:D}
\psi(\vec{r},t) = |\psi|(\vec{r},t)e^{iS(\vec{r},t)} = \sqrt{\frac{\rho(\vec{r},t) }{m}}e^{iS(\vec{r},t)},
\end{equation}
where 
\begin{equation}
\label{eq:MD}
\rho(\vec{r},t) = m|\psi|^2
\end{equation}
is the mass density.
Through the current
\begin{equation}
\label{eq:j}
\vec{j}(\vec{r},t) = \frac{\hbar}{2im}(\psi^*\vec{\nabla} \psi -\psi\vec{\nabla} \psi^* ) = n\frac{\hbar}{m}\vec{\nabla} S
\end{equation}
and the identification of
\begin{equation}
\label{eq:n}
\vec{j} = n  \vec{v},
\end{equation}
we can define the bulk (or flow) velocity field as the phase gradient 
\begin{equation}
\label{eq:v}
\vec{v} = \frac{\hbar}{m}\vec{\nabla} S.
\end{equation}
Assuming particle number is conserved, we identify an Euler-like equation of motion,
\begin{equation}
\label{eq:euler}
\rho \frac{\partial \vec{v}}{\partial t} + \rho (\vec{v} \cdot \vec{\nabla})\vec{v} = -\rho \vec{\nabla} Q - \rho \vec{\nabla} \Phi - \vec{\nabla} P_{SI}\ ,
\end{equation}
and a continuity equation,
\begin{equation}
\label{eq:cont}
\frac{\partial \rho}{\partial t} + \vec{\nabla} \cdot (\rho \vec{v})=0\ .
\end{equation}
The so-called Bohm quantum potential, defined as
\begin{equation}
\label{eq:Q}
Q = -\frac{\hbar^2}{2m^2}\frac{\Delta \sqrt{\rho}}{\sqrt{\rho}},
\end{equation}
gives rise to what is often referred to as 'quantum pressure'. In addition, SI gives rise to a pressure of polytropic form, 
\begin{equation}
\label{eq:PSI}
P_{SI} = \frac{g}{2m^2}\rho^2 = K_p\rho^{1+1/n}\ ,
\end{equation}
with $n=1$ and the polytropic constant $K_p$ depends upon the DM particle parameters $m$ and $g$.  
 
In the spirit of RS12, we will also restrict our analysis to stationary systems and their corresponding energy, i.e. to the time-independent GP equation including the chemical potential $\mu$,
\begin{equation} \label{eq:stationary}
    \mu \psi_s (\vec{r}) = -\frac{\hbar^2}{2m}\Delta \psi_s (\vec{r}) + (m\Phi + g |\psi_s|^2)\psi_s (\vec{r})\ .
\end{equation}
This time-independent GP equation can be obtained from the time-dependent equation (\ref{eq:GP}) by inserting the state
\begin{equation}
\label{eq: stat-sol}
    \psi (\vec{r},t) = \psi_s (\vec{r}) \text{e}^{-i\mu t / \hbar}\ .
\end{equation}
Stationary states have this form of wavefunction, which evolves harmonically in time and yields the time-independent density $\rho = m |\psi_s|^2 $ and hence a time-independent gravitational potential. Of course, $\psi_s$ itself can be decomposed analogously to ansatz (\ref{eq:D}) as 
\begin{equation}
\label{eq: decom}
    \psi_s(\vec{r}) = |\psi_s|(\vec{r}) \text{e}^{iS_s(\vec{r})}\ ,
\end{equation}
where from now on the subscript $s$ will be omitted. The GP energy functional is given by
\begin{equation}
\label{eq:GPstat}
    E[\psi] = \int_V \left[ \frac{\hbar^2}{2m} |\nabla \psi|^2 + \frac{m}{2} \Phi |\psi|^2 + \frac{g}{2}|\psi|^4 \right] \text{d}^3r\ .
\end{equation}
By means of decomposition (\ref{eq: decom}) we can write the total energy,
\begin{equation} \label{eq:energytot}
E = K + W + U_{SI}    
\end{equation}
as a sum of the total kinetic energy 
\begin{eqnarray}
K &\equiv & \int_V  \frac{\hbar^2}{2m} |\nabla \psi|^2 \text{d}^3r \\
\label{eq:KQ}
&=& \int_V \frac{\hbar^2}{2m^2} (\nabla \sqrt{\rho})^2 \text{d}^3r +  \int_V \frac{\rho}{2} \vec{v}^2 \text{d}^3r \\
&\equiv& K_Q + T, 
\end{eqnarray}
the gravitational potential energy,
\begin{equation}
\label{eq: stat-W}
    W \equiv \int_V \frac{\rho}{2} \Phi \text{d}^3r\ ,
\end{equation}
and the internal energy,
\begin{equation}
    U_{SI} \equiv \int_V \frac{g}{2m^2} \rho^2 \text{d}^3r\ .
\end{equation}
$K_Q$ is the part of the kinetic energy accounting for the quantum-like phenomena; it is absent in classical galactic dynamics. $T$ describes the kinetic energy of the system due to bulk motions, like rotation or other internal fluid motions. $U_{SI} = \int P_{SI} \text{d}V$ is determined by the SI pressure given in (\ref{eq:PSI}).
Since we are interested in $g=0$, we will have that $U_{SI} = 0$.  However, in the course of our analysis we will see that, under certain assumptions, $K_Q$ can be approximated by an energy which \textit{formally} looks like the internal energy of a polytrope: the internal energy which arises from \textit{any} polytropic pressure, $P = K_p \rho^{1+ 1/n}$, is given as
\begin{equation}
\label{eq:uint}
    U = K_p n \int \rho^{1+ 1/n} \text{d}^3r \ .
\end{equation}
Here, we still leave the general dependency on $n$.
Using all these energy contributions, we can write the equilibrium scalar virial theorem as
\begin{equation}
\label{eq:virial_sf}
    2K + W + \frac{3}{n}U = 0\ ,
\end{equation}
see e.g. \cite{Wang01} for a derivation.
We will assume that gravitationally bound FDM halo cores in equilibrium fulfil (approximate) virial equilibrium.

\subsection{Vorticity}\label{subsec:vortex}

The GP framework was originally designed to account specifically for quantized vorticity in BECs. Indeed, the Madelung transformation unveils the superfluid character of this dissipation-free mean-field formulation. 
A zero-viscosity (i.e. dissipation-free) fluid has a conservative velocity field, i.e. a gradient flow $\Vec{v} \propto \Vec{\nabla} S$, which implies irrotationality, i.e. $\omega = \Vec{\nabla} \times \Vec{\nabla} S = 0$\ . However, it is only at first glance, that the formulation leaves no room for fluid vorticity. Wherever the mass density $\rho = m |\psi|^2$ vanishes, the phase $S$ and hence the velocity flow $\Vec{v}$ become discontinuous and ill-defined. The implication of irrotationality would only hold true if the phase function $S$ had continuous first and second derivatives everywhere. However, this is not any longer the case, once a vortex line appears, where $\Vec{v}$ diverges along its centre. (Since the density goes to zero towards the centre, there are no particles moving with infinite speed inside the vortex core.)  It turns out that the phase function along vortex lines has a non-trivial winding and, as a result, the vorticity there does not vanish: Since the wavefunction is required to be singly-valued, a circulation along a contour $C$ enclosing a vortex may not change $\psi$. Hence, $S$ can vary at most by $2\pi d$, where $d$ is the winding number, also called the vortex charge. As a consequence, this circulation is an integer multiple of the quantum of circulation $\kappa = h/m$,
\begin{equation}
\label{eq:circ-cond}
    \Gamma = \oint_C \text{d}\Vec{r} \cdot \Vec{v} = d\frac{2\pi \hbar}{m} = d\frac{h}{m}\ .
\end{equation}
 The wavefunction of an axially-symmetric vortex aligned with the $z$-direction at the origin in cylindrical coordinates is a stationary solution and has the form
\begin{equation}
    \label{eq:vortex1}
    \psi(\Vec{r}) = \psi(r,z,\phi) = |\psi|(r,z)\text{e}^{id\phi}\ .
\end{equation}
The velocity field around this vortex has a form given by the Madelung transformation;
\begin{equation}
\label{eq: velocityfield}
    \Vec{v} = \frac{\hbar}{m} \Vec{\nabla} S = \frac{\hbar}{m} \frac{1}{r} \frac{\partial S}{\partial \phi} \Vec{\phi} = \frac{\hbar}{m} \frac{d}{r} \Vec{\phi}\ ,
\end{equation}
directed along the azimuthal direction. 
It turns out that the above vortex wavefunction is an angular momentum eigenstate, where the $z$-component of the angular momentum is given by $l_z = d \hbar$ and thus yields the total angular momentum
\begin{equation}
\label{eq:LQM}
    L_z = d N \hbar \equiv d L_{QM}\ .
\end{equation}
$L_{QM}$ denotes the angular momentum required to sustain a singly-quantized (or singly-charged) vortex within an object composed of $N$ bosons.
There are many perspectives on such a vortex; it can be seen as a perturbation, topological defect, or excitation with a potentially higher energy than the quantum ground state.  For simplicity, we will focus in this paper on singly-quantized vortices with $d=1$, which tend to be energetically more favoured than vortices with $d>1$.

\section{The fuzzy regime}\label{sec:fuzzy}

Considering the three terms on the right-hand side of the Euler-like equation of motion (\ref{eq:euler}), two regimes can be distinguished. On the one hand, there is the regime where SI, or in other words the scattering of the DM particles is entirely neglected and hence solely quantum pressure works against gravity and prevents gravitational collapse. This regime is referred to as fuzzy regime, in conjunction with the requirement of Eq. (\ref{eq:lam-as}). In RS12, this regime was called BEC-CDM of type I. On the other hand, there is the regime of strongly interacting particles, where SI pressure dominates and balances gravity. This regime is the TF regime (called type II BEC-CDM in RS12), and its characteristic length scale is usually also required to be of order $\lesssim 1$ kpc, in order to resolve the CDM small-scale crisis.

The regime of interest in this work is the fuzzy regime, so it is important to review a few of its properties that are necessary in order to justify our choice of approximate models in the forthcoming.

The fuzzy regime effectively amounts to the limit of no SI ($g = 0$). Thus, we are interested in solutions of the time-independent GPP system in (\ref{eq:stationary}) and (\ref{eq:P}) without SI, namely 
\begin{eqnarray}
\label{eq:SP-1}
\mu \psi (\vec{r}) &=& -\frac{\hbar^2}{2m}\Delta \psi (\vec{r}) + m\Phi(\vec{r}) \psi (\vec{r})\ , \\
\label{eq:SP-2}
\Delta \Phi(\vec{r}) &=& 4\pi G m |\psi (\vec{r})|^2\ .
\end{eqnarray}
This system of equations has been also known in the literature as the Schr\"odinger - Poisson (\textbf{SP}) or Schr\"odinger - Newton equations, see e.g. \cite*{1998CQGra..15.2733M}, \cite{1999Nonli..12..201T}, provided we identify the eigenenergy $E$ with the chemical potential $\mu$.

Without loss of generality, \cite{1999Nonli..12..201T} assume that $\psi$ is real and they introduce the non-dimensionalized functions\footnote{\cite{1999Nonli..12..201T} use the notation $S$ for $\Sigma$, which we avoid, because $S$ in our paper refers to the phase function.} $\Sigma$ and $V$ by setting \footnote{Notice that their function $U$ has the dimension of gravitational energy and corresponds to $m\Phi$.}
\begin{equation}
\label{eq:def}
\psi = \left( \frac{\hbar^2}{8 \pi G m^3} \right)^{1/2} \Sigma, \ \ \ \ E - m\Phi = \frac{\hbar^2}{2m}V\ .
\end{equation}
This change of variables, together with the assumption of spherical symmetry, simplifies the equations to a new set,
\begin{equation}
\label{eq:SNneu1}
\frac{1}{r} (r\Sigma)'' = -\Sigma V 
\end{equation}
\begin{equation}
\label{eq:SNneu2}    
\frac{1}{r} (rV)'' = -\Sigma^2\ .
\end{equation}
The prime denotes differentiation with respect to the spherical radial coordinate $r$. The SP system admits a specific scaling invariance: given a solution $(\Sigma(r),V(r))$, for any arbitrary real $\gamma$, there is another solution of the form
\begin{equation}
\label{eq:scaling}
    \hat{\Sigma}(r)= \gamma^2 \Sigma(\gamma r)\ , \ \ \ \ \hat{V}(r)= \gamma^2V(\gamma r)\ .
\end{equation}
Their analysis yields that there exists a discrete family of finite, smooth, normalizable solutions where the $j$th solution ($j \in \mathbb{N}$) has $j-1$ zeros and that the energy eigenvalues increase monotonically towards $0$ with increasing $j$.
In the case of spherical symmetry, two relevant asymptotic forms are derived:
The solutions admit power series expansions near $r=0$,
\begin{equation}
\label{eq:expan}
\Sigma = \Sigma_0 - \frac{1}{6}\Sigma_0V_0r^2+ \mathcal{O}(r^4)\ \ \text{and} \  \ V = V_0 - \frac{1}{6}\Sigma_0^2r^2+ \mathcal{O}(r^4)\ ,
\end{equation}
where $\Sigma_0 = \Sigma(0), V_0 = V(0)$.
On the other hand, for large $r$ and real constants $A$,$B$ and $k$, there are solutions looking like
\begin{equation}
\label{eq:expanlar}
\Sigma = \frac{A}{r}\text{e}^{-kr}+...\ \ \text{and} \  \ V = -k^2 + \frac{B}{r}+...\ \ .
\end{equation}
In later sections of their paper, they show that the bound state solutions - those are the ones we are interested in -, have exactly these asymptotic forms.

In principle, the above differential equations (\ref{eq:SP-1})-(\ref{eq:SP-2}) or (\ref{eq:SNneu1})-(\ref{eq:SNneu2}), respectively, can be solved by looking for regular and finite solutions for the variables $\Sigma$ (resp. $\psi$) and $V$ (resp. $\Phi$) of the SP system, reducing to an eigenvalue problem which can be solved numerically. Many authors, such as \cite{1968PhRv..172.1331K}, \cite{1969PhRv..187.1767R}, \cite*{1989PhRvA..39.4207M}, \cite{1994PhRvL..72.2516S}, \cite{2004PhRvD..69l4033G}, or \cite{2017PhRvD..95d3541H}, have numerically calculated eigenstates of the SP system in spherical symmetry, e.g. assuming isolated systems, i.e. that $\psi$ and $\Phi$ approach $0$ as $r$ goes to infinity and that they are regular at the origin. The scaling invariance means that a solution level $j$ forms a one-parameter family which can be specified by the total mass $M$. Then, the central density is given by
\begin{equation}
 \label{eq: rc}
    \rho_c = \left( \frac{Gm^2}{\hbar^2}\right)^3M^4 \rho_j\ ,
\end{equation}
where the dimensionless constant $\rho_j$ (depending on the eigenstate label $j$) is calculated numerically. The densest state is the ground state, $j = 1$, from where the central density strongly decreases with increasing level number $j$. The $j = 1$ "soliton" state is a long-term attractor for bound, isolated FDM objects, since excited states decay to the soliton state, through dispersion of probability density to infinity, a process known as "gravitational cooling" (\cite{1994PhRvL..72.2516S}).  

The total mass $M = Nm$ of the soliton is conserved and finite, however it has no compact support and thus has to be cut off artificially, if we want to have a finite size. It is customary to pick a radius which includes $99\%$ of the mass. The resulting mass-radius relation of the bound ground-state has been calculated in \cite*{1989PhRvA..39.4207M},
\begin{equation}
\label{eq:R99-1}
    R_{99,S} = 9.946 \left(\frac{\hbar}{m}\right)^2\frac{1}{GM}\ ,
\end{equation}
where "S" stands for the numerically calculated "soliton".  This result leads up to a general discussion of characteristic length scales, before we present our models.

The analysis of vortex formation in RS12 focused on the TF regime. While conditions were established which determine when a halo (core) is described in either one of the regimes, the focus was targeted on haloes in the TF regime, i.e., the quantum-pressure term, eq.(\ref{eq:Q}), was neglected at the scales of the halo (core), but not at the scale of vortices within such haloes! There is a characteristic length scale introduced in the classic literature on BECs, for example \cite{pethick_smith_2008}, namely the so-called healing length\footnote{The literature, incl. RS12, use the notation $\xi$ for the healing length, which we avoid, because later we will use the notation $\xi$ in a different context.} $\ell$. This length describes the distance over which the wavefunction tends to its background value when subjected to a localized perturbation and is the result of balancing the quantum kinetic term with SI,
\begin{equation}
    0 = -\frac{\hbar^2}{2m}\frac{\psi}{\ell^2} + g |\psi|^2\psi\ ,
\end{equation}
yielding
\begin{equation}
    \label{eq:xi}
    \ell = \frac{\hbar}{\sqrt{2\Bar{\rho}g}}\ ,
\end{equation}
with $\Bar{\rho} = 3M/(4 \pi R^3)$, $M$ and $R$ are the total mass and radius of the object and $\psi$ has dimension [Length$^{-3/2}$].
To put it another way, it takes the BEC this distance to "heal" a local disturbance. By estimating the forces associated with the quantum-kinetic term and the SI term in the same way, RS12 conclude that in the TF regime the characteristic size of the system - the (halo) core size -  is much larger than both, the healing length and the de Broglie wavelength, i.e.
\begin{equation} \label{eq:tf}
    R \gg \ell, \ \ \ \ \ \ R \gg \lambda_{\text{deB}}.
\end{equation}
Moreover, it is shown that $\lambda_{\text{deB}} \sim 4.3\ \ell$.

However, in the fuzzy regime we have no SI. By replacing the SI energy by the gravitational energy, we introduce the \textit{gravitational healing length}, $\ell_{\rm{grav}}$, and derive it by setting the quantum kinetic energy equal to the gravitational energy, see the GP energy functional (\ref{eq:GPstat}),
\begin{equation}
    \frac{\hbar ^2}{2m}|\nabla \psi|^2 = \frac{m}{2}\Phi | \psi|^2.
\end{equation}
The kinetic energy is of order $\hbar ^2 /(2m \ell_{\rm{grav}}^2)$ and the gravitational energy is of order $mGM/(2\ell_{\rm{grav}})$, which yields
\begin{equation}
\label{eq:xi_G-1}
    \ell_{\rm{grav}} = \frac{\hbar^2 }{m^2GM},
\end{equation}
where, again, $M$ and $R$ are the total mass and radius of the object. Approximating the spatial dimension in $\Phi$ with $\ell_{\rm{grav}}$, i.e. with the same length scale used to approximate the Laplace operator on the left-hand side, corresponds to the notion of a Jeans analysis, where $\ell_{\rm{grav}}$ can be understood as the smallest length scale for bound structures, and also as the length scale of local perturbations. (Another possibility would have been to choose the gravitational energy to be of order $mGM/(2R)$, which represents global properties as opposed to the local sensitivity of the Laplace operator.).

The forthcoming energy analysis in the fuzzy regime gives rise to a quantity with the dimension of mass, which we define to be the \textit{characteristic particle mass}, $m_c$, whose dimensional combination is
\begin{equation} \label{charm}
    m_c^2 = \frac{\hbar^2 }{RGM}.
\end{equation}
As a result, we have
\begin{equation}
\label{eq:xi_G-m}
    \ell_{\rm{grav}} = \left(\frac{m_c}{m}\right)^2R\ .
\end{equation}
Previous works used a slightly different definition of the characteristic particle mass, e.g. $m_H$ in RS12. The comparison yields
\begin{equation}
    m_H \equiv \frac{2}{\sqrt{3}}\frac{\hbar}{\sqrt{RGM}} = \frac{2}{\sqrt{3}} m_c \approx 1.155\ m_c.
\end{equation}
In BEC-DM without SI (i.e. FDM), any gravitationally bound, ground-state solution of equ.(\ref{eq:SP-1})-(\ref{eq:SP-2}) has a spatial extent of order $\lambda_{\text{deB}}$, and this is also true for the solitonic cores found in simulations of FDM. If we require
\begin{equation}
\label{eq:lam}
    \lambda_{\text{deB}} \lesssim R,
\end{equation}
with
\begin{equation}
\label{eq:deb}
    \lambda_{\text{deB}} = \frac{h}{mv} \approx \frac{h}{m} \sqrt{\frac{R}{GM}},
\end{equation}
where we use $v_\text{core} \approx v_\text{circ} = (GM/R)^{1/2}$, we have 
\begin{equation}
\label{eq: mbound}
    2\pi \leq \frac{m}{m_c}.
\end{equation}
On the other hand, using $R_{99,S}$ of equ.(\ref{eq:R99-1}) in the definition (\ref{charm}) yields
\begin{equation}\label{massmem}
 \frac{m}{m_c} = \sqrt{9.946} \approx 3.154.
\end{equation}
However, we can explicitely see again, what we have pointed out before, namely that (\ref{eq:R99-1}) and (\ref{eq:xi_G-1}) are within a factor $10$ of similar order of magnitude, in stark contrast to (\ref{eq:tf}) of the TF regime. Therefore, it follows from (\ref{eq:xi_G-m}) that $m$ has to be of the same order as $m_c$, or in other words, there is only a relatively small range allowed for $m/m_c$ in the fuzzy regime.  
In order to factor in the imprecision in all these estimates, and in light of our models discussed below, we will consider a fiducial choice of range for that ratio in the forthcoming analysis, namely
\begin{equation}\label{massratio}
2 < \frac{m}{m_c} < 10.    
\end{equation}
There are two interpretations for our characteristic particle mass $m_c$. On the one hand, it is the mass which results from requiring
\begin{equation}
    \ell_{\rm{grav}} = \frac{\lambda_{\text{deB}}}{2 \pi} = R\ . 
\end{equation}
(It is precisely the fact that all length scales in this regime are of similar order, which makes the fuzzy limit of the GPP framework a challenging ground for analytic theory.) 
On the other hand, just as in RS12, we note another meaning for $m_c$ by observing that $m = (2/ \sqrt{3}) m_c = m_H$, if the characteristic gravitational angular frequency
\begin{equation}
    \label{eq: omg}
    \Omega_{\text{grav}} = \sqrt{\pi G \Bar{\rho}}
\end{equation}
equals the angular frequency 
\begin{equation}
    \Omega_{QM} = \frac{\hbar}{mR^2}
\end{equation}
of a uniformly rotating object with mass $M$ and angular momentum 
\begin{equation} 
    L = MR^2 \Omega_{QM} \stackrel{!}{=} L_{QM} = N \hbar\ ,
\end{equation}
where $L_{QM}$ is given by (\ref{eq:LQM}).

We stress that the range in (\ref{massratio}) applies only to the ratio $m/m_c$, and is not a bound per se on the particle mass $m$. As long as the ratio stays within that range, it doesn't matter whether we consider ultralight bosons, or bosons with the mass of the QCD axion, because $R_{99,S}$ in (\ref{eq:R99-1}), as well as the approximate mass-radius relations derived below, apply to all boson masses $m$. However, that size of the soliton decreases for increasing $m$, in accordance with the discussion of de Broglie scales in \autoref{sec:intro}.

\subsection{Approximations to the FDM ground state density distribution}

We have already mentioned that the SP system can only be solved by numerical means, yielding the respective wavefunction $\psi$ and the gravitational potential $\Phi$ coupled to it, and that the ground state solution, the soliton, is an attractor of a system in isolation.  However, during FDM halo formation, multiple mergers of solitons can form bigger haloes, whose envelopes are the result of complicated wave dynamics (\cite*{2014NatPh..10..496S,2016PhRvD..94d3513S}; \cite{2017MNRAS.471.4559M}). Still, the centres of these bigger haloes remain soliton-like and that central profile has been "empirically" fit to the outcome of merger simulations.  
\cite{2014PhRvL.113z1302S} introduced a function of the form
\begin{equation}\label{eq: schive}
    \rho = \rho_{c} \left( 1+ (r/c)^2\right)^{-8}\ ,
\end{equation}
with the central density $\rho_{c}$, the core radius $c$, and $r$ the distance from the centre, in order to fit that central core region made up by the "central soliton". 
In fact, the core size parameterized by $c$ is of order $\lambda_{\text{deB}}$, if the virial ($\approx$ circular) velocity of the core is used in the formula for $\lambda_{\text{deB}}$ in (\ref{eq:deb}).

Although the Schive profile in (\ref{eq: schive}) was determined to be a good fit to the central regions of virialized haloes, it is also a good fit to the numerical profile of the SP ground-state solution in isolation, which was discussed before. However, other good analytical fits to this SP ground-state include the Gaussian profile and the $(n=2)$-polytrope described below. We will use these two approximate models, instead of (\ref{eq: schive}), for our analytical treatment of the question of vortices in self-gravitating FDM halo cores, because i) the Gaussian has very nice analytical properties, and ii) the $(n=2)$-polytrope will be used in an extension to ellipsoidal haloes, where we can rely on global energy terms which have been already derived in the literature.


\subsubsection{The Gaussian sphere}\label{subsec:gaussian}

\begin{figure}
	\includegraphics[width=\columnwidth]{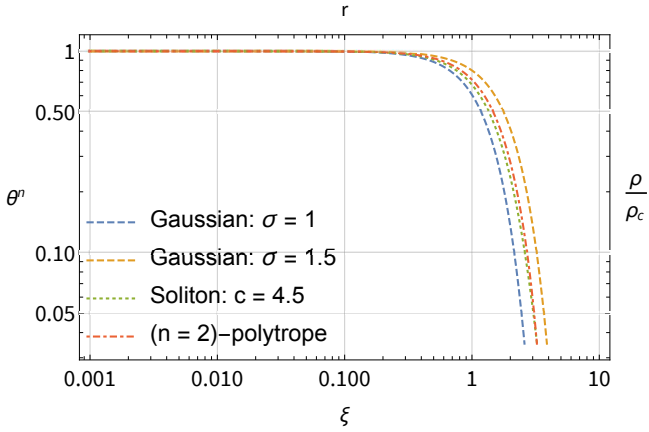}
    \caption{Approximations for FDM halo core density profiles in a double-logarithmic plot. Some illustrative examples for the same value of the central density $\rho_c$:
    The dashed curves correspond to the Gaussian density profile (\ref{eq:toy})-(\ref{eq:a}) with $\sigma = 1$ and $\sigma = 1.5$, respectively. The dashed-dotted curve corresponds to $\theta^2$, the density profile (\ref{eq: polymodel}) of an $(n=2)$-polytrope in units of the central density. For comparison, the solitonic profile (\ref{eq: schive}) for $c=4.5$ corresponds to the dotted curve. We can see that all these profiles are qualitatively very similar.}
    \label{fig:gauss-polytrope-soliton}
\end{figure}

Our first approximate density model is inspired by the well-known Gaussian "wave-packet" of quantum mechanics.  
It is described as
$|\psi|^2 = \rho_0/m$ with
\begin{equation}
\label{eq:toy}
\rho_0(r) =  \rho_c \text{e}^{-ar^2}\ 
\end{equation}
and
\begin{equation}
\label{eq:a}
a = \frac{1}{2\sigma^2},
\end{equation}
where $\sigma^2$ is the variance of the Gaussian, see \autoref{fig:gauss-polytrope-soliton} for example plots.
This model will serve as one of our approximations for the density profile of a vortex-free, bound FDM halo core in spherical symmetry, where $\rho_0$ denotes the (vortex-free) matter density, $\rho_c$ is the central density of the system and $r$ is the radial distance in spherical coordinates.  Similar to the numerical profile, this Gaussian approximation has no compact support. However, it is normalized as 
\begin{equation}
    \int_0^{\infty} \rho_c \text{e}^{-ar^2} 4 \pi r^2 dr = M = N m,
\end{equation}
which yields
\begin{equation}
    \label{eq:rhoc}
    \rho_c = \frac{N m}{\sigma^3 (2 \pi)^{3/2}}\ .
\end{equation}
The Gaussian profile (or ansatz) finds its justification in various previous analysis, where the Gaussian has been used in variational calculations, see e.g. \cite*{1996PhRvL..76....6B}, \cite{2011PhRvD..84d3531C}, or \cite{Schiappacasse_2018}, as well as a fit to the numerical soliton, e.g. in \cite*{2015MNRAS.451.2479M} or \cite{2018PhRvD..97k6003G}. Also, the asymptotic behaviour of $\Sigma$ in (\ref{eq:expanlar}), derived by \cite{1999Nonli..12..201T}, 
shows that the exponential drop-off is very much faster
than formula (\ref{eq: schive}) would provide, i.e. the sharper fall-off
is much better captured by the Gaussian. This has also been shown by \cite{PhysRevD.103.063012}, see their figure 7, where the soliton is compared to the Gaussian and the Schive profile.
That paper shows more generally the 
usefulness and appropriateness of the Gaussian ansatz as an approximate model for the numerical soliton solution, also in the context of core-halo relationships and implications thereof.
 
In the forthcoming, we will cut off the Gaussian profile at the radius $R=R_{99,G}$, which includes $99\%$ of its mass. This procedure is motivated by (\ref{eq:R99-1}), and we will neglect the error introduced by effectively replacing $M$ by $M_{99}$, i.e. $99\%$ of the total mass of the Gaussian. 
Given the finite size and in order to highlight the geometry that we will adopt, we will refer to this model as the "Gaussian sphere".
Now, the mass-radius relationship of the Gaussian sphere has been derived in \cite{2011PhRvD..84d3531C} and \cite{PhysRevD.103.063012}, and it follows easily, once we apply the virial condition (\ref{eq:virial_sf}) to FDM,
\begin{equation}
    2 K_Q + W = 0,
\end{equation}
implying 
\begin{equation} \label{r99G}
    R_{99,G} = 5.419 \left(\frac{\hbar}{m}\right)^2\frac{1}{GM}\ ,
\end{equation}
which differs only by a factor of $\approx 1.835$ from (\ref{eq:R99-1}). 
Of course, $R_{99,G}$ is not the same as $R_{99,S}$: the former includes $99\%$ of the total mass of the Gaussian, the latter includes $99\%$ of the total mass of the numerical soliton, and so we will highlight the distinction by subscripts.
Now, if we insert $R_{99,G}$ into (\ref{charm}), we have
\begin{equation}
    \frac{m}{m_c} = 2.328\ ,
\end{equation}
thus smaller than (\ref{massmem}); therefore we consider a range, see (\ref{massratio}).

In later parts of this paper, it will be useful to express spatial variables of the Gaussian model in units of $\sigma$, and the dimensionless variables which result carry a tilde, i.e.
\begin{displaymath}
  \tilde{x} \equiv \frac{x}{\sigma}, 
\end{displaymath}
where $x$ stands for any spatial variable. For the cutoff radius $R_{99,G}$ we write
\begin{equation}
\label{eq:99halosize}
\Tilde{R} = \frac{R_{99,G}}{\sigma} =  2.575829 \approx 2.576\ .
\end{equation}

\subsubsection{The $(n=2)$-polytrope}

The second approximate density model to substitute for the numerical soliton is inspired by considerations of \cite{2019PhRvD.100l3506C}, where he presents the following argument as to why the density profile of an ($n = 2$)-polytrope is a particular solution to the hydrostatic version of Eq. (\ref{eq:GP}) in the fuzzy limit, and the Poisson equation (\ref{eq:P}), 
\begin{equation}
\label{eq:KQpoiss}
     \frac{\hbar^2}{2m^2} \Delta \left( \frac{\Delta\sqrt{\rho}}{\sqrt{\rho}}\right) = 4\pi G \rho\ .
\end{equation}
Before introducing the conventional dimensionless variables $\theta$ and $\xi$ and considering spherically symmetric configurations, the Lane-Emden equation (see appendix \ref{appendix-polytrope}) takes the form
\begin{equation}
\label{eq: prelane}
    K_p(n+1) \Delta \rho^{1/n} = -4\pi G \rho \ .
\end{equation}
Setting $n= 2$ in Eq. (\ref{eq: prelane}), dividing it by $\sqrt{\rho}$, applying the Laplacian operator $\Delta$ and then substituting $\Delta \sqrt{\rho}$ on the right-hand side again through Eq. (\ref{eq: prelane}) itself, yields
\begin{equation}
\label{eq: lane-GP}
      \Delta \left( \frac{\Delta\sqrt{\rho}}{\sqrt{\rho}}\right) = \left(\frac{4\pi G}{3K_p}\right)^{2} \rho\ \ .
\end{equation}
Now, this equation coincides with (\ref{eq:KQpoiss}), provided that
\begin{equation}
\label{eq:kfix}
    K_p = \left(\frac{2\pi G \hbar^2}{9m^2}\right)^{1/2} = \frac{\sqrt{2\pi G}}{3}\frac{\hbar}{m} \ .
\end{equation}
However, this result should be handled with care since Eq. (\ref{eq: prelane}) implies (\ref{eq: lane-GP}), but (\ref{eq: lane-GP}) does not imply (\ref{eq: prelane}) due to the nature of operations between them. Owing to this non-equivalency, the polytrope of index $n = 2$ with fixed polytropic constant in (\ref{eq:kfix}), in particular its density profile
\begin{equation}
\label{eq: polymodel}
    \rho(r) = \rho_c \theta(\xi)^2, \ \ \ \ \xi = r \left(\frac{ \hbar^2}{8 \pi G m^2 \rho_c}\right)^{-1/4},
\end{equation}
is a valid approximation for our purpose, but it is not an exact solution of the GPP system in the fuzzy limit. (Appendix \ref{appendix-polytrope} gives an overview on polytropic spheres and on the numerical approach to solve the Lane-Emden equation.) \autoref{fig:gauss-polytrope-soliton} shows a plot of the $(n=2)$-polytropic density profile (\ref{eq: polymodel}). Note that we use again the notation $\rho_c$ for the central density of the FDM core profile, like with the Gaussian sphere or Schive profile before. 

In addition to the above argument, we remind the reader that the mass-radius relationship in
Eq. (\ref{eq:R99-1}) is -up to prefactors- the same as the one for an ($n=2$)-polytropic sphere; denoting its radius as $R_0$ and using equ.(\ref{eq:kfix}), we arrive at
\begin{equation} \label{polyrad}
    R_0 = 5.247 \left(\frac{\hbar}{m}\right)^2\frac{1}{GM}\ 
\end{equation}
(see also eq.(\ref{eq:polyrad}) or appendix \ref{appendix-polytrope}, eq.(\ref{A12})). Thus, we find the same mass-radius relationship as before, where the prefactor differs only by $\approx 1.033$ from (\ref{r99G}), i.e. also in this respect are the Gaussian sphere and the $(n=2)$-polytropic sphere very similar to each other.

A closer look at the energies of a self-gravitating FDM object and the energy of an $(n = 2)$-polytropic sphere shows the following difference, see \cite{2019PhRvD.100l3506C}. We have seen that the quantum kinetic energy term is given by (\ref{eq:KQ}). Integration by parts yields
\begin{equation}
    K_{Q} = \frac{\hbar^2}{4m^2} \int \nabla \rho \cdot  \,\text{d}s - \frac{\hbar^2}{2m^2}\int \sqrt{\rho} \Delta \sqrt{\rho}\  \text{d}^3r\ .
\end{equation}
The surface term vanishes since (\ref{eq: polymodel}) indicates $\rho'|_{r = R} \propto \theta \theta' |_{\xi = \xi_1}$ and therefore $\rho'(R) = 0$ at the surface of the complete polytrope. By inserting (\ref{eq: prelane}) for $n = 2$, the quantum kinetic term takes the form
\begin{equation}\label{eq:K_Q-U}
    K_Q = 3K_p \int \rho^{3/2} \text{d}^3r = \frac{3}{2}U\ .
\end{equation}
The last equality follows from (\ref{eq:uint}), the internal energy that arises from a polytropic pressure of index $n=2$. In other words, we know that in the fuzzy limit there is no SI which would lead to an internal energy arising from a polytropic SI pressure. Moreover, we know that the total energy of a standard polytropic sphere, i.e. a spherical system in hydrostatic equilibrium with a density profile according to the solutions of the Lane-Emden equation (see appendix \ref{appendix-polytrope}), does not incorporate a "quantum" kinetic energy term. However, we have seen that certain operations and identifications allow to associate a polytropic density profile in the form of (\ref{eq: polymodel}) with an FDM halo core density profile. Hence, one might guess that the quantum kinetic term corresponds to an internal energy of the form of eq.(\ref{eq:uint}) and, according to \cite{2019PhRvD.100l3506C}, this is true up to a factor of $3/2$.

Unlike the numerical "soliton" or the Gaussian profile, the density profile (\ref{eq: polymodel}) has a compact support from the outset ("complete polytrope"), i.e. the density becomes zero at a finite radius. Furthermore, unlike e.g. the polytrope of index $n = 1$, used in the TF regime, the $(n=2)$-polytrope cannot be represented in a closed-form expression, which seems counterproductive at first sight. However, we can benefit from previous calculations on approximate equilibrium solutions for uniformly and non-uniformly rotating polytropes carried out by \cite*{1993ApJS...88..205L} (henceforth abbreviated \textbf{LRS93}) when constructing our second halo core model. Thus, we can exploit the sole fact that our second halo core model is based upon an $(n=2)$-polytrope, in order to derive global energy expressions of uniformly (in the sense of constant angular velocity $\Omega$) and non-uniformly (in the sense of superposed velocity fields) rotating FDM halo cores. This is described in the next section.

\section{Virialized FDM halo cores with angular momentum}\label{sec:virial-halos}

\subsection{Preliminaries}

From a cosmological point of view, we expect angular momentum to play a crucial role in halo dynamics. During their formation, haloes are tidally torqued by
larger-scale structure, which helps them acquire angular momentum in their early phases. While acquisition of angular momentum by BEC-DM haloes 
during their formation has not yet been studied, 
BEC-DM is expected to exhibit structure formation similar to standard CDM on scales much larger than the de Broglie wavelength (see e.g. \cite*{2014NatPh..10..496S}). Hence, we expect
that the same large-scale 
process by which angular momentum
is acquired by CDM haloes during their formation
also applies to the BEC-DM haloes of interest to
us here.  As in RS12, we shall express the amount
of angular momentum of a given BEC-DM 
halo in terms of
the so-called spin parameter which is the dimensionless ratio
\begin{equation}
\label{eq:lambda}
    \lambda = \frac{L |E|^{1/2}}{GM^{5/2}},
\end{equation}
where $E$ denotes the total energy, $L$ the angular momentum and $M$ the total mass of the object under question. 
  Cosmological N-body simulations of standard CDM have been analysed, using this spin-parameter to quantify the degree of rotational support of a CDM halo with net angular momentum $L$. Typical values for $\lambda$ are in the range $[0.01,0.1]$ (see e.g. \cite{2001ApJ...555..240B}; \cite{2010MNRAS.407.1338A}).
  In contrast, a self-gravitating, rotationally-supported disk has $\lambda \sim 0.4$.
  Now, by the above argument, we expect FDM haloes to have spin parameters similar to the values typical for CDM, as given in the range above. However, our analysis concerns the cores of FDM haloes and, as in CDM, the question arises as to whether there is a radial dependence of $\lambda$ within haloes. \cite{2001ApJ...555..240B} studied this question for CDM, finding that radial profiles for $\lambda$ range from flat (i.e. no spatial dependence) to the case where the $\lambda$ of the central parts is somewhat smaller than $\lambda$ for the halo at large\footnote{One can
  see that flat $\lambda$-profiles occur 
  among the simulated halo examples in Figure 15 of \cite{2001ApJ...555..240B} 
  if one approximates the halo mass profile as
  that of an isothermal sphere, for simplicity. 
  It can be argued that FDM halo envelopes are close to isothermal profiles given simulation findings, hence the flat case applies, and there is no harm in assuming the same value of $\lambda$ for the core and the halo at large.}. If the $\lambda$ of the core is smaller than that of the halo, our study will provide conservative bounds, because vortex formation will be increasingly favoured for higher $\lambda$. 
  
In any case, it follows that FDM haloes and their cores should not be rotationally supported. 
Nevertheless, given our ignorance and to provide conservative bounds, we consider a somewhat broader range in $\lambda$ than is typical for CDM, namely $\lambda \in [0.01, 0.2].$
In addition, and related to the idea that cores should be in approximate virial equilibrium, we require that the so-called $t$-parameter, $t \equiv T/|W|$, which measures the importance of the gravitational potential energy $W$ over the energy of bulk motions $T$, stays below a certain threshold for our models which we choose to be
$< 0.3$.

Now, the way we will assign a bulk angular momentum to our equilibrium models is to let them rotate with a constant angular velocity $\Omega$, and the associated angular momentum will be used in eq.(\ref{eq:lambda}). 
BEC-DM haloes experience superfluid currents once $\Omega \neq 0$, which will manifest themselves by a non-trivial phase function $S$ (see also RS12). As long as no defects like vortices appear in the velocity flow, $S$ is a smooth function which we will denote $S_0$ and the respective velocity is $\vec{v}=\hbar \nabla S_0/m$ (see Eq. (\ref{eq:v})). 

One key point is that we study configurations in equilibrium. In other words, we consider stationary solutions given by (\ref{eq: stat-sol}) and (\ref{eq: decom}), i.e. quantum states with time-independant observables which are eigenstates of the Hamiltonian, and add the distinction between different reference frames to our considerations. Rotating configurations at constant $\vec{\Omega}$ in the rest frame of the object correspond to stationary solutions in the co-rotating frame. In this co-rotating frame, the bulk velocity is given by
\begin{equation}
    \Vec{v}^{\ '} = \Vec{v}- \Vec{\Omega} \times \Vec{r},
\end{equation}
where primed quantities and variables denote those in the co-rotating frame. 

In the next section, we will consider this co-rotating frame, assuming without loss of generality that the objects rotate about the $z$-axis, i.e. $\vec{\Omega} = (0,0,\Omega)$. We will analyse whether and under what conditions vortex formation in these rotating, self-gravitating objects is energetically favoured. Our analysis requires a complete model of the equilibrium object, i.e. a geometry, density and velocity profile consistent with the demands of the GPP framework. In the following two subsections, we present the two complete models upon which our investigation is based, adopting a rotating Gaussian sphere in the first case, and a rotating, $(n=2)$-polytropic, irrotational Riemann-S ellipsoid in the second case, and in the course of that we calculate the angular momenta $L$ of objects which rotate at constant $\vec{\Omega}$, and derive useful relations which will be required in our energy analysis of the next section.

\subsection{FDM cores as rotating Gaussian spheres}\label{subsec:haloA-gauss}

This model approximates the object under question as a sphere with radius $R_{99,G}=2.576\sigma$, rotating with constant angular velocity $\vec{\Omega}=(0,0,\Omega)$ about the $z$-axis. The density profile is chosen according to the first density model (\ref{eq:toy})-(\ref{eq:a}), the Gaussian. As of now, we choose the notation $(r_s,\theta, \phi)$ for spherical coordinates, in order not to confuse them with cylindrical coordinates introduced below.
The Gaussian sphere has a bulk rotation in the rest frame, $\vec{\Omega} \times \vec{r}$, hence it shows no net velocity in the co-rotating frame,
\begin{equation}
\label{eq:corotating-vel-undisturbed}
    \Vec{v}^{\ '} = \frac{\hbar}{m}\vec{\nabla}S' = 0 \ .
\end{equation}
Now, the total angular momentum of a uniformly rotating system whose axis of rotation coincides with an axis of symmetry, is generally given by
\begin{equation}
    \vec{L} = I \vec{\Omega}\ ,
\end{equation}
where $I$ denotes the moment of inertia. For a continuous body rotating about a specified axis, it can be written as
\begin{equation}
\label{eq:inertia}
    I = \int_V \vec{r}_{\perp}^2 \rho(\vec{r})\text{d}V\ ,
\end{equation}
if we decompose the position vector into a component parallel to the axis of rotation and perpendicular to the axis of rotation. The radial distance between each mass element and the axis of rotation is given by $r_{\perp}$, provided that the centre of the mass distribution is located at the origin of the coordinate system. Thus, in spherical coordinates 
\begin{equation}
    r_{\perp}^2 = r_s^2 \sin^2 \theta\ .
\end{equation}
The total angular momentum of a uniformly rotating sphere filled with matter distributed according to our Gaussian density profile (\ref{eq:toy})-(\ref{eq:a}) is then given by
\begin{equation}
    L = |\vec{L}| = \Omega \int_V r_s^2 \sin^2 \theta\  \rho_c \text{e}^{-ar_s^2} r_s^2 \sin \theta \text{d}\theta \text{d}\phi \text{d}r_s\ .
\end{equation}
This yields
\begin{equation}
\label{eq:l_sphere}
    L = \Omega I = \Omega  \frac{8\pi}{3}\rho_c B,
\end{equation}
where
\begin{equation}
\label{eq:B-def}
 B = -\frac{R_{99,G}}{4a^2}\text{e}^{-aR_{99,G}^2}(3+2aR_{99,G}^2)+\frac{3 \sqrt{\pi}}{8 a^{5/2}}\text{Erf}\left(\sqrt{a}R_{99,G}\right),
\end{equation}
is an expression with dimension [$\text{Length}^5$]
and $\text{Erf}(x)$ denotes the Gauss error function given by the integral
\begin{equation}
\label{eq:erf-function}
    \text{Erf}(x) = \frac{2}{\sqrt{\pi}} \int_0^x \text{e}^{-t^2}dt\ .
\end{equation}

\subsubsection{Comparison to the $\lambda$-spin parameter}

 We will use the spin parameter $\lambda$, Eq. (\ref{eq:lambda}), in order to derive meaningful angular velocities for our FDM halo cores for given $\lambda \in [0.01,0.2]$. To this aim, we will express $\lambda$ as a function of the angular velocity and the DM particle mass, $\lambda = \lambda(\Omega, m)$. We use global quantities in the rest frame of the rotating object, see eq.(\ref{eq:energytot}-\ref{eq: stat-W}). The total energy is  
\begin{equation}
E = K + W\ ,   
\end{equation}
with
\begin{displaymath}
K = \int_V  \frac{\hbar^2}{2m} |\nabla \psi_0|^2 \text{d}^3r 
\end{displaymath}
\begin{equation}
\label{eq:KQ-neu}
= \int_V \frac{\hbar^2}{2m^2} (\nabla \sqrt{\rho_0})^2 \text{d}^3r +  \int_V \frac{\rho_0}{2} \vec{v}^2 \text{d}^3r \equiv K_Q + T 
\end{equation}
and
\begin{equation} \label{eq:W}
    W = \int_V \frac{\rho_0}{2} \Phi_0 \text{d}^3r\ .
\end{equation}
Given the Gaussian profile in (\ref{eq:toy}) and the definition in Eq. (\ref{eq:B-def}), the quantum kinetic energy term $K_Q$ amounts to
\begin{equation}
\label{eq:KQ-novortex}
 K_Q = \frac{\hbar^2}{m^2} \rho_c a^2 2 \pi B = \frac{\hbar^2}{m^2} \frac{3}{4} a^2 I\ .
 \end{equation}
The rotational kinetic energy incorporates the square of the bulk velocity in the rest frame, i.e.
\begin{equation}
    \vec{v}^2 = (\vec{\Omega} \times \vec{r})\ \cdot \ (\vec{\Omega} \times \vec{r}) 
    = \Omega^2r^2 - (\vec{\Omega} \cdot \vec{r})^2 \nonumber 
    = \Omega^2\  \vec{r}^{\ 2}_{\perp} \ .
\end{equation}
This yields
\begin{equation}\label{eq:Tnovortex}
    T = \Omega^2 \frac{1}{2} \int_V \rho_0 r^2_{\perp} \text{d}V =\frac{\Omega^2}{2} I\ .
\end{equation}
The moment of inertia $I$, identified by expression (\ref{eq:inertia}), is given on the right-hand side (divided by $\Omega$) of (\ref{eq:l_sphere}). 

Now, we need to calculate the gravitational potential energy $W$ in (\ref{eq:W}), for which we need to determine the gravitational potential $\Phi_0$ first. Given our Gaussian ansatz for the density, we are required to solve the Poisson equation (\ref{eq:P}), thus $\Phi_0$ will be the solution of the following boundary value problem in spherical coordinates:
\begin{subequations}
\label{eqn:alles}
\begin{eqnarray}
    \Delta \Phi_0(r_s) &=& 4\pi G \rho_c \text{e}^{-ar_s^2} \label{eqn:1}\\
\nabla \Phi_0(0) &=& 0 \label{eqn:2}\\
\Phi_0(0) &=& A, \label{eqn:3}
\end{eqnarray}
\end{subequations}
where A is a constant. Imposing regularity, Eq. (\ref{eqn:2}), is effectively equivalent to requiring the absence of gravitational force at the centre. The calculation yields
\begin{equation}
\label{eq:sol1}
    \Phi_0(r_s) = A + \frac{2\pi G \rho_c}{a}-G\rho_c \left(\frac{\pi}{a}\right)^{3/2}\frac{\text{Erf}(\sqrt{a}r_s)}{r_s},
\end{equation}
as the solution to system (\ref{eqn:alles}). Requiring a bound, isolated configuration, i.e.
\begin{equation}
\label{eq:lim}
    \lim \limits_{r_s \to \infty} \Phi_0(r_s) = 0,
\end{equation}
sets $A = -2\pi G \rho_c/a$ and yields
\begin{equation}
\label{eq:sol2}
    \Phi_0(r_s) = -G\rho_c \left(\frac{\pi}{a}\right)^{3/2}\frac{\text{Erf}(\sqrt{a}r_s)}{r_s}\ .
\end{equation}
Inserting this solution along with the Gaussian density in (\ref{eq:W}) results in  
\begin{eqnarray}
\label{eq:term2}
  W &=& 4 \pi \frac{\rho_c}{2}\int_0^{R_{99,G}} \text{e}^{-ar_s^2} \left(-G\rho_c \left(\frac{\pi}{a}\right)^{3/2}\frac{\text{Erf}(\sqrt{a}r_s)}{r_s} \right) r_s^2 \text{d}r_s \nonumber \\
   & = & -2 \pi G \rho_c^2  \left(\frac{\pi}{a}\right)^{3/2} \int_0^{R_{99,G}} \text{e}^{-ar_s^2} \text{Erf}(\sqrt{a}r_s) r_s \text{d}r_s. 
\end{eqnarray}
Collecting the expressions of (\ref{eq:l_sphere}), (\ref{eq:KQ-novortex}), (\ref{eq:Tnovortex}) and (\ref{eq:term2}), we get for the spin-parameter,
\begin{displaymath}
    \lambda = \frac{\Omega I}{GM^{5/2}}~ \times
\end{displaymath}    
\begin{equation}\label{eq:lambda2}    
    \left|\frac{\hbar^2}{4m^2} a^2 3I+\frac{\Omega^2}{2} I- 2 \pi G \rho_c^2 \left(\frac{\pi}{a}\right)^{3/2} \int_0^{R_{99,G}} \text{e}^{-ar_s^2} \text{Erf}(\sqrt{a}r_s) r_s \text{d}r_s\right|^{1/2} .
\end{equation}
The square of the spin-parameter can be written as
\begin{displaymath}
\lambda^2 = \left( \frac{\Omega}{\Omega_{\text{grav}}}\right)^2 ~\times
\end{displaymath}
\begin{equation}
\frac{9|I|^3}{16M^3 R_{99,G}^6}\left|\left( \frac{\Omega_{QM}}{\Omega_{\text{grav}}}\right)^2 \Tilde{R}^4 \frac{3}{2^4} +\frac{1}{2} \left( \frac{\Omega}{\Omega_{\text{grav}}}\right)^2  - \frac{R_{99,G}^3\sigma^2}{B}C\right|,
\end{equation}
with the gravitational angular velocity\footnote{In RS12, $\Omega_{\text{grav}}$ was denoted as $\Omega_G$.} defined as
\begin{equation}\label{ograv1}
    \Omega_{\text{grav}} \equiv \sqrt{\frac{3GM}{4R_{99,G}^3}} 
\end{equation}
and $M$ and $R_{99,G}$ are the mass and radius of the rotating Gaussian sphere and $\Tilde{R}$ in (\ref{eq:99halosize}).
Furthermore,
\begin{equation}
    C= \int_0^{\Tilde{R}}\text{e}^{-\frac{\Tilde{r}_s^2}{2}} \text{Erf}\left(\frac{\Tilde{r}_s}{ \sqrt{2}}\right)\Tilde{r}_s \text{d}\Tilde{r}_s
\end{equation}
denotes a dimensionless quantity and $B$ is given in Eq. (\ref{eq:B-def}). Inserting
\begin{equation}
    \frac{\Omega_{QM}}{\Omega_{\text{grav}}} = \frac{\hbar}{m} \frac{2}{\sqrt{3GMR_{99,G}}} = \frac{m_H}{m} = \frac{2}{\sqrt{3}}\frac{m_c}{m}\ ,
\end{equation}
and defining
\begin{equation} \label{eq:barOmega}
    \Bar{\Omega} \equiv \frac{\Omega}{\Omega_{QM}}
\end{equation}
finally yields
\begin{eqnarray}\label{eq:lambda-lang}
\lambda^2 &=& \Bar{\Omega}^2 \left(\frac{m_c}{m} \right)^2 \frac{4 \sqrt{2}}{9\pi^{3/2}} \times \nonumber \\ & &\left|-\frac{3+\Tilde{R}^2}{\Tilde{R}}\exp(-\Tilde{R}^2/2)+\frac{3\sqrt{\pi}}{8}2^{5/2} \frac{\text{Erf}(\Tilde{R}/ \sqrt{2})}{\Tilde{R}^2} \right|^3 \times \nonumber \\
& & \left|\left(\frac{m_c}{m} \right)^2 \frac{\Tilde{R}^4}{4}+ \Bar{\Omega}^2 \frac{2}{3}\left(\frac{m_c}{m} \right)^2 -R_{99,G}^3 \sigma^2\frac{C}{B}\right|.
\end{eqnarray}
We can see that $\lambda$ is a cumbersome function of $(\Bar{\Omega},\Tilde{R}, m/m_c) $. Since the final goal is to derive values for $\Omega$ from given $\lambda$-values in the range $[0.01,0.2]$, we calculated the roots of Eq. (\ref{eq:lambda-lang}) for given $\lambda^2$ and $m/ m_c$. Since Eq. (\ref{eq:lambda-lang}) is a non-linear function in $\Bar{\Omega}$, several values for $\Bar{\Omega}$ may correspond to one $\lambda$-value, depending on $m/m_c$. We deal with this multitude of solutions by selecting only those whose $t$-parameters fulfill $t < 0.3$. As an illustration for the numbers we get, let us consider e.g. $m/m_c = 2\pi$, see (\ref{eq: mbound}). Then, values of $\lambda = (0.01, 0.05, 0.1, 0.15, 0.2)$ correspond to angular velocities $\Bar{\Omega} = (0.34654, 1.74409, 3.56718, 5.60416, 8.34664)$, respectively. The corresponding $t$-parameters are $t=(4.991\cdot 10^{-4}, 0.0126421, 0.052885, 0.130528, 0.289537)$, respectively.
For the sake of an easier comparison later, we also quote here the corresponding angular velocities in units of $\Omega_\text{grav}$: $\Omega/\Omega_{\rm grav} = (0.06368, 0.32051, 0.65554, 1.02988, 1.53386)$.

\subsection{FDM cores as rotating, $(n=2)$-polytropic, irrotational Riemann-S ellipsoids}\label{subsec:riemann}

DM haloes presumably acquire angular momentum by tidal torques caused by large-scale structure, during their formation. \textit{Non-spherical} equilibrium configurations are thus expected. In general, ellipsoidal shapes are expected for any object with non-vanishing angular momentum, yet spherical models are often employed due to their simplicity.
Indeed, the rotating Gaussian sphere of our first model has the great advantage of being an analytical model for the ground-state, and this will be of substantial help in the next section, when we employ the energy analysis of the perturbed system with vortex.  However, the Gaussian sphere suffers not only that is has a spherical shape, but also that it is not strictly irrotational in the rest frame. Therefore, we consider a further model, namely the irrotational Riemann-S ellipsoid, which is strictly irrotational in the rest frame (prior to any possible vortex formation), because on top of a uniform rotation, an internal velocity field is superposed, which combine to yield a vanishing net vorticity in the rest frame.

Exact solutions exist for Riemann-S ellipsoids
only for the case of uniform density, but fortunately, as we
shall see below, LRS93 developed compressible (non-uniform) generalizations of Riemann-S ellipsoids (and other classical figures of rotation) by using their "ellipsoidal approximation", see appendix \ref{appendix-riemann}. In fact, their approximate solutions for the compressible cases agree well with the true equilibria. 

Irrotational Riemann-S ellipsoids have been applied to haloes in RS12 for the first time.
RS12 showed that, in general, a rotating ellipsoidal halo
cannot be both axisymmetric and irrotational, if it is non-singular at the origin, and it was this insight which prompted their application of the irrotational Riemann-S ellipsoid.
As an equation-of-state for this Riemann-S ellipsoid, RS12 used an $(n=1)$-polytrope, which is appropriate for the TF regime, where the pressure due to SI (see Eq.(\ref{eq:PSI})) is the dominant source for hydrostatic equilibrium.
The results of LRS93 were then used in order to calculate the global energies of such ellipsoids.

We have already established in the previous section, that the $(n = 2)$-polytropic density profile is an appropriate approximation to the actual "soliton" in the fuzzy regime, hence we paved the way for applying the results of LRS93 for $(n=2)$-polytropic, irrotational Riemann-S ellipsoids, in this work.
However, by nature the classical study of ellipsoidal equilibrium figures by LRS93 does not include the quantum-kinetic energy $K_Q$. Therefore, we require some extra care by adopting Eq. (\ref{eq:K_Q-U}) as will be seen in \autoref{subsec:B-riemann}.

\cite{1969efe..book.....C} revisits and elaborates on the incompressible Riemann ellipsoids - homogeneous fluid masses which maintain their ellipsoidal configuration under rotation and under their own gravity at all times with a velocity field that is given by the sum of a uniform rotation with angular velocity $\Vec{\Omega}$ and internal motions with uniform vorticity $\Vec{\zeta}^{\ '}$. In the case of so-called Riemann-S ellipsoids, the vectors $\Vec{\Omega}$ and $\Vec{\zeta}^{\ '}$ are both oriented along the same principal axis and one can define sequences along which the ratio
\begin{equation}
    \label{eq: l-frac}
    f_R = \frac{\zeta'}{\Omega}\ 
\end{equation}
is constant,
where $\Omega = |\Vec{\Omega}|$ and $\zeta' = |\Vec{\zeta}^{\ '}|$. The geometry of this non-axisymmetric object is given by its three semi-axes $(a_1,a_2,a_3)$ along Cartesian coordinates $(x,y,z)$, or equally by its eccentricities
\begin{equation}
        \label{eq:eccen-def}
        e_1 = \sqrt{1-(a_2/a_1)^2}\ \ \ \text{and}\ \ \ e_2= \sqrt{1-(a_3/a_1)^2}\ .
\end{equation}
Also, these ellipsoids fulfil $a_1 \geq a_3 \geq a_2$ , i.e. they are all
prolate bodies (see also \autoref{fig:ellipsoids} in appendix \ref{appendix-riemann}).

Now, by applying an energy variational method and their ellipsoidal approximation (see appendix \ref{appendix-riemann} for a summary), LRS93 developed generalized Riemann-S ellipsoids in the sense that they considered density and pressure profiles according to those of a polytrope of index $n$. As a result, the approximate internal and gravitational potential energy of polytropic Riemann-S ellipsoids can be written as
\begin{equation}
\label{eq:u-sperical}
    U = k_1 K_p \rho_c^{1/n}M
\end{equation}
and
\begin{equation}
\label{eq:W-riemann}
     W = - \frac{3}{5-n} \frac{GM^2}{R_R} f = -k_2 \rho_c^{1/3}GM^{5/3} f \ ,
\end{equation}
respectively, where the constants $k_1$, $k_2$ (depending upon $n$) and the dimensionless ratio $f$ (depending upon the geometry) are given in appendix \ref{appendix-riemann}, see eq.(\ref{eq:k1}, \ref{eq:k2}) and (\ref{eq: f-ratio-riemann}). Here, $R_R$ denotes the mean radius of the ellipsoid\footnote{The subscript $R$ stands for "Riemann" and emphasizes that the system is being described by the mean radius defined in Eq. (\ref{eq:Rmean}), as opposed to the radius $R_{99,G}$ of the Gaussian sphere.}, given by
\begin{equation}
    \label{eq:Rmean}
        R_R = (a_1a_2a_3)^{1/3}\ .
\end{equation}
The geometry and velocity field of this non-axisymmetric Riemann-S ellipsoid in equilibrium are closely related as follows. Let us denote unit vectors $\Vec{e}_1, \Vec{e}_2, \Vec{e}_3$ along $(x,y,z)$. One starts with an object that rotates rigidly with constant angular velocity $\Vec{\Omega} = \Omega \Vec{e}_3$ about the $z$-axis. Then, one superposes an internal velocity field with uniform vorticity parallel to $\Vec{\Omega}$,
\begin{equation}
\label{eq:vel-1}
    \zeta' \equiv (\Vec{\nabla}^{\ '} \times \Vec{v}^{\ '}) \cdot \Vec{e}_3\ ,
\end{equation}
specified by the requirement that the resulting velocity vector at any point shall be tangent to the isodensity surface at that point. From this follows that the fluid velocity in the rest frame can be written as
\begin{equation}
\label{eq:vel-2}
    \Vec{v} = \Vec{v}^{\ '} + \Vec{\Omega} \times \Vec{r}\ ,
\end{equation}
where
\begin{equation}
\label{eq:internal-vel-riem}
    \Vec{v}^{\ '}  = - \frac{a_1^2}{a_1^2+a_2^2} \zeta'\  y \Vec{e}_1 + \frac{a_2^2}{a_1^2+a_2^2}\zeta'\  x \Vec{e}_2\ .
\end{equation}
In addition, there is a relation between the angular frequency of the internal motions $\Lambda$ and the vorticity,
\begin{equation}
\label{eq:xi-lambda}
    \zeta' = - \frac{a_1^2+a_2^2}{a_1a_2} \Lambda\ .
\end{equation}
By means of the above relations, and assuming polytropes of index $n$, LRS93 find for the angular momentum $\Vec{L}$ and rotational kinetic energy $T$,
\begin{equation}
\label{eq:L-riemann}
    \Vec{L} = \int \Vec{r} \times \Vec{v} \rho \ \text{d}^3r = \left(I\Omega -\frac{2}{5}\kappa_n M a_1 a_2 \Lambda \right) \Vec{e}_3
\end{equation}
and
\begin{eqnarray}
\label{eq:T-riemann}
    T &=& \frac{1}{2} 
    \int \Vec{v} \cdot \Vec{v} \rho~ \text{d}^3r \\
    &=& \frac{\kappa_n}{20}M \left((a_1 -a_2)^2(\Omega+\Lambda)^2+(a_1 +a_2)^2(\Omega-\Lambda)^2\right)\ , \nonumber
\end{eqnarray}
respectively. The moment of inertia $I$ is given by
\begin{equation}
\label{eq:I-riemann}
    I = \frac{\kappa_n}{5}M (a_1^2 +a_2^2)
\end{equation}
and the definition of the constant $\kappa_n$ and its values for $n \in \{1,2\} $ can be found in Eq. (\ref{eq:kappan}).

A bound BEC-DM object described by the GPP framework is irrotational, whenever vortices are not present. In fact, the construction of Riemann-S ellipsoids allows to guarantee irrotationality of the object in its rest frame, given a certain choice of the parameter $f_R$ in eq.(\ref{eq: l-frac}). This can be shown by introducing the circulation along the equator,
\begin{equation}
    \Gamma_{\text{equator}}  \equiv \oint_{\text{equator}} \Vec{v} \cdot \text{d}\Vec{l} = \pi (2+f_R)a_1a_2 \Omega\ ,
\end{equation}
and the vorticity in the rest frame
\begin{equation}
    \zeta \equiv (\Vec{\nabla} \times \Vec{v}) \cdot \Vec{e}_3 = (2+f_R)\Omega\ .
\end{equation}
Along the so-called \textit{irrotational} Riemann-S sequence the ratio $f_R = -2$ yields $\zeta = 0 = \Gamma_{\text{equator}}$, as required. 
\autoref{fig:velocity-fields} in appendix \ref{appendix-riemann} illustrates how the defining internal velocity field of the irrotational Riemann-S ellipsoid with $n=2$ guarantees zero vorticity in the rest frame. RS12 provide similar visualisations of the velocity fields (their figure 2) for polytropic index $n=1$.

Two crucial equilibrium conditions can be obtained upon
extremizing the total energy $E = U + T + W$ of the Riemann ellipsoid with respect to all variations of the central value of the polytropic density profile and the axis ratios. For $f_R = -2$, these conditions read
\begin{equation*}
     \frac{4(a_2/a_1)^2}{(1+(a_2/a_1)^2)^2}-
\end{equation*}
\begin{equation}
\label{eq: axis-rel-2}
    \frac{4B_{12}(a_2/a_1)^2}{(a_3/a_1)^2A_3-(a_2/a_1)^2\frac{A_1-A_2}{(a_2/a_1)^2-1}}\frac{1}{1+(a_2/a_1)^2}+1 = 0
\end{equation}
and
\begin{equation}
\label{eq:omega-grav}
    \Tilde{\Omega} = \left(\frac{2B_{12}}{q_n}\right)^{1/2} \left( 1+ \frac{4a_1^2a_2^2}{(a_1^2+a_2^2)^2}\right)^{-1/2}\ .
\end{equation}
Here, we defined 
\begin{equation}\label{eq:tom}
\Tilde{\Omega} = \frac{\Omega}{\Omega_{\text{grav},R}}\ ,
\end{equation} 
where the gravitational angular velocity of the ellipsoid is also defined as above,
\begin{equation}\label{eq:omgrav}
    \Omega_{\text{grav},R} \equiv \sqrt{\pi G \Bar{\rho}} = \sqrt{\frac{3GM}{4R_R^3}}\ ,
\end{equation}
but now using the mean radius $R_R$ in (\ref{eq:Rmean}).
The geometry-dependent quantities $A_1, A_2, A_3, B_{12}$ and $q_n$ (depending upon $n$) are given in the appendices, (\ref{eq:B12}) and (\ref{eq:qnvalues}). Thus, in case of equilibrium, one axis ratio determines the other one, thereby immediately fixing the geometry and furthermore $\Tilde{\Omega}$. Moreover, equilibrium yields an important relation between the radius of the non-rotating spherical polytrope of \textit{same mass} $M$, $n$ and $K_p$ as the Riemann ellipsoid, which is
\begin{equation} \label{eq:polyrad}
    R_0 = \xi_1 (\xi_1^2|\theta_1'|)^{-\frac{1-n}{3-n}}\left( \frac{M}{4\pi}\right)^{\frac{1-n}{3-n}} \left( \frac{(n+1)K_p}{4\pi G}\right)^{\frac{n}{3-n}} \ ,
\end{equation}
(see also (\ref{A12})),
and the mean radius in (\ref{eq:Rmean}). The relationship between the two radii reads
\begin{equation}
    R_R = R_0 \left[ f\left( 1- 2\frac{T}{|W|}\right) \right]^{-n/(3-n)}\ .
\end{equation}
A third equilibrium condition is the virial relation Eq. (\ref{eq:virial_sf}) which yields the total equilibrium energy
\begin{equation}
\label{eq: E-equi}
    E_{eq} = \frac{3-n}{n} W \left( 1- \frac{3-2n}{3-n}\frac{T}{|W|}\right)\ ,
\end{equation}
see LRS93 for details.

Inserting $n=2$, the mean radius of the ellipsoid is
\begin{equation} \label{eq:meanrad}
    R_R = R_0 \left[ f\left( 1- 2\frac{T}{|W|}\right) \right]^{-2} \equiv R_0 g(e_1,e_2)^{-2}\ ,
\end{equation}
where $g(e_1,e_2) \equiv f(e_1,e_2)(1-2T/|W|)$, since $f$ in (\ref{eq: f-ratio-riemann}) can be written as a function of the axis ratios, or equivalently as a function of the eccentricities. 
For increasing angular momentum (parameterized using $\lambda$ below), the mean radius $R_R$ in (\ref{eq:meanrad}) increases with respect to the spherical radius $R_0$ (see also \autoref{tab:lambda-table}). \\
On the other hand, the total equilibrium energy for $n=2$ is 
\begin{equation}
    E_{eq} = \frac{1}{2} W \left( 1 + \frac{T}{|W|}\right)\ .
\end{equation}

\subsubsection{Comparison to the $\lambda$-spin parameter}

Now, we have everything in place to make the connection between Riemann-S ellipsoid and spin parameter. We can write $\lambda$ as
\begin{equation}
    \lambda = \frac{L |W|^{1/2}}{GM^{5/2}} \left|\frac{E_{eq}}{W}\right|^{1/2}\ ,
\end{equation}
where 
\begin{equation}
\label{eq:LW}
    \frac{L |W|^{1/2}}{GM^{5/2}} = \frac{\kappa_2}{5}(\Omega(a_1^2+a_2^2)-2a_1a_2\Lambda)\left( \frac{|f(e_1,e_2)|}{GMR_R}\right)^{1/2}
\end{equation}
follows from inserting (\ref{eq:W-riemann}) and (\ref{eq:L-riemann}). A division of $E_{eq}$ by the gravitational potential energy amounts to
\begin{equation}
    \label{eq:E-W}
    \left|\frac{E_{eq}}{W}\right|^{1/2} = \left(\frac{1}{2}(1+t)\right)^{1/2}\ ,
\end{equation}
and again, $t \equiv T/|W|$. Rewriting and multiplying the expressions (\ref{eq:LW}) and (\ref{eq:E-W}) yields
\begin{equation*}
   \lambda = \frac{\Tilde{\Omega} \kappa_2 \sqrt{3}}{10} \left(\frac{1}{2}(1+t)\right)^{\frac{1}{2}}\left|f(e_1, e_2)\right|^{\frac{1}{2}} \times
\end{equation*}
\begin{equation}
\label{eq:lambda-final}
  \ \ \ \frac{(1-(a_2/a_1)^2)^2}{1+(a_2/a_1)^2}\left(\frac{a_2}{a_1}\right)^{-\frac{2}{3}}\left(\frac{a_3}{a_1}\right)^{-\frac{2}{3}},
\end{equation}
with
\begin{equation}
   \label{eq:t-final}
   t = \kappa_2 \frac{3}{40}|f(e_1,e_2)|^{-1}\frac{e_1^4}{(1-e_1^2)^{1/3}(1-e_2^2)^{1/3}(2-e_1^2)}\ . 
\end{equation}
Thus, specifying the spin-parameter $\lambda$ allows us to determine \textit{both} axis ratios self-consistently by solving a system of equations consisting of relation (\ref{eq: axis-rel-2}) and (\ref{eq:lambda-final})-(\ref{eq:t-final}).
Appendix \ref{appendix-riemann} demonstrates how setting a value for the spin-parameter $\lambda$ effectively fixes the geometry of the ellipsoid and furthermore several dimensionless global quantities such as $\Tilde{\Omega}$ in Eq. (\ref{eq:omega-grav}). Thereby, $\lambda = (0.01, 0.05, 0.1, 0.15, 0.2)$ correspond to angular velocities $\Tilde{\Omega} = (0.55513, 0.55659, 0.55266, 0.54376, 0.53113)$, respectively (see \autoref{tab:lambda-table} for the corresponding $t$-parameters and other quantities).
Note that these numbers vary not as much as the corresponding numbers $\Omega/\Omega_{\rm{grav}}$ of the Gaussian sphere model at the end of \autoref{subsec:haloA-gauss}; moreover, they are not monotonic, due to the nature of the irrotational Riemann-S ellipsoid.

\section{Stability of rotating FDM halo cores to vortex formation}
\label{sec:energyanalysis}

This section includes the most important parts of this paper,
namely the study of the conditions for vortex formation within rotating FDM halo cores.
We study the necessary and sufficient conditions for vortex formation, and we will show that vortices will \textit{not} form for any FDM parameters. The failure to fulfil the necessary condition excludes already a substantial part of the FDM parameter space, while the rest is excluded by the failure to fulfil the sufficient condition. We study these conditions for two different kinds of equilibrium models, and our conclusions are the same: solitonic cores are not subject to vortex formation.  We will discuss in \autoref{sec:con} why this result has been anticipated and that it is in accordance with (and not in contradiction to) simulation results in the literature.

\subsection{Necessary condition for vortex formation: $L \geq L_{QM}$}
\label{sec:neccond}

It is known from laboratory BECs and superfluids that a vortex will arise only, once an applied rotation surpasses a critical value. In our context, there is a minimum amount of angular momentum necessary for one singly-quantized vortex to appear, namely $L_{QM}$ in equ.(\ref{eq:LQM}).
As in RS12, we first derive a relationship between the angular momentum $L$ of our models and $L_{QM}$. 

If we divide $L$ in equ.(\ref{eq:l_sphere}) by $L_{QM}$, we get the following relationship for the rotating Gaussian sphere
\begin{equation}\label{Lgauss}
\frac{L}{L_{QM}} = l(\Tilde{R})\ \Bar{\Omega} = \frac{\sqrt{3}}{2} l(\Tilde{R}) \frac{m}{m_c} \frac{\Omega}{\Omega_{\text{grav}}}\ , 
\end{equation}
where
\begin{equation} 
l(\Tilde{R}) = \frac{1}{3}\sqrt{\frac{8}{\pi}}  
\left[-\frac{1}{\Tilde{R}}\text{e}^{-\Tilde{R}^2/2} (3+\Tilde{R}^2) +\frac{3 \sqrt{\pi}}{8\Tilde{R}^{2}} 2^{5/2} \text{Erf}\left(\frac{\Tilde{R}}{\sqrt{2}}\right) \right]\ ,
\end{equation}
with $\Bar{\Omega}$ and $\Omega_{\text{grav}}$ defined in (\ref{eq:barOmega}) and (\ref{ograv1}), respectively.  
The value of $l(\Tilde{R}) = l(2.576) = 0.226298$ is fixed by our choice of $R_{99,G}$ in (\ref{eq:99halosize}). Therefore, the ratio of $L/L_{QM}$ is determined solely by $\Bar{\Omega}$, or equivalently by the product of $m \Omega/(m_c \Omega_{\text{grav}})$. 
Requiring 
\begin{equation}
\frac{L}{L_{QM}} \geq 1 
\end{equation} 
implies a lower bound on $\Bar{\Omega}$, namely
\begin{equation}
\label{eq:lower-omega}
    4.41894 = \frac{1}{l(2.576)} \leq \Bar{\Omega}\ .
\end{equation}
Let us put this in perspective with the above values for $\Bar{\Omega}$, for $m/m_c = 2\pi$ and a range of spin-parameters, at the end of \autoref{subsec:haloA-gauss}. The corresponding angular momenta are $L/L_{QM} = (0.0784214,0.394685,0.807247,1.26821,1.88883)$.
We can see that $\lambda > 0.1$ is required in order to fulfill the necessary condition for vortex formation for this choice of $m/m_c$. 

However, in general, the higher $m/m_c$, the higher is $L/L_{QM}$ for a fixed $\lambda$, and the necessary condition for vortex formation will be increasingly fulfilled.    

\begin{figure}
    \includegraphics[width=0.9\columnwidth]{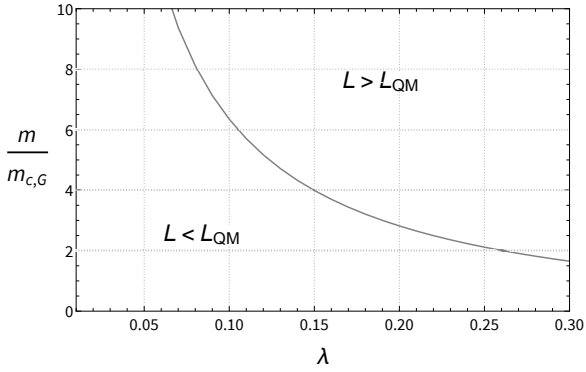}
    \centering
	\caption{Particle mass $m$ in units of $m_{c,G}$ according to (\ref{massratio}), as a function of the spin-parameter $\lambda$ for Riemann-S models with fixed angular momentum $L = L_{QM}$, according to eq.(\ref{eq:total-l-lqm}). Models with parameter combinations below the curve have $L < L_{QM}$, hence do not fulfill the necessary condition for vortex formation, while models with parameter combinations above the curve fulfil $L > L_{QM}$. (The reader may compare this plot with the left-hand panel of figure 3 of RS12, but note the difference between linear and logarithmic scaling between the figures; also $m/m_H \approx 0.866\  m/m_{c,G}$.)}
	\label{fig:m-lambda}
\end{figure}

For the Riemann-S ellipsoid, we can proceed in an analogous manner: combining the relations (\ref{eq:L-riemann}), (\ref{eq:I-riemann}) and setting $f_R = -2$ yields $L$; dividing it by $L_{QM}$ results in the expression
\begin{equation}
\label{eq:total-l-lqm}
    \frac{L}{L_{QM}} = \frac{\kappa_2}{5} \frac{m}{m_{c,R}} \frac{\sqrt{3}}{2} \Tilde{\Omega} \frac{e_1^4}{(1-e_1^2)^{1/3}(1-e_2^2)^{1/3}(2-e_1^2)}\ ,
\end{equation}
with $\Tilde{\Omega}$ in (\ref{eq:omega-grav}-\ref{eq:tom}).
It shows that for fixed eccentricities - which also fix $\Tilde{\Omega}$ -, and fixed polytropic index (here $n=2$), the amount of angular momentum (in units of $L_{QM}$) depends only on the DM particle mass. Here, we expressed the DM mass in units
of $m_{c,R}$, defined as
\begin{equation}
\label{eq:mcrcg}
m_{c,R} = \frac{\hbar}{\sqrt{R_{99,G} GM}}g(e_1,e_2) = m_{c,G} g(e_1,e_2) ,
\end{equation}
where $R_{99,G}$ and $m_{c,G}$ refer to the radius and particle mass of the Gaussian values before, in order to compare on an equal footing, i.e. $m_{c,G}$ just corresponds to $m_c$.  This way, (\ref{eq:total-l-lqm}) contains the same quantity than (\ref{Lgauss}), modified by a geometry-dependent factor $g(e_1,e_2)$. 
Different from the Gaussian sphere, the geometry of the Riemann-S ellipsoid is fixed by $\lambda$ and therefore, only the DM particle mass is left to determine $L/L_{QM}$. The latter grows for increasing $m/m_{c,R}$ or $m/m_{c,G}$, respectively, i.e.
a sufficiently high particle mass ratio is required to sustain at least one vortex. For even higher ratios,
more angular momentum than needed for one vortex can be provided. Of course, this general trend is the same as found in the case of the $(n=1)$-polytropic Riemann-S ellipsoid of RS12. In fact, if we were to plot equ.(\ref{eq:total-l-lqm}) for different values of $L/L_{QM} = (1,10,100)$ with $m/m_{c,G}$ as a function of $\lambda$, it would look almost indistinguishable from figure 3 (left-hand panel) in RS12, because $L/L_{QM}$ depends only very weakly on the polytrope index $n$. Thus, a range of high $m/m_{c,G}$ (equivalently high $m/m_{c,R}$) would correspond to a range of high $m/m_{H}$ in the notation of RS12. In fact, the analysis in RS12 is valid only for high numbers of $m/m_{H}$ (or $m/m_{c,G}$, respectively), but this is not the case anymore for the FDM regime which we study here. We have to content to small range in $m/m_{c,G}$, see (\ref{massratio}), which is reflected in the plots of our figures. 
 As an illustration, we show in \autoref{fig:m-lambda} the ratio $m/m_{c,G}$ as a function of $\lambda$ for which $L/L_{QM} = 1$, according to equ.(\ref{eq:total-l-lqm}-\ref{eq:mcrcg}). The curve which results yields a lower bound on the particle mass, for a given $\lambda$-value of the Riemann-S ellipsoid, \textit{above of which} the necessary condition for vortex formation is fulfilled.

In short, if the size of the halo core is of the order of the equilibrium radius in the absence of rotation, according to (\ref{eq:R99-1}), (\ref{r99G}) or (\ref{polyrad}), as we expect for small $\lambda$-values of interest for FDM haloes, then $m/m_c$ is limited to the lower end of the range in (\ref{massratio}) and in that case $L/L_{QM} < 1$, implying that the necessary condition for vortex formation is not met. Indeed, the smaller $\lambda$ is, the larger is the required value of $m/m_c$ in order to make $L = L_{QM}$ and we get to the upper end of (\ref{massratio}). The detailed thresholds for given $\lambda$ depend upon the model (Gaussian sphere versus Riemann-S ellipsoid), but
the trend is the same. This behaviour is a fundamental reflection of
the nature of FDM cores and distinguishes them from the cores
studied in RS12 for the TF regime, for which the size was related, instead, to the equilibrium radius for the ($n = 1$)-polytrope when SI pressure supports the cores against gravity, instead of the quantum pressure of FDM. In the TF regime, the corresponding ratio $m/m_c$ was required to be much higher than the small range of (\ref{massratio}) for FDM.

\subsection{Energy analysis}

Our energy analysis is based upon the comparison between the total energy of the unperturbed system (without a vortex) and the system with a vortex, all in the co-rotating frame. More precisely, the question will be whether the amount of angular momentum quantified by those spin-parameter values, which fulfill the necessary condition of vortex formation, is sufficient to make vortex formation also energetically favoured.  Again, we stress that the vortex is required to be energetically favoured in the co-rotating frame of reference, where the wavefunction with vortex is stationary.

In the co-rotating frame, the angular momentum is given by
\begin{equation}
    \Vec{L}^{\ '} = -i\hbar \Vec{r}^{\ '} \times \Vec{\nabla}^{\ '}
\end{equation}
and the GP energy functional in the fuzzy limit ($g = 0$) is
\begin{equation}
\label{eq:GPstatrot}
    E'[\psi'] =
\end{equation}
\begin{equation*}
     \int_V \left[ \frac{\hbar^2}{2m} |\Vec{\nabla}^{\ '} \psi'|^2 + \frac{m}{2} \Phi |\psi'|^2 +i \hbar \psi'^* \Vec{\nabla}^{\ '}\psi' \cdot (\Vec{\Omega} \times \Vec{r}^{\ '} ) \right] \text{d}^3r' 
\end{equation*}
(compare this to eq.(\ref{eq:GPstat}) for the rest frame).
In the course of the following energy analysis, we will use this expression for the energy in the co-rotating frame. For brevity, the primes on variables indicating the co-rotating frame will be omitted, except on phase functions and energies.

\subsection{Halo model A: The perturbed Gaussian sphere}\label{sec:sphere}

\subsubsection{Energy splitting and vortex ansatz}

For this model, we construct a wavefunction which accounts for a vortex which disturbs the original Gaussian sphere, calculate the total energy arising from this wavefunction by means of the GP energy functional in the fuzzy limit (\ref{eq:GPstatrot}), identify those energy terms that arise due to the vortex and determine whether they lower or raise the total energy for given parameters of the system and the DM particle. We thereby assume that the Gaussian sphere with its vortex remains a viable approximate solution for the SP system. We call it halo model A.

To this aim, the wavefunction $\psi$ of the object in equilibrium is decomposed into an unperturbed and a vortex-part, as follows
\begin{equation}
\label{eq:wave-ansatz}
    \psi = \psi_0 w = |\psi_0| |w| \text{e}^{iS_0'+S_1'},
\end{equation}
where "0"-indices indicate variables of the unperturbed system, $\psi_0 = |\psi_0|\text{e}^{iS_0' }$, and the vortex is included by means of the ansatz $w = |w|\text{e}^{iS_1' }$. Of course, the appearance of a vortex affects the density and the gravitational potential of the initially unperturbed (vortex-free) system. This raises the question how to identify the density and gravitational potential associated with the perturbation due to the vortex. The total density may be decomposed into
\begin{equation}
    \rho = \rho_0 + \rho_1\ ,
\end{equation} 
and thanks to the linearity of the Laplace operator in the Poisson equation, we have
\begin{equation}
\label{eq:potentiale}
    \Delta\Phi =  \Delta\Phi_0 +  \Delta\Phi_1 = 4\pi G (\rho_0 + \rho_1)\ .
\end{equation} 
The fact that the density of the unperturbed halo core is given by
\begin{equation}
    \rho_0 = m |\psi_0|^2\ ,
\end{equation}
yields
\begin{equation}
    \rho_1 = \rho - \rho_0 = \rho_0(|w|^2-1)
\end{equation}
and hence
\begin{equation}
    \Delta\Phi_1 = 4\pi G \rho_0(|w|^2-1)
\end{equation}
must be solved for the gravitational potential associated to the distortion of the density, brought about by the vortex.

We apply the same method as in RS12 (derived in detail in their Appendix B) to arrive at a convenient splitting of the energy contributions into vortex-free and vortex-carrying parts. According to that method, inserting the ansatz (\ref{eq:wave-ansatz}) into the energy functional (\ref{eq:GPstatrot}) yields
\begin{eqnarray}
\label{eq:energy-split}
E'[\psi] &=& E'[\psi_0] + G_{\rho_0}'[w]-R_{\rho_0}'[w],
\end{eqnarray}
where
\begin{displaymath} 
E'[\psi_0] = \int_V \left[ \frac{\hbar^2}{2m} (\Vec{\nabla} |\psi_0|)^2 + \frac{m}{2} \Phi_0 |\psi_0|^2 \right]\text{d}V
\end{displaymath}
\begin{equation} \label{eq:E0_split}
    + \int_V \left[ \frac{\hbar^2}{2m} |\psi_0|^2 \Vec{\nabla} S_0' \cdot  \left(\Vec{\nabla} S_0'-\frac{2m}{\hbar} \Vec{\Omega} \times \Vec{r}\right) \right] \text{d}V\ , 
    \end{equation}
is the vortex-free energy, while
\begin{displaymath}
    G_{\rho_0}'[w] = \int_V \left[ \frac{\hbar^2}{2m^2}  \rho_0 |\Vec{\nabla} w|^2+ \frac{\rho_0}{2} \Phi_0 \right] \text{d}V 
    \end{displaymath}
\begin{equation} \label{eq:Gf}
    +\int_V \left[  -\frac{\rho_0}{2} \Phi_0|w|^2+ \frac{\rho_0}{2} \Phi_1|w|^2 \right] \text{d}V
\end{equation}
and
\begin{equation}
    \label{eq:Rf}
    R_{\rho_0}'[w] = \int_V  \frac{\hbar^2}{2m^2}  \rho_0\ i\ w^*\ \Vec{\nabla} w \cdot \left(\Vec{\nabla} S_0'-\frac{m}{\hbar} \Vec{\Omega} \times \Vec{r}\right)  \text{d}V\ 
\end{equation}
are the energies due to the vortex.
An FDM halo core with one central vortex is energetically favoured, as compared to an unperturbed, vortex-free one, if
\begin{equation} \label{eq:endiff}
    \delta E' \equiv G_{\rho_0}'[w] -R_{\rho_0}'[w] 
\end{equation}
is negative, $\delta E' < 0$,
i.e. the total energy of the system (\ref{eq:energy-split}) is reduced through the vortex. Our aim is to calculate $\delta E'$ as a function of the parameters which define this model, and in order to do so, we apply our density models and a vortex wavefunction. 
For this halo model A, we choose the Gaussian density (\ref{eq:toy})-(\ref{eq:a}) as the unperturbed background in (\ref{eq:wave-ansatz}), namely
\begin{equation}
        \label{eq:toy2-calc}
        |\psi_0|^2 = \frac{\rho_0}{m} =  \frac{\rho_c }{m} \text{e}^{-ar_s^2} = \frac{\rho_c }{m} \text{e}^{-\frac{(r^2+z^2)}{2\sigma^2}} \ .
\end{equation}
For the vortex in (\ref{eq:wave-ansatz}), we pick the ansatz in cylindrical\footnote{The reader might wonder about the choice of notation in this paper, considering the radial lengths in spherical and cylindrical coordinates, $r_s$ and $r$, respectively. For brevity, we have chosen $r$ to denote the radial length in cylindrical coordinates in this section, because most integrations will be done in cylindrical coordinates, due to the ubiquity of the vortex wavefunction $w(r,\phi)$.} coordinates $(r,\phi,z)$,
\begin{equation}
    \label{eq:vortex2}
    w(r,\phi) = |w|(r)\text{e}^{id\phi},
\end{equation}
with amplitude 
\begin{equation}
    \label{eq: amplitude}
    |w|(r) =
    \begin{cases}
 1 & \text{for } r \geq s\ , \\
C_n\left( \frac{r}{s}\right) & \text{otherwise }\ . \\
\end{cases}
\end{equation}
 This ansatz corresponds to a cylindrical-symmetric vortex along the axis of rotation (which we choose to be the $z$-axis) with vortex core radius $s$ and winding number $d$. The amplitude is dimensionless and the constant $C_n$ will be given by a normalization condition\footnote{The subscript $n$ in $C_n$ refers to "normalization" and is not to be confused with the polytropic index.}. The amplitude function (\ref{eq: amplitude}) guarantees that the real and imaginary parts of the wavefunction tend to zero for $r \to 0$, hence the overall density $|\psi|^2$ then also goes to zero.  
The form of the vortex ansatz (\ref{eq:vortex2})-(\ref{eq: amplitude}) has some implications. First, it reflects the property that outside of the vortex, i.e. abruptly at $r = s$, the density is simply given by the unperturbed profile (\ref{eq:toy2-calc}), and that there is a discontinuity. Moreover, each integration which includes the vortex wavefunction has to be split accordingly at $r = s$. 

The above ansatz is the same as in RS12; see also their figure 4 for $d=1$.
However, the vortex ansatz does not depend upon the regime we consider, i.e. it is valid in the fuzzy regime, as well as in the TF regime studied in RS12.  The physics of the regime only enters, once we replace the characteristic vortex core radius $s$, e.g. with the healing length\footnote{For $r \gg s$, the wavefunction tends to its asymptotic background value, while for $r \ll s$ the centrifugal force dominates and the wavefunction is proportional to $r$ (for $d=1$), corresponding to a "free" particle with angular momentum $\hbar$, turning around the $z$-axis. The vortex size is of order the system's healing length.}. We will discuss this point in due course.
From now on, we will set $d=1$.

Despite the fact that the global shape of halo model A is a sphere, the parametrization of the vortex demands cylindrical coordinates. As a result, the spherical domain over which the integration is performed is defined by the following intervals for the three cylindrical coordinates:
\begin{eqnarray}
r &\in & [0,s]\ \ \text{and} \ \ [s,\sqrt{R_{99,G}^2-z^2}]\ , \label{eq:rdomain}\\
\phi &\in & [0,2\pi]\ , \\
z &\in & [0,R_{99,G}]\ ,
\end{eqnarray}
where $R_{99,G}$ is the radius of the Gaussian sphere, set by (\ref{eq:99halosize}). 

First, the dimensionless constant $C_n$ in Eq. (\ref{eq: amplitude}) can be determined via the normalization condition of the unpertubed system, i.e. the vortex shall not change the overall mass of the system,
\begin{equation}
    \int_V |\psi|^2 dV = \int_V \frac{\rho_0}{m} |w|^2 dV = N\ .
\end{equation}
Using (\ref{eq:rhoc}), (\ref{eq:toy2-calc}) and (\ref{eq: amplitude}), we get\footnote{Remember that the subscript $n$ in $C_n$ and $K_n$ refers to "normalization" and is not to be confused with the polytropic index.}
 \begin{eqnarray}
 C_n^2 &=& \frac{\Tilde{s}^2}{2} \frac{1-\text{e}^{-\frac{\Tilde{s}^2}{2}}\text{Erf}\left(\frac{\Tilde{R}}{\sqrt{2}}\right) +\text{e}^{-\frac{\Tilde{R}^2}{2}} \sqrt{\frac{2}{\pi}} \Tilde{R}}{\left(1 - \exp\left(-\frac{\Tilde{s}^2}{2}\right)\left(1+\frac{\Tilde{s}^2}{2}\right) \right)\text{Erf}\left(\frac{\Tilde{R}}{\sqrt{2}}\right)} \label{eq:kn-def}\\
 \label{eq:Kn}
 &\equiv & \frac{\Tilde{s}^2}{2} K_n^2\ ,
\end{eqnarray}
where $\text{Erf}(x)$ is given in Eq. (\ref{eq:erf-function}).
It turns out that it will be useful in the course of calculating $\delta E'$ to write $C_n^2$ in the form of (\ref{eq:Kn}).

\begin{figure*}
     \begin{minipage}[b]{0.5\linewidth}
      \centering\includegraphics[width=6cm]{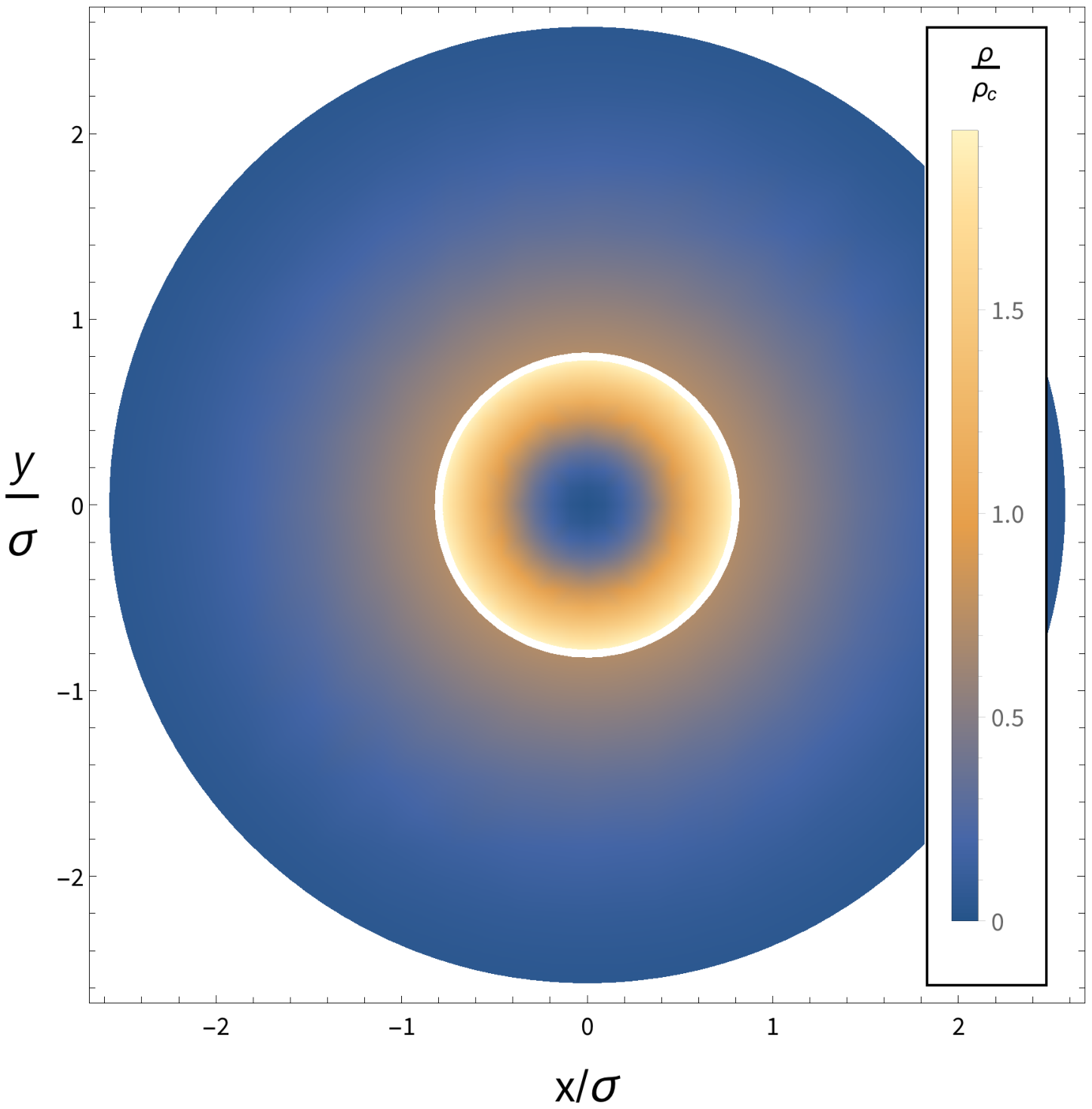}
     \hspace{0.1cm}
    \end{minipage}%
 \begin{minipage}[b]{0.5\linewidth}
      \centering\includegraphics[width=6cm]{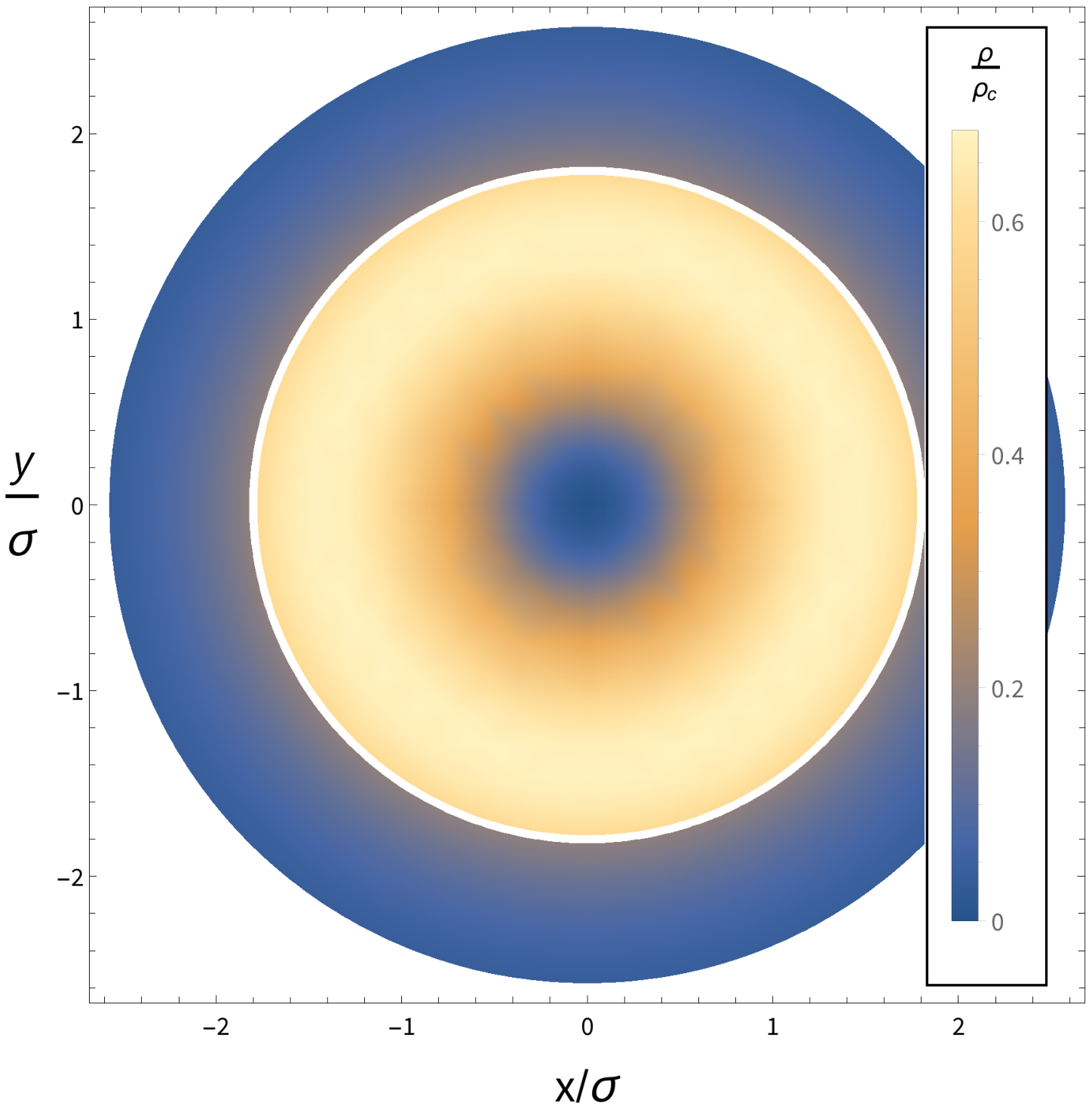}
     \hspace{0.1cm}
    \end{minipage}
 \caption{Density profile $\rho = \rho_0|w|^2$ in units of $\rho_c$ plotted for an FDM halo core as a rotating Gaussian sphere with radius $R_{99,G}/\sigma = \Tilde{R} = 2.576$ and vortex core radius $s/\sigma = \Tilde{s} = 0.8$ (left-hand-plot) and $s/\sigma = \Tilde{s} = 1.8$ (right-hand-plot), respectively.}
 \label{fig:gauss-vortex-08-18}
\end{figure*}

Now, with the normalization determined, the density of an FDM halo core with a singly-quantized vortex in its centre is then given by
\begin{equation} \label{eq:profil-case2}
    \rho = \rho_0|w|^2 = \begin{cases}
 \rho_0  =  \rho_c \text{e}^{-a(r^2+z^2)} & \text{for } r \geq s\ , \\
 
\rho_0\ C_n^2 \left( \frac{r}{s}\right)^2  = \rho_c \text{e}^{-a(r^2+z^2)}\ C_n^2 \left( \frac{r}{s}\right)^2 & \text{otherwise}\ ,
\end{cases}
\end{equation}
with $C_n^2$ in (\ref{eq:kn-def}). As an illustration for FDM halo cores with vortex, we plot two-dimensional density profiles $m|\psi|^2 = \rho$ in units of $\rho_c$ for two different vortex core radii, $\Tilde{s} = s/ \sigma = 0.8$ and $\Tilde{s} = s/ \sigma = 1.8$ in \autoref{fig:gauss-vortex-08-18}. They show that the vortex eats up the density in the very centre of the halo core, as expected from the vortex wavefunction. The white ring at $\Tilde{r} = \Tilde{s}$ indicates a discontinuity of the overall density profile. This discontinuity can be understood by considering both one-sided limits, $\lim_{r\to s
^{\pm}} \rho$ approaching $r=s$ from above or below. According to Eq. (\ref{eq:profil-case2}), approaching $\rho$ at $r=s$ from below yields $\rho_c \text{e}^{-a(s^2+z^2)}\ C_n^2$ and from above yields $\rho_c \text{e}^{-a(s^2+z^2)}$. These two expressions differ by $C_n^2$, which itself is a function of the vortex core radius $s$. 
However, this discontinuity poses no further problem since we will consider global energies only, in contrast to a dynamical analysis which may be affected by the detailed transition between vortex and outer region.

Next, looking at Eq. (\ref{eq:Gf}), we see that we require to calculate the gravitational potential $\Phi_1$ associated with the perturbation of the density caused by the vortex $\rho_1$.

\begin{figure}
    \includegraphics[width=0.9\columnwidth]{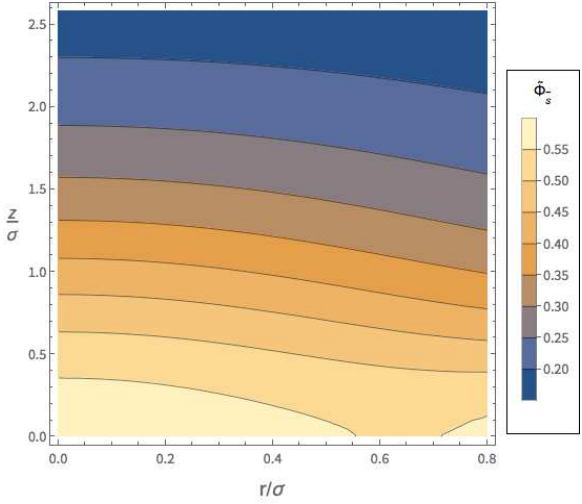}
    \centering
	\caption{Contour plot of $\Tilde{\Phi}_{\Tilde{s}}$ as a function of $\Tilde{r}$ and $\Tilde{z}$. $\Tilde{R}$ was set according to ($\ref{eq:99halosize}$) and $\Tilde{s} = 0.8$}
	\label{fig:potential}
\end{figure}

The density perturbation due to the central singly-quantized vortex,
\begin{eqnarray}
\label{eq:rho-disturbed}
\rho_1 &=& \rho - \rho_0 = m\frac{\rho_0}{m}|w|^2-\rho_0 \nonumber \\
&=& \rho_0(|w|^2-1) \nonumber \\
&=& \begin{cases}
0 & \text{for } r \geq s \\
\rho_0 \left( C_n^2 \left( \frac{r}{s}\right)^{2} -1\right) < 0 & \text{otherwise}\ ,
\end{cases} 
\end{eqnarray}
is the source of the gravitational potential $\Phi_1$. Hence, 
\begin{equation}
\label{eq:phi-1-diff-equ}
\Delta \Phi_1
= \begin{cases}
\Delta \Phi_1^{(o)} = 0 & \text{for } r \geq s \\
\Delta \Phi_1^{(i)} = 4 \pi G \rho_c \text{e}^{-a(r^2+z^2)}\left( C_n^2 \left( \frac{r}{s}\right)^{2} -1\right)& \text{otherwise.}
\end{cases} 
\end{equation}
In fact, obtaining $\Phi_1$ is more complicated than was the determination of $\Phi_0$, the potential of the vortex-free Gaussian sphere in (\ref{eqn:1})-(\ref{eqn:3}) with result in (\ref{eq:sol2}), because now we are looking for two different functions, $\Phi_1^{(o)}$ and 
$\Phi_1^{(i)}$. The former is the solution to the following Laplace equation,
\begin{equation}
\label{eq:poto}
    \Delta \Phi_1^{(o)}(r,z) = 0\ \ \text{for}\ \ r \geq s,
\end{equation}
only valid outside the vortex (hence $"(o)"$). The general solution to Eq. (\ref{eq:poto}) in cylindrical coordinates is
\begin{equation}
\label{eq:sol5}
    \Phi_1^{(o)}(r,z) = \frac{C_2-C_1}{\sqrt{r^2+z^2}},
\end{equation}
where $C_1$ and $C_2$ are integration constants. We follow the same line of argument as in RS12 by imposing that the solution approaches a point-mass potential for large $r$ at fixed $z$ and for large $z$ at fixed $r$. This yields
\begin{equation}
\label{eq:cond}
    C_2-C_1 = -G M_{\text{source}},
\end{equation}
where $M_{\text{source}}$ is the mass of the assumed point-source. Basically, the idea is such that if we are far enough away from the source, i.e. the region in which the vortex-perturbation acts, its gravitational potential "feels" like a point-mass potential. However, how can $M_{\text{source}}$ be determined? It is important to keep in mind, that the perturbed matter density inside the vortex region (\ref{eq:rho-disturbed}) is negative, which is comprehensible since the vortex removes, or rather redistributes the initial matter away from the vortex core region. Since this should be reflected by $\Phi_1^{(o)}$, we will set
\begin{equation}
    M_{\text{source}} = -M_i,
\end{equation}
where $M_i$ is the mass inside the vortex core. In other words, the source of the outer-vortex potential associated with the perturbation of the density due to the vortex is set to be the negative vortex core mass. The mass inside the vortex core is given by
\begin{eqnarray}
\label{eq:massinsidevortex}
&M_i = \int_{z=-R_{99,G}}^{R_{99,G}} \int_{\phi=0}^{2\pi} \int_{r=0}^s \rho_0 C_n^2 \left( \frac{r}{s}\right)^{2} r\text{d}r \text{d}\phi \text{d}z \nonumber \\
&= Nm\ \left( 1-\text{e}^{-\frac{\Tilde{s}^2}{2}}\text{Erf}\left(\frac{\Tilde{R}}{\sqrt{2}}\right)  +\text{e}^{-\frac{\Tilde{R}^2}{2}} \sqrt{\frac{2}{\pi}} \Tilde{R} \right)
\end{eqnarray}
and finally the outer-vortex potential is given by
\begin{equation}
\label{eq:phi-aussen}
    \Phi_1^{(o)}(r,z) = \frac{GM_i}{\sqrt{r^2+z^2}}\ .
\end{equation}
Finding an analytical inner-vortex solution for $\Phi_1^{(i)}$ (hence "(i)") posed a hard problem for our analysis. 
The corresponding partial differential equation
\begin{equation}
\label{eq:phi-diff-innen}
    \Delta \Phi_1^{(i)} = 4 \pi G \rho_c \text{e}^{-ar^2} \text{e}^{-az^2} \left( C_n^2 \left( \frac{r}{s}\right)^{2} -1\right)\ \ \text{for } r < s
\end{equation}
(or equivalently (\ref{eq:phi-diff-dimless}) below)
seems to admit no closed-form solution as our attempts to find one has failed, or rather we found only a very approximate solution, based upon a multipole expansion, but its accuracy was
determined to be too insufficient, so we will not reproduce it here. Therefore, we have solved the Poisson equation (\ref{eq:phi-diff-innen}) numerically, and used that numerical solution in the forthcoming energy integrals. By multiplying both sides of Eq. (\ref{eq:phi-diff-innen}) with 
\[ (4 \pi G)^{-1} \rho_c^{-1} = \frac{\sigma^3 (2\pi)^{3/2}}{Nm4\pi G}\]
and introducing the dimensionless variable
\begin{equation}
\label{eq:phi_dimless}
    \Tilde{\Phi}_1^{(i)} = \Phi_1^{(i)} \frac{ (2\pi)^{3/2}}{4\pi} \frac{\sigma}{NmG } =  \Phi_1^{(i)} \frac{1}{4\pi G \rho_c \sigma^2}\ ,
\end{equation}
the Poisson equation in question can be written as
\begin{equation}
\label{eq:phi-diff-dimless}
    \left[ \frac{\partial^2}{\partial \Tilde{z}^2} + \frac{1}{\Tilde{r}} \frac{\partial}{\partial \Tilde{r}} +  \frac{\partial^2}{\partial \Tilde{r}^2}\right] \Tilde{\Phi}_1^{(i)}(\Tilde{r},\Tilde{z}) =  \text{e}^{-\Tilde{r}^2/2-\Tilde{z}^2/2} \left( K_n^2  \frac{\Tilde{r}^2}{2} -1\right),
\end{equation}
with 
\begin{equation}
\label{eq:K_n_s}
    K_n^2 = K_n^2(\Tilde{s},\Tilde{R})
\end{equation}
given in (\ref{eq:Kn}). Eq. (\ref{eq:phi-diff-dimless}) has to be solved for $\Tilde{\Phi}_1^{(i)}$ inside the vortex region, i.e. in the region given by
\[ \Tilde{r} \in [0, \Tilde{s}]\ \text{and}\ \Tilde{z} \in [-\Tilde{R},\Tilde{R}]\ .\]
However, since the density is an even function of $\Tilde{z}$, the potential will be too, and therefore we solved the system in the region
\[ \Tilde{r} \in [0, \Tilde{s}]\ \text{and}\ \Tilde{z} \in [0,\Tilde{R}]\ .\]
The boundary of that rectangular domain of integration in the $(\Tilde{r},\Tilde{z})$-plane consists of four line segments connecting the four vertices $(0,0),(\Tilde{s},0),(0,\Tilde{R}),(\Tilde{s},\Tilde{R})$. There are two line segments connecting the vertices $(0,0)$ and $(\Tilde{s},0)$, and $(0,0)$ and $(0,\Tilde{R})$ respectively, that lie within the vortex volume. Von Neumann boundary conditions, setting the normal derivative of $\Tilde{\Phi}_1^{(i)}$ to zero, were imposed on the differential equation along these line segments. Continuity requires that the solution of this partial differential equation matches with the analytical (closed-form) expression for the outer-vortex solution $\Phi_1^{(o)}(r,z)$ in Eq. (\ref{eq:phi-aussen}) at the respective line segments connecting the vertices $(\Tilde{s},0)$ and $(\Tilde{s},\Tilde{R})$, and $(0,\Tilde{R})$ and $(\Tilde{s},\Tilde{R})$ respectively, i.e. those line segments at the boundary of the vortex volume. Hence, we imposed the Dirichlet boundary condition
\begin{displaymath}
 \Tilde{\Phi}_1^{(i)}(\Tilde{r},\Tilde{z}) \stackrel{!}{=} \Phi_1^{(o)} \frac{ (2\pi)^{3/2}}{4\pi} \frac{\sigma}{NmG}
\end{displaymath}
\begin{displaymath}
        = \frac{1}{\sqrt{\Tilde{r}^2+\Tilde{z}^2}} \sqrt{\frac{\pi}{2}} ~\times 
\end{displaymath}
\begin{equation}
        \left(1-\exp\left(-\frac{\Tilde{s}^2}{2}\right)\text{Erf}\left(\frac{\Tilde{R}}{\sqrt{2}}\right) +\exp\left(-\frac{\Tilde{R}^2}{2}\right) \sqrt{\frac{2}{\pi}} \Tilde{R} \right) \label{eq: dirichlet}
\end{equation}
along those two line segments. To arrive at Eq. (\ref{eq: dirichlet}), we have used equations (\ref{eq:massinsidevortex}) and (\ref{eq:phi-aussen}). The implementation of these boundary conditions together with an implicit Runge-Kutta-solver returned a dimensionless, numerically interpolating function for our solution that we denote as $\Tilde{\Phi}_{\Tilde{s}} = \Tilde{\Phi}_{\Tilde{s}}(\Tilde{r}, \Tilde{z})$. A contour plot of this function can be seen in \autoref{fig:potential}. However, the calculation of the gravitational energy below requires the dimensional form, namely
\begin{equation}
   \Phi_{\Tilde{s}} = \Tilde{\Phi}_{\Tilde{s}} 4\pi G \rho_c \sigma^2\ .
\end{equation}

\subsubsection{Vortex energy}

Now, the way is clear to proceed with the heart of the calculation, namely calculating $\delta E'$ in Eq. (\ref{eq:endiff}) and determining whether it is negative or positive. We will consider each term in (\ref{eq:Gf}) and (\ref{eq:Rf}) separately.

The quantum-kinetic term,
\begin{equation}
\label{eq:term1}
   \int_V \frac{\hbar^2}{2m^2}  \rho_0 |\Vec{\nabla} w|^2 \text{d}V\ ,
\end{equation}
obviously requires to calculate the square of the absolute value of the gradient of the vortex wavefunction $w$,
\begin{eqnarray}
\label{eq:wgrad}
   \Vec{\nabla} w &=&
   \begin{cases}
   \frac{i}{r}\text{e}^{i\phi} \hat{\textbf{e}}_{\phi} & r \geq s \\
   C_n \frac{i}{s}\text{e}^{i\phi} \hat{\textbf{e}}_{\phi} + C_n \frac{1}{s}\text{e}^{i\phi} \hat{\textbf{e}}_{r} & \text{otherwise}, 
   \end{cases}
\end{eqnarray}
where $\hat{\textbf{e}}_{\phi}$ and $\hat{\textbf{e}}_{r}$ denote the azimuthal and radial unit vectors along the cylindrical coordinate directions $\phi$ and $r$, respectively. As a result, the absolute value of the complex vector field $\Vec{\nabla} w$ can be written as
\begin{eqnarray}
\label{eq:wbetrag}
   |\Vec{\nabla} w | &=& \sqrt{(\Vec{\nabla} w)^* \cdot \Vec{\nabla} w} \nonumber \\
   &=& \begin{cases}
   \frac{1}{r} & r \geq s \\
   \sqrt{2}\frac{C_n}{s} & \text{otherwise}. 
   \end{cases}
\end{eqnarray}
After splitting the axisymmetric integration domain according to (\ref{eq:rdomain}), the quantum-kinetic term yields
\begin{displaymath}
 \int_V \frac{\hbar^2}{2m^2}  \rho_0 |\Vec{\nabla} w|^2 \text{d}V  = 
\end{displaymath}
\begin{displaymath}
    = \frac{\hbar^2}{2m^2} \rho_c  \int_{-R_{99,G}}^{R_{99,G}} \int_0^{2\pi} \int_s^{\sqrt{R_{99,G}^2-z^2}}\text{e}^{-ar^2}\text{e}^{-az^2}\frac{1}{r^2}r \text{d}r\text{d}\phi \text{d}z 
    \end{displaymath}
\begin{displaymath}
   +\ \frac{\hbar^2}{2m^2} \rho_c  \int_{-R_{99,G}}^{R_{99,G}} \int_0^{2\pi} \int_0^{s}\text{e}^{-ar^2}\text{e}^{-az^2}2 \frac{C_n^2}{s^2}r \text{d}r\text{d}\phi \text{d}z 
\end{displaymath}
\begin{displaymath}
   = \frac{\hbar^2\rho_c \pi}{2m^2}  \times
   \left[  \int_{-R_{99,G}}^{R_{99,G}} \text{e}^{-az^2} \text{Ei}(-a(R_{99,G}^2-z^2)) \text{d}z \right.
   \end{displaymath}
   \begin{displaymath}
   \left.-\sqrt{\frac{\pi}{a}} \text{Erf}(\sqrt{a}R_{99,G}) \text{Ei}(-as^2) \right] 
\end{displaymath}
\begin{equation}\label{eq:quantum-term1}
     +\ \frac{\hbar^2\rho_c C_n^2  \pi }{m^2 s^2 a} (1- \text{e}^{-as^2}) \sqrt{\frac{\pi}{a}} \text{Erf}(\sqrt{a}R_{99,G}), 
\end{equation}
with the integral function 
\begin{displaymath}
\text{Ei}(x) = -\int_{-x}^{\infty}\frac{\text{e}^{-t}}{t}\text{d}t 
\end{displaymath}
(denoted ExpIntegralEi[$x$] in Mathematica). Let us briefly discuss this result.
In the TF regime of strong SI, it is known that the quantum-kinetic energy due to the vortex always has a form
$\sim \ln{(Z/s)}$, where $Z$ stands for a characteristic global length scale of the system. Likewise, the quantum-kinetic term in RS12, who incorporated a homogeneous Maclaurin spheroid for the halo (or halo core) with vortex, yielded a logarithmic term of that form, where $Z$ was the length of the equatorial semi-axis of the Maclaurin spheroid, and $s$ was the vortex core radius. 
In addition, $Z/s \gg 1$ in the TF regime (compare also to Eq. (\ref{eq:tf})), because $s$ is of the order of the healing length $\ell$. As a result, the quantum-kinetic energy due to the vortex is of leading order in the total energy, as can be shown.
Now, if we expand the result in (\ref{eq:quantum-term1}) for large $R_{99,G}/s$ in a series representation, we would recover a similar logarithmic term, by means of the contribution of 
\begin{equation}
    -\text{Ei}(-x^2) = -\gamma +\frac{\ln(-1/x^2)}{2} - \frac{\ln(-x^2)}{2} - \sum_{k=1}^{\infty} \frac{(-x^2)^k}{k\ k!}\ ,
\end{equation}
where $\gamma$ is the Euler-Mascheroni constant. 
However, unlike in the TF regime, the quantum-kinetic term due to the vortex is not anymore of leading order in the energy in the fuzzy regime, because $R_{99,G}/s$ is not any longer much larger than one\footnote{E.g. the cases shown in \autoref{fig:gauss-vortex-08-18} have $R_{99,G}/s = \tilde{R}/\tilde{s}$ of $3.22$ and $1.431\dot{1}$, respectively. }, an important feature of vortices in FDM that we will discuss in \autoref{sec:con}.

The second term of $G_{\rho_0}'[w]$ in Eq. (\ref{eq:Gf}) is a gravitational potential energy term, which by itself does not include information about the vortex, and has already been calculated in \autoref{subsec:haloA-gauss}, given by Eq. (\ref{eq:term2}). However, the integration domain chosen for the calculation of that energy term is not exactly the same as the integration domain of all the other terms in this section. 
The second term involves no vortex-disturbed profile. Hence, there is no need to split the axisymmetric integration domain according to the vortex ansatz, as is the case for all the other terms. However, in most integrations here we integrate over a cylinder with height $2R_{99,G}$ and radius $s$ and in addition over a domain, which can be visualized by imagining the result of shooting this cylinder through the centre of a sphere. With increasing vortex core radius $s$, the sum of the volume of this specific domain and the volume of the cylinder differs more and more from the integration volume of a sphere with radius $R_{99,G}$.

The third term of $G_{\rho_0}'[w]$ in Eq. (\ref{eq:Gf}) amounts to
\begin{displaymath} 
     - \int_V \frac{\rho_0}{2} \Phi_0|w|^2 \text{d}V = 
\end{displaymath}
\begin{displaymath}
  = \frac{C_n^2}{s^2}\rho_c^2 G \pi  \left(\frac{\pi}{a}\right)^{3/2} \times
  \end{displaymath}
  \begin{displaymath}
  \times \int_{-R_{99,G}}^{R_{99,G}} \int_0^s \text{e}^{-ar^2-az^2}  \frac{\text{Erf}\left(\sqrt{a(r^2+z^2)}\right)r^3}{\sqrt{r^2+z^2}} \text{d}r \text{d}z 
  \end{displaymath}
 \begin{displaymath}  
 +  \rho_c^2G \pi \left(\frac{\pi}{a}\right)^{3/2} \times
 \end{displaymath}
 \begin{equation} \label{eq:term3}
 \times \int_{-R_{99,G}}^{R_{99,G}} \int_s^{\sqrt{R_{99,G}^2-z^2}} \text{e}^{-ar^2-az^2}  \frac{\text{Erf}\left(\sqrt{a(r^2+z^2)}\right)r}{\sqrt{r^2+z^2}} \text{d}r \text{d}z.
\end{equation}
The fourth term of $G_{\rho_0}'[w]$ contains the gravitational potential $\Phi_1$, associated with the distortion of the density by the vortex, and it is
\begin{displaymath}
 \int_V \frac{\rho_0}{2} \Phi_1|w|^2 \text{d}V = 
\end{displaymath}
\begin{displaymath}
    \frac{\rho_c}{2}  \int_{-R_{99,G}}^{R_{99,G}} \int_0^{2\pi}\int_0^s \text{e}^{-ar^2-az^2}C_n^2 \left(\frac{r}{s}\right)^2 \Tilde{\Phi}_{\Tilde{s}} (\Tilde{r},\Tilde{z}) 4\pi G\ \rho_c \sigma^2\  r\text{d}r \text{d}\phi \text{d}z  
\end{displaymath}
\begin{equation} \label{eq:term4}
     + \frac{\rho_c}{2}   \int_{-R_{99,G}}^{R_{99,G}} \int_0^{2\pi} \int_s^{\sqrt{R_{99,G}^2-z^2}} \text{e}^{-ar^2} \text{e}^{-az^2} \frac{GM_i}{\sqrt{r^2+z^2}} r\text{d}r \text{d}\phi \text{d}z.
\end{equation}
Now, we turn to the rotational energy in Eq. (\ref{eq:Rf}). First, we have already established via Eq. (\ref{eq:corotating-vel-undisturbed}) that the unperturbed sphere shows no net velocity in the rotating frame, i.e. $\vec{\nabla}S_0' = 0$. Finally, from (\ref{eq:wgrad}) and $\vec{\Omega} \times \vec{r} = \Omega r \hat{\textbf{e}}_{\phi}$ follows that
\begin{displaymath}
   -R_{\rho_0}'[w] = \frac{\hbar}{m} \rho_c \int_V i  \text{e}^{-ar^2}\text{e}^{-az^2} w ^* \vec{\nabla}  w  \cdot ( \vec{\Omega} \times \vec{r}) \text{d}V 
 \end{displaymath}
 \begin{displaymath}
    = - \frac{\hbar}{m} \rho_c \int_{-R_{99,G}}^{R_{99,G}} \int_0^{2\pi} \int_0^s \text{e}^{-ar^2}\text{e}^{-az^2} \frac{C_n^2}{s^2}\  \Omega\  r^3 \text{d}r \text{d}\phi \text{d}z -
\end{displaymath}
\begin{displaymath}
     - \frac{\hbar}{m} \rho_c \int_{-R_{99,G}}^{R_{99,G}} \int_0^{2\pi} \int_s^{\sqrt{R_{99,G}^2-z^2}} \text{e}^{-ar^2}\text{e}^{-az^2}  \Omega r \text{d}r \text{d}\phi \text{d}z 
\end{displaymath}
\begin{displaymath}
   =  - \Omega \frac{\hbar}{m} \rho_c  \pi   \frac{C_n^2}{s^2}  \frac{1-\text{e}^{-as^2}(1+as^2)}{a^2} \sqrt{\frac{\pi}{a}}\text{Erf}(\sqrt{a}R_{99,G}) -
   \end{displaymath}
  \begin{displaymath} 
    - \Omega \frac{\hbar}{m} \rho_c  \frac{\pi}{a} \left[ -2R_{99,G} \text{e}^{-aR_{99,G}^2} + \text{e}^{-as^2} \sqrt{\frac{\pi}{a}}\text{Erf}(\sqrt{a}R_{99,G}) \right]
\end{displaymath}    
\begin{equation} \label{eq:term5}
  =  - \hbar N \Omega\ ,
\end{equation}
where we have inserted the expressions for $\rho_c$ and $C_n^2$, (\ref{eq:rhoc}) and (\ref{eq:kn-def}) respectively, in order to arrive at the last equality. 

In conclusion, we have derived the energy difference in the co-rotating frame between the total energy of the halo core with vortex $E'[\psi]$ and the total energy of the halo core without vortex $E'[\psi_0]$, i.e. $\delta E'$ in Eq. (\ref{eq:endiff}). In units of
\begin{equation}
\label{eq:oqmlgm}
\Omega_{QM}L_{QM} = \frac{N \hbar^2}{mR^2}
\end{equation}
and using the characteristic mass introduced in (\ref{charm}), that energy difference reads as
\begin{eqnarray}
\label{eq:deltaELO}
  && \frac{\delta E'}{\Omega_{QM}L_{QM}} = \nonumber \\
  &=&  \Tilde{R}^2 2^{-5/2}\pi^{-1/2}\sqrt{2\pi} \int_{-\Tilde{R}}^{\Tilde{R}}\text{e}^{-\Tilde{z}^2/2}\text{Ei}(-\frac{1}{2}(\Tilde{R^2}-\Tilde{z}^2))\text{d}\Tilde{z} \nonumber\\
  && -\Tilde{R}^2 2^{-5/2}\pi^{-1/2}\sqrt{2\pi}\ \text{Ei}(-\frac{\Tilde{s}^2}{2})\text{Erf}\left(\frac{\Tilde{R}}{ \sqrt{2}}\right) \nonumber \\
   & & +K_n^2\  \Tilde{R}^2\  \frac{1}{2}(1- \text{e}^{-\frac{\Tilde{s}^2}{2}})\text{Erf}\left(\frac{\Tilde{R}}{ \sqrt{2}}\right) \nonumber \\
   & & -\left(\frac{m}{m_c}\right)^2 \frac{\Tilde{R}}{\sqrt{2\pi}}\int_0^{\Tilde{R}}\text{e}^{-\frac{\Tilde{r}_s^2}{2}} \text{Erf}\left(\frac{\Tilde{r}_s}{ \sqrt{2}}\right)\Tilde{r}_s \text{d}\Tilde{r}_s \nonumber \\
   & & +K_n^2 \left(\frac{m}{m_c}\right)^2 2^{-5/2}\pi^{-1/2} \Tilde{R}~\times \nonumber\\
   && \int_{-\Tilde{R}}^{\Tilde{R}} \int_0^{\Tilde{s}}\text{e}^{-\frac{\Tilde{r}^2}{2}} \text{e}^{-\frac{\Tilde{z}^2}{2}} \text{Erf}\left(\sqrt{(\Tilde{r}^2+\Tilde{z}^2)/2}\right) \frac{\Tilde{r}^3\text{d}\Tilde{r} \text{d}\Tilde{z}}{\sqrt{\Tilde{r}^2+\Tilde{z}^2}} \nonumber \\
   & &  + \left(\frac{m}{m_c}\right)^2 2^{-3/2}\pi^{-1/2} \Tilde{R}  \times \nonumber\\ 
   && \int_{-\Tilde{R}}^{\Tilde{R}} \int_{\Tilde{s}}^{\sqrt{\Tilde{R}^2-\Tilde{z}^2}}\text{e}^{-\frac{\Tilde{r}^2}{2}} \text{e}^{-\frac{\Tilde{z}^2}{2}} \text{Erf}\left(\sqrt{(\Tilde{r}^2+\Tilde{z}^2)/2}\right) \frac{\Tilde{r}\text{d}\Tilde{r} \text{d}\Tilde{z}}{\sqrt{\Tilde{r}^2+\Tilde{z}^2}} \nonumber \\
   & & + \left(\frac{m}{m_c}\right)^2 \Tilde{R}\  \frac{K_n^2}{4\pi}  \int_{-\Tilde{R}}^{\Tilde{R}} \int_0^{\Tilde{s}}\text{e}^{-\frac{\Tilde{r}^2}{2}} \text{e}^{-\frac{\Tilde{z}^2}{2}} \Tilde{r}^3\  \Tilde{\Phi}_{\Tilde{s}} (\Tilde{r},\Tilde{z}) \text{d}r  \text{d}z  \nonumber\\
    & & + \left(\frac{m}{m_c}\right)^2   \frac{\Tilde{R}\ K_n^2}{\sqrt{2^3\pi}}\left(1-  \text{e}^{-\frac{\Tilde{s}^2}{2}}(1+\frac{\Tilde{s}^2}{2})\right)\text{Erf}\left(\frac{\Tilde{R}}{ \sqrt{2}}\right)~\times \nonumber\\
    && \int_{-\Tilde{R}}^{\Tilde{R}} \int_{\Tilde{s}}^{\sqrt{\Tilde{R}^2-\Tilde{z}^2}}\text{e}^{-\frac{\Tilde{r}^2}{2}-\frac{\Tilde{z}^2}{2}} \frac{\Tilde{r}\ \text{d}\Tilde{r} \text{d}\Tilde{z}}{\sqrt{\Tilde{r}^2+\Tilde{z}^2}} - \Bar{\Omega}\  
\end{eqnarray}
with $\Bar{\Omega}$ defined in Eq. (\ref{eq:barOmega}). We can see that the vortex energy is a function of several parameters, DM particle mass and halo model parameters, i.e.
\begin{equation}
    \frac{\delta E'}{\Omega_{QM}L_{QM}} = \frac{\delta E'}{\Omega_{QM}L_{QM}} \left(\frac{m}{m_c}, \Tilde{R},\Tilde{s},\Bar{\Omega}\right)\ .
\end{equation}
Understanding the implications of this result requires some intuition, regarding the variables upon which $\delta E'/(\Omega_{QM}L_{QM})$ depends.

\begin{figure*}
     \begin{minipage}[b]{0.5\linewidth}
      \centering\includegraphics[width=8cm]{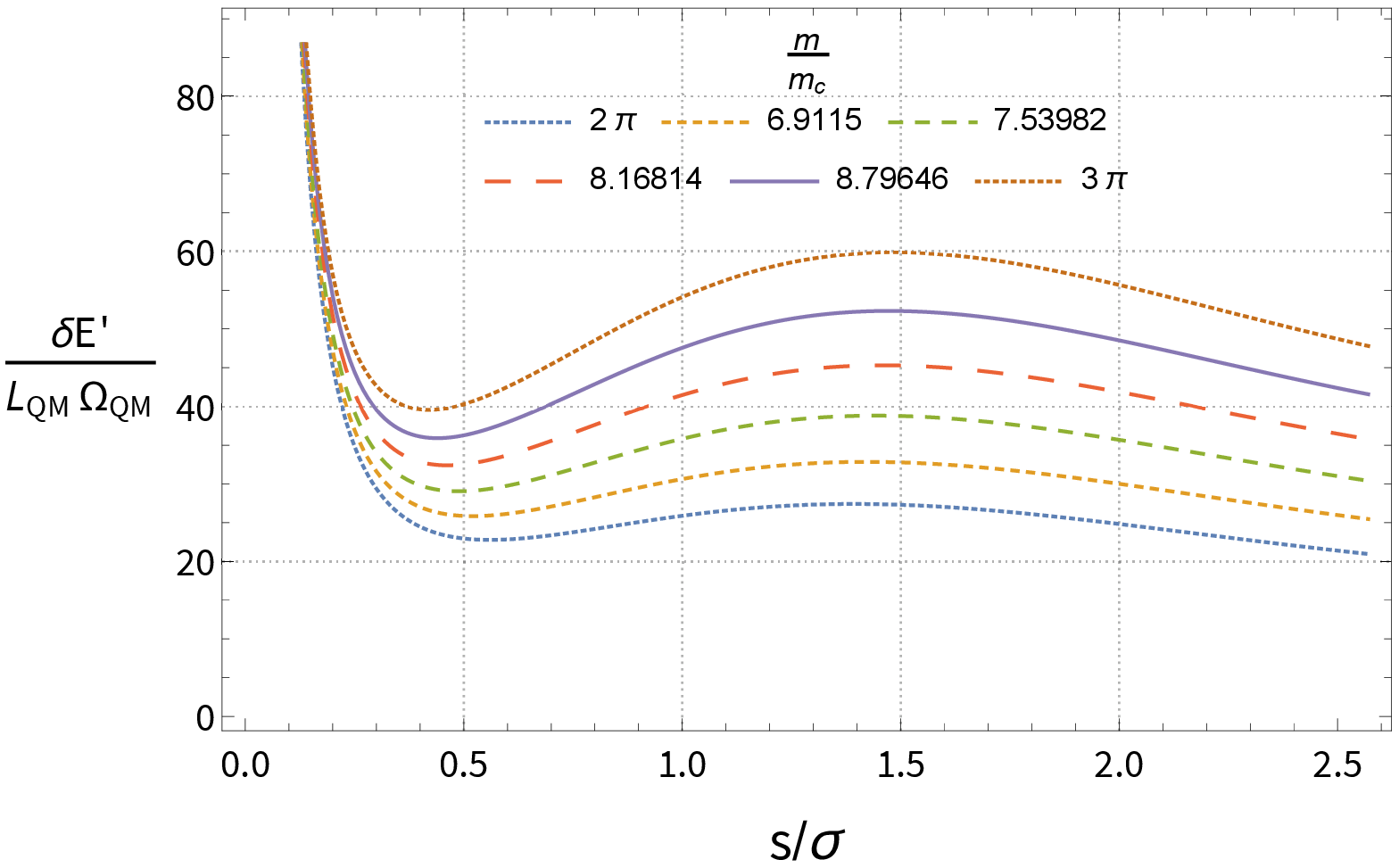}
     \hspace{0.1cm}
    \end{minipage}%
 \begin{minipage}[b]{0.5\linewidth}
      \centering\includegraphics[width=8cm]{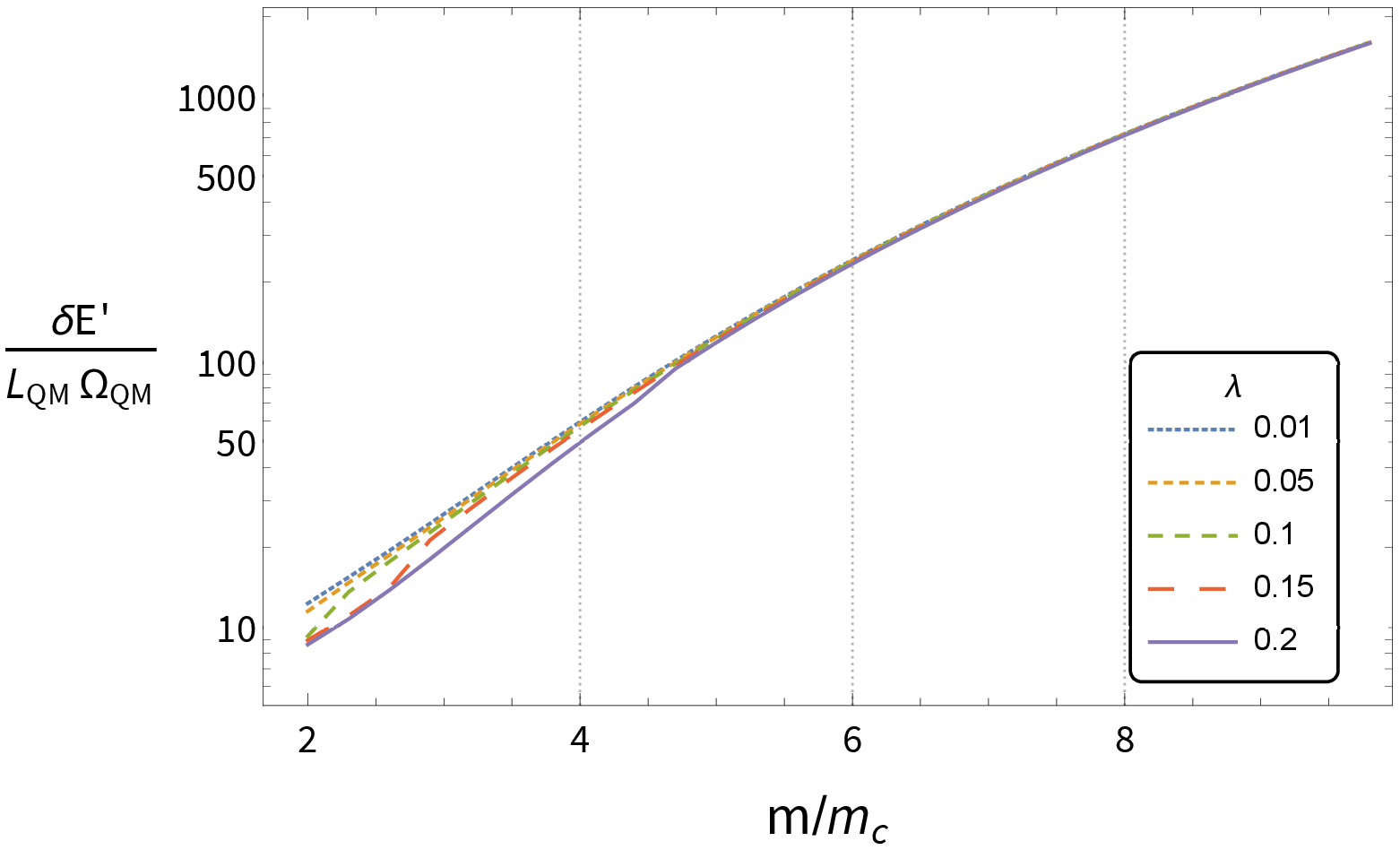}
     \hspace{0.1cm}
    \end{minipage}
 \caption{Left panel: Vortex energy $\delta E'$ in the co-rotating frame, in units of $(\Omega_{QM}L_{QM})$, plotted as a function of $\tilde{s}$ for different particle masses $m/m_c$ within the range in (\ref{massratio}). The dimensionless angular velocity is set to its minimum value, $\Bar{\Omega} = \Omega / \Omega_{QM} = 4.41894$, corresponding to $L=L_{QM}$. Right panel: Vortex energy $\delta E'$ in the co-rotating frame, in units of $(\Omega_{QM}L_{QM})$, logarithmically plotted as a function of the particle mass $m/ m_c$ for different spin-parameter values $\lambda$. The angular velocity is determined according to Eq. (\ref{eq:lambda-lang}) for fixed $\lambda$ and $m/m_c$.}
 \label{fig:delta-E-s-lambda}
\end{figure*}

As we have already emphasized in \autoref{sec:intro}, in the fuzzy regime the vortex could in principle take up the whole halo, i.e. $s \lesssim R$ or in dimensionless notation $\Tilde{s} \lesssim \Tilde{R}$.
Therefore, we will study $\delta E'$ for the entire range $0 \leq \Tilde{s} \leq \Tilde{R}$. In addition, plotting the vortex energy for different particle masses provides even further insight, see \autoref{fig:delta-E-s-lambda}, left panel, where the contributions of the central singly-quantized vortex to the energy of the system is plotted as a function of the vortex core radius for different DM particle masses $m$ for the lower bound on angular velocity given in (\ref{eq:lower-omega}), i.e. $\Bar{\Omega} = \Omega / \Omega_{QM} = 4.41894$, corresponding to $L=L_{QM}$.
It is evident that for this minimum $\Bar{\Omega}$, FDM is not able to form a central vortex in the context of halo model A for any of the considered parameter values, since $\delta E'/(\Omega_{QM}L_{QM}) > 0$ everywhere. 
  Before we proceed in adopting values for $\Bar{\Omega}$, which correspond to spin parameters of interest, let us examine first the features of the vortex energy curves: The divergence for $\Tilde{s} \rightarrow 0$ is due to the quantum-kinetic energy term (\ref{eq:quantum-term1}) and it is only weakly dependent on $m/m_c$. The $m$-dependent gravitational potential energy contribution, which does not include any information on the vortex in Eq. (\ref{eq:term2}), merely fans out the individual curves; this term just adds an $m$-dependent constant value to the energy. The third term of $G_{\rho_0}'[w]$, which combines the unperturbed gravitational potential with the total density $\rho_0 |w|^2$ does not show a strong dependence on the vortex core radius $s$ in the given range. Its effect on the vortex energy $\delta E'/(\Omega_{QM}L_{QM})$ amounts to a slight decrease with increasing $s$. The rough overall shape of the curves and their dependence on the vortex core radius $s$ is to a great extent the result of the quantum-kinetic energy of the vortex and the gravitational potential energy due to the vortex potential $\Phi_1$. If we were to plot their sum,
\begin{equation}
    \int_V \frac{\hbar^2}{2m^2}  \rho_0 |\Vec{\nabla} w|^2 \text{d}V + \int_V \frac{\rho_0}{2} \Phi_1|w|^2 \text{d}V,
\end{equation}
in units of $(\Omega_{QM}L_{QM})$, as a function of $\Tilde{s}=s/ \sigma$ for different $m/m_c$, and compared it to the left panel of \autoref{fig:delta-E-s-lambda}, it would reveal that these figures look quite the same; the disregarded terms only yield a slight horizontal stretching of the energy curves.
Therefore, we established that, in the fuzzy regime, the vortex energy is strongly dominated by the quantum-kinetic \textit{and} the gravitational potential energy due to the vortex.

\begin{figure*}
     \begin{minipage}[b]{0.5\linewidth}
      \centering\includegraphics[width=7cm]{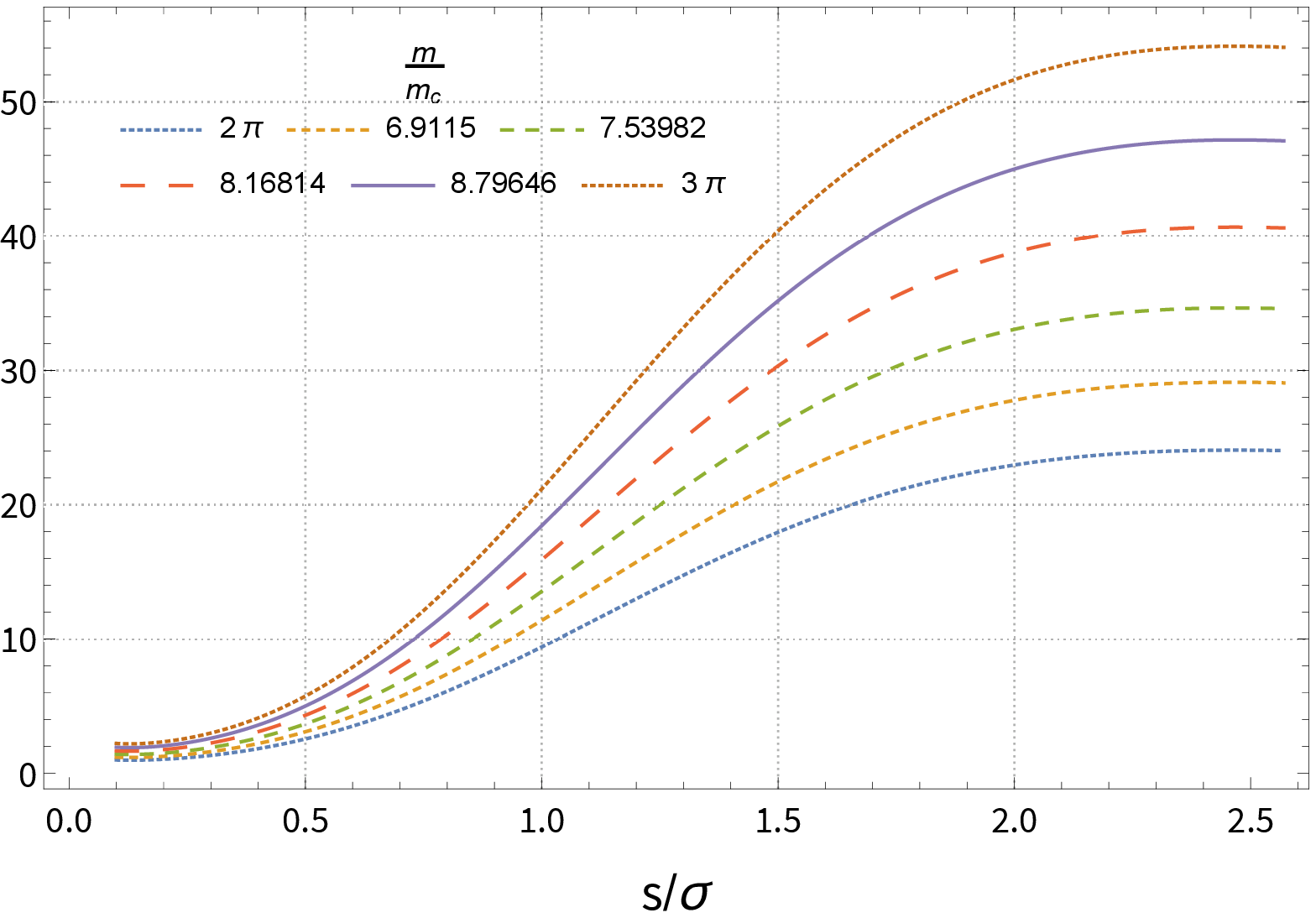}
     \hspace{0.1cm}
    \end{minipage}%
 \begin{minipage}[b]{0.5\linewidth}
      \centering\includegraphics[width=7cm]{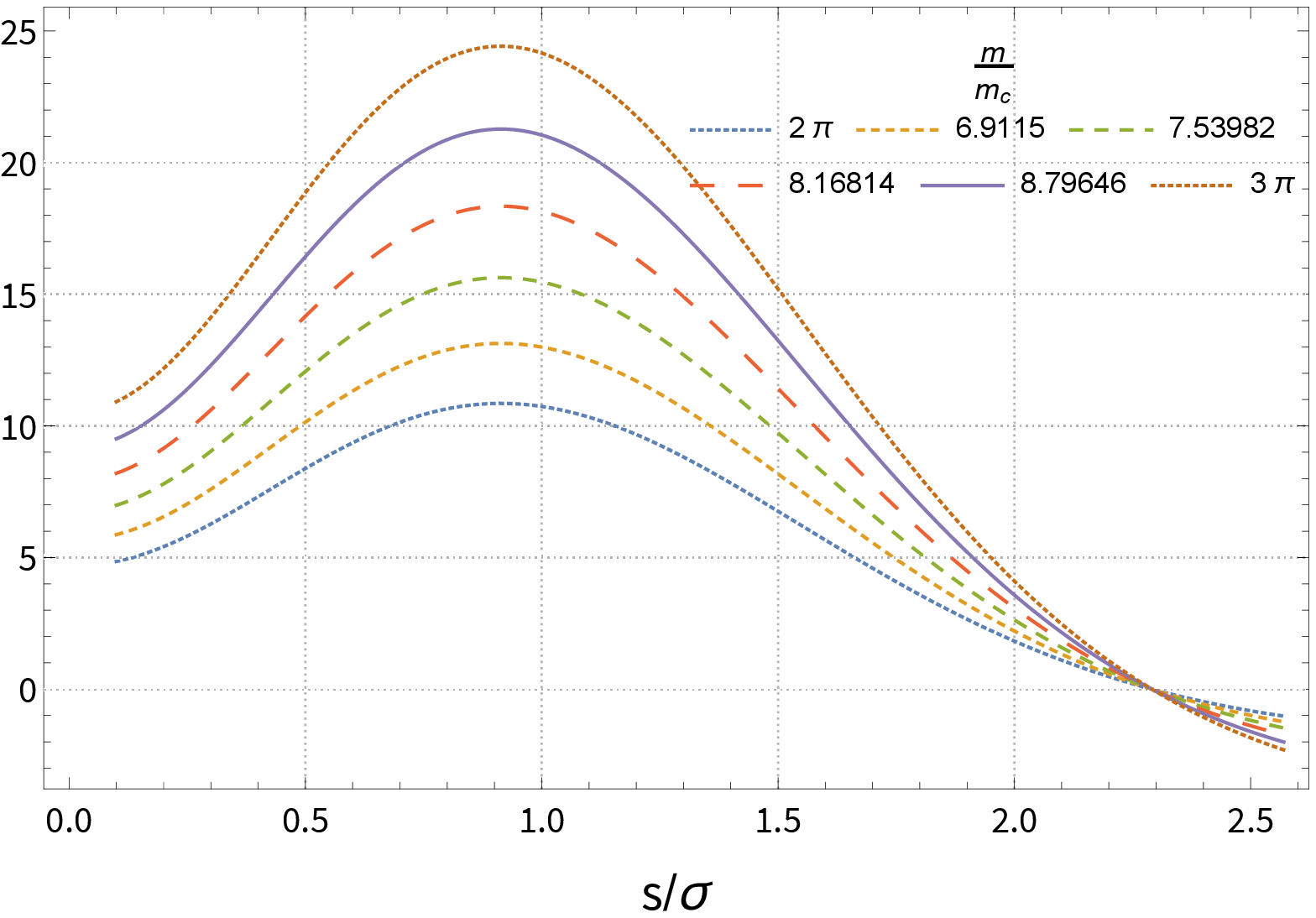}
     \hspace{0.1cm}
    \end{minipage}
 \caption{Left panel: The first integral in expression (\ref{eq:term4}) (inner-vortex contribution), in units of $(\Omega_{QM}L_{QM})$, plotted as a function of $\tilde{s}$ for different particle masses $m/m_c$ within the range in (\ref{massratio}). It is reasonable that the contribution to the vortex gravitational potential energy (\ref{eq:term4}) from inside the vortex increases for halo cores with a larger central vortex. Right panel: The second integral in expression (\ref{eq:term4}) (outer-vortex contribution), in units of $(\Omega_{QM}L_{QM})$, again plotted as a function of $\tilde{s}$ for different $m/m_c$. This term shows the most peculiar dependency on the vortex configuration. More explanations can be found in the main text.}
 \label{fig:delta-E-grav-innen-außen}
\end{figure*}

Now, the striking dependence of the gravitational potential energy generated by the vortex potential $\Phi_1$ on the vortex core radius itself, see Eq. (\ref{eq:term4}), is shown in \autoref{fig:delta-E-grav-innen-außen}, where the two integrals corresponding to the inner- and outer-vortex contributions are plotted separately. While the monotonic increase of the inner-vortex contribution (left panel), first integral in (\ref{eq:term4}), for configurations with a larger and larger central vortex is easily comprehensible, the $s$-dependence of the outer-vortex contribution (right panel), second integral in (\ref{eq:term4}), is quite striking. In principle, this second integral should decrease with increasing $\Tilde{s}$ since in that case the integration domain decreases, given a fixed halo core size $\Tilde{R}$. However, the $s$-dependence of the inner vortex mass (\ref{eq:massinsidevortex}) counteracts this general trend initially, yielding a local maximum before the curves decrease for increasing $\tilde{s}$.

Last but not least, there is the rotational energy term (\ref{eq:term5}) which, in units of $(\Omega_{QM}L_{QM})$, yields the term $-\Bar{\Omega}$ in our energy difference (\ref{eq:deltaELO}). This term is responsible for a global shift of the energy curves in the left panel of \autoref{fig:delta-E-s-lambda}, downwards to lower energy values. This makes sense, because increasing angular velocities should lower the vortex energy, hence make a vortex more favourable, in principle. Vortex formation would be shown to be energetically favoured, as soon as one of the curves would cross the abscissa. Therefore, we need now to consider angular velocities which correspond to spin parameters of interest, in order to determine whether this is the case. 

But first, we revisit considerations of \autoref{sec:fuzzy} regarding the gravitational healing length $\ell_{\rm{grav}}$. 
We have already mentioned that $\ell_{\rm{grav}}$, as well as the corresponding healing length $\ell$ in the TF regime, can be regarded as the distance over which the wavefunction tends to its background value when subjected to a localized perturbation. For this reason, RS12 proceed in the course of their vortex energy analysis in the TF regime by replacing the vortex core radius $s$ with $\ell$, based upon the assumption that $s$ is very close to the healing length $\ell$, motivated by the vortex ansatz. In just the same spirit, we may proceed with the assumption that the vortex core radius in the fuzzy regime is well approximated by the gravitational healing length $\ell_{\rm{grav}}$, given in (\ref{eq:xi_G-1}) and (\ref{eq:xi_G-m}). 
That is,
\begin{equation}
\label{eq: s-ersatz}
    \frac{s}{\sigma} = \Tilde{s} \approx \Tilde{\ell}_{\rm{grav}} = \left(\frac{m_c}{m}\right)^2 \Tilde{R}\ ,
\end{equation}
i.e. for given $\Tilde{R} = 2.576$, the vortex core radius is now a function of the boson mass, more precisely the ratio $m/m_c$. Now, replacing $\Tilde{s}$ in (\ref{eq:deltaELO}), according to (\ref{eq: s-ersatz}), constitutes the first step in order to arrive at the right panel of \autoref{fig:delta-E-s-lambda}. 
In \autoref{subsec:haloA-gauss}, we have calculated the relationship between the spin parameter $\lambda$ and $\Bar{\Omega}$, see Eq. (\ref{eq:lambda-lang}). Once we employ that relationship, the angular velocity $\Omega$ corresponding to a given $\lambda$ now depends on the particle mass $m$. I.e. now we calculate the vortex energy $\delta E'/(\Omega_{QM}L_{QM})$ as a function of $m/m_c$ by setting $\Tilde{s} = \Tilde{\ell}_{\rm{grav}}$, fixing the value of $\lambda$ and calculating $\Bar{\Omega}$ for every $m/m_c$ in the considered range. The right panel of \autoref{fig:delta-E-s-lambda} finally shows the result of this procedure for $\lambda$-values of $(0.01,0.05,0.1,0.15,0.2)$. We can immediately see that $\delta E'/(\Omega_{QM}L_{QM}) > 0$ for all parameters considered. 
The plot also shows that vortex formation is even less energetically favoured, the higher $m/m_c$. In short, while the necessary condition for vortex formation is better and better fulfilled, the higher $m/m_c$ (i.e. the necessary condition is fulfilled for an increasing range of $\lambda$), vortex formation is \textit{less energetically favoured}, the higher $m/m_c$ gets: no parameter space remains where vortices will arise.

Now, we checked whether a change of the definition of our gravitational healing length in equ.(\ref{eq:xi_G-1}) affects our conclusion. The definition in (\ref{eq:xi_G-1}) already picks the higher-end value, so we repeated our calculation for values of $\ell_{\rm{grav}}$ which are a factor of $0.5$ and $0.1$ smaller, respectively. As a result, the associated vortex core radii $s$ are assigned smaller values, and we are effectively getting closer to the asymptotic energy range of small $s$ with its quantum-kinetic energy, which diverges for small $s$; see left panel in \autoref{fig:delta-E-s-lambda}. This implies even much larger (positive) values for the vortex energy $\delta E'/(\Omega_{QM}L_{QM})$, moving it orders of magnitude to higher values than those shown in the right panel of \autoref{fig:delta-E-s-lambda}. 

Therefore, we can safely conclude that FDM halo cores as rotating Gaussian spheres are \textit{not} subject to vortex formation for any value of allowed spin parameter $\lambda$.

\subsection{Halo model B: irrotational Riemann-S ellipsoid transitions to Gaussian sphere with  vortex}\label{subsec:B-riemann}

In this section, we will also employ an energy analysis in order to verify whether the improved modelling of the rotating equilibrium object would change the conclusion of the last section. In fact, our conclusion does not change: vortex formation is not favoured.
 To this aim, we will apply an energy argument, which is similar but not equivalent to the approach described in the previous section, however, it is also inspired by a similar consideration of RS12.  
 
 Let us consider some unspecified dynamical process in the course of which the halo core may form one singly-quantized vortex in its centre. The actual occurrence of this vortex creation requires that the process in question transforms the initial state of the halo core into a final state with lower total energy compared to the initial state. As a result, we are going to look at two "snapshots" of that transformation, namely the initial and the final configuration, calculate their total energies and compare them for a given set of parameter ranges. The underlying halo core in the initial state will be modeled by an irrotational Riemann-S ellipsoid along the lines of \autoref{subsec:riemann}. We will assume that the envisaged process transforms the halo core from a vortex-free configuration to a configuration with one central vortex: the vortex takes up all of the angular momentum of the system, once it has formed. Thereby, the initial ellipsoidal shape of the halo core disappears, it becomes spherical, allowing us to draw on the perturbed Gaussian sphere with vortex as a model for the second snapshot or final state. We call it halo model B.
 In the course of the following calculations, it is important to distinguish clearly between quantities and global properties of the initial Riemann-S ellipsoid, which will be denoted by the index $"R"$, and the final vortex-carrying Gaussian sphere, denoted by the index $"G"$.

We start the analysis by considering the total energy of the vortex-free initial configuration \text{in the rest frame} which can be written as
\begin{equation}
\label{eq:ER}
    E_R = K_Q + T + W\ ,
\end{equation}
where $T$ and $W$ are given by (\ref{eq:T-riemann}) and (\ref{eq:W-riemann}), respectively. On the other hand, $K_Q$ has no classical counterpart and is therefore absent in the studies of \cite{1969efe..book.....C} or LRS93. However, the internal energy of a Riemann-S ellipsoid is given by (\ref{eq:u-sperical}) and can be written as
\begin{equation}
\label{eq: u-riemann}
    U = k_1 K_p \rho_{c,R}^{1/n}M\ ,
\end{equation}
due to the ellipsoidal approximation, where $\rho_{c,R}$ is the central density of the initial Riemann-S ellipsoid incorporating a polytropic density profile. We have already seen that the internal energy of a sphere arising from an $(n=2)$-polytrope is related to the quantum-kinetic energy via Eq.(\ref{eq:K_Q-U}),
given the density profile (\ref{eq: polymodel}) in \autoref{sec:fuzzy}. Owing to the ellipsoidal approximation of LRS93, the total internal energy of a rotating polytrope is identical to that of a spherical one with the same central density. This implies that the quantum-kinetic energy can be written as
\begin{equation} \label{kqriemann}
    K_Q = \frac{3}{2} U = \frac{3}{2} k_1 K_p \rho_{c,R}^{1/2}M\ ,
\end{equation}
with $K_p$ given by (\ref{eq:kfix}) and $n=2$.
Before we can continue with the calculation of the total energy $E_R$, the current analysis requires to establish relations between several quantities of the initial and final state of the halo core. While the underlying initial halo core configuration is modeled as an irrotational, $(n=2)$-polytropic Riemann-S ellipsoid, whose respective energy terms LRS93 derive from initially spherical polytropes modified by rotation, we will set the final halo core to be a sphere - a Gaussian sphere, not a polytropic sphere! However, LRS93 provide us with relation (\ref{eq:meanrad}), i.e. the process we consider shall transform the system in such a way that it settles with the cutoff radius of the final Gaussian sphere, $R_{99,G}$, which we associate with the radius of the $(n=2)$-polytropic sphere, $R_0$. This is justified in light of (\ref{r99G}) and (\ref{polyrad}), which established that $R_{99,G} \approx 1.033 ~R_0$. 
Thus, we demand
\begin{equation}
\label{eq:RRR0}
    R_R \stackrel{!}{=} R_{99,G}\ g(e_1,e_2)^{-2}\ .
\end{equation}
What about the central densities of the initial and final configurations? Given the central density $\rho_{c,s}$ of the spherical polytrope with radius $R_0$, we have the relation between the central and mean density of a polytrope,
\begin{equation}
\label{eq:rhos-riem}
    \Bar{\rho}_s = \frac{3M}{4 \pi R_0^3} = 3 \rho_{c,s} \frac{|\theta_1'|}{\xi_1}\ ,
\end{equation}
see appendix \ref{appendix-polytrope}. The analogue for the polytropic Riemann ellipsoid can be written as
\begin{equation}
\label{eq:rhoR-riem}
    \Bar{\rho}_R = \frac{3M}{4 \pi R_R^3} = 3 \rho_{c,R} \frac{|\theta_1'|}{\xi_1}\ .
\end{equation}
The halo core mass $M$ is required to be conserved during this transition that we consider. Thus, dividing Eq. (\ref{eq:rhoR-riem}) by Eq. (\ref{eq:rhos-riem}) yields
\begin{equation}
    \label{eq:rhoRS}
    \Bar{\rho}_R = \Bar{\rho}_s g(e_1,e_2)^6\ \ \text{and}\ \ \rho_{c,R} = \rho_{c,s} g(e_1,e_2)^6\ ,
\end{equation}
where we have used relation (\ref{eq:RRR0}). In the context of this halo model B, the vortex-free Riemann-S ellipsoid shall transition to a Gaussian sphere with vortex. Hence, we set
\begin{equation}
    \rho_{c,s} \stackrel{!}{=} \rho_{c,G} = \frac{M}{\sigma^3 (2\pi)^{3/2}}\ ,
\end{equation}
although, unlike the spherical polytrope, the Gaussian profile has no compact support, i.e. $ \rho_{c,G} $ is the central density of an infinitely extended system which we cut off at $R_{99,G}$. Again, we rely on the presumption that the two density models are equally appropriate and closely related.
As a result, the total internal energy of the Riemann-S ellipsoid (\ref{eq: u-riemann}) can be written as
\begin{equation}
    U = k_1 \left( \frac{2\pi}{9}\right)^{1/2} G^{1/2} \frac{\hbar}{m} \rho_{c,G}^{1/2}  g(e_1,e_2)^3 M\ ,
\end{equation}
where we have used (\ref{eq:kfix}) and $k_1$ in (\ref{eq:k1}) has to be evaluated for $n=2$. Consequently, the quantum-kinetic energy of the FDM halo core, as a Riemann-S ellipsoid (\ref{kqriemann}), in units of $\Omega_{QM,R}L_{QM}$, is given by
\begin{equation}
\label{eq:K_qzeile1}
\frac{K_Q}{\Omega_{QM,R}L_{QM}}
= \frac{k_1}{2(2\pi)^{1/4}} \frac{m}{m_{c,R}}\Tilde{R}^{3/2}\ ,
\end{equation}
where we have used (\ref{eq:omgrav}), (\ref{eq:mcrcg}), and (\ref{eq:RRR0}), as well as
\begin{eqnarray}
\Omega_{QM,R} &=& \frac{\hbar}{\sqrt{R_R G M}}\ , \\ 
\frac{\Omega_{\text{grav},R}}{\Omega_{QM,R}} &=& \frac{\sqrt{3}}{2}\frac{m}{m_{c,R}} = \frac{\sqrt{3}}{2}\frac{m}{m_{c,G}}g(e_1,e_2)^{-1} \ .
\end{eqnarray}
By means of these relations, the gravitational potential energy (\ref{eq:W-riemann}) for $n=2$, in units of $\Omega_{QM,R}L_{QM}$, amounts to
\begin{eqnarray}
\frac{W}{\Omega_{QM,R}L_{QM}} &=& - \frac{4\Omega_{\text{grav},R}^2MR_R^2}{3}\frac{f(e_1,e_2)}{\Omega_{QM,R}N \hbar} \\
&=& - \left( \frac{m}{m_{c,R}}\right)^2 f(e_1,e_2)\ .
\end{eqnarray}
Finally, the rotational kinetic energy (\ref{eq:T-riemann}) reads as
\begin{equation*}
\frac{T}{\Omega_{QM,R}L_{QM}} =\frac{\kappa_2m\Omega^2}{20 \Omega_{QM,R}  \hbar} \times
\end{equation*}
\begin{equation*}
 \left[ (a_1-a_2)^2\left(1+\frac{1}{h(e_1)}\right)^2+(a_1+a_2)^2\left(1-\frac{1}{h(e_1)}\right)^2\right] 
\end{equation*}
\begin{equation}
    = \left(\frac{\Omega}{\Omega_{QM,R}}\right)^2 \frac{e_1^4}{(1-e_1^2)^{1/3}(1-e_2^2)^{1/3}(2-e_1^2)}\ ,
\end{equation}
where $\kappa_2$ can be found in appendix \ref{appendix-polytrope}, Eq.(\ref{eq:kappan}), and the definition of the dimensionless factor $h(e_1)$ arises from setting $f_R = -2$ and combining the relations (\ref{eq: l-frac}) and (\ref{eq:xi-lambda}):
\begin{equation}
\label{eq:h1}
    h(e_1) = \frac{2-e_1^2}{2\sqrt{1-e_1^2}} = \frac{1}{2}\left(\frac{a_1}{a_2}+\frac{a_2}{a_1}\right) = \frac{\Omega}{\Lambda}\ .
\end{equation}
The vortex-free irrotational Riemann-S ellipsoid shall transition to a final state that consists of a Gaussian sphere, hosting a central singly-quantized vortex. In other words, the final state corresponds to our halo model A. Thus, the total energy \textit{in the co-rotating frame} of the final configuration amounts to
\begin{equation}
    E_G' = E_G'[\psi_0] + G_{\rho_0}'[w]-R_{\rho_0}'[w] = E_G'[\psi_0]+ \delta E_G'\ ,
\end{equation}
where the total energy of the vortex-free system $E_G'[\psi_0]$ is given by (\ref{eq:E0_split}) and the energy terms associated with the vortex $\delta E_G'$ are given by (\ref{eq:Gf}-\ref{eq:endiff}). Since $\Vec{\nabla} S_0'$ vanishes, $E_G'[\psi_0]$ is given by the sum of the two expressions (\ref{eq:KQ-novortex}) and (\ref{eq:term2}). $\delta E_G'$ is given by Eq. (\ref{eq:deltaELO}), except for the fact that Eq. (\ref{eq:deltaELO}) already includes a division by $\Omega_{QM,G}L_{QM}$. However, in order to compare the total energies of the initial and final state on an equal footing, we need to express $E_G'[\psi_0]$ as well as $\delta E_G'$ in units of $\Omega_{QM,R}L_{QM}$ which, due to $\Omega_{QM,R} = g(e_1,e_2)^{4}\Omega_{QM,G} $, yields in the case of $\delta E_G'$,
\begin{equation}
    \frac{\delta E_G'}{\Omega_{QM,R}L_{QM}} = g(e_1,e_2)^{-4} \frac{\delta E_G'}{\Omega_{QM,G}L_{QM}}\ ,
\end{equation}
where $\delta E_G'/ (\Omega_{QM,G}L_{QM})$ is given in (\ref{eq:deltaELO}).
The energy difference between the initial vortex-free Riemann-S ellipsoidal state and the final spherical state with a vortex in the centre, in the frame co-rotating with angular velocity $\Omega$, will be denoted as $\delta E_{RG}'$, and it can be written as
\begin{equation}
    \delta E_{RG}' = E_G' - E_R' = (E_G'[\psi_0]+\delta E_G') - (E_R-\Omega L)\ ,
\end{equation}
where we are still missing the expression for $\Omega L$ in units of $\Omega_{QM,R}L_{QM}$, namely
\begin{equation}
    \frac{\Omega L}{\Omega_{QM,R}L_{QM}} = \Tilde{\Omega}^2\frac{3\kappa_2}{20} \left(\frac{m}{m_{c,G}}\right)^2\frac{g(e_1,e_2)^{-2}e_1^4}{(1-e_1^2)^{1/3}(1-e_2^2)^{1/3}(2-e_1^2)}\ ,
\end{equation}
with $\Tilde{\Omega}$ in (\ref{eq:omega-grav}-\ref{eq:tom}) and using
(\ref{eq:total-l-lqm}).

In summary, the energy difference in units of $\Omega_{QM,R}L_{QM}$ between the initial state and final state amounts to
\begin{eqnarray}
\label{eq:total-riemann}
   && \frac{\delta E_{RG}'}{\Omega_{QM,R}L_{QM}} = g(e_1,e_2)^{-4}\Tilde{R}^2 \frac{\pi}{2 (2\pi)^{3/2}} \times \nonumber \\
   & & \left[-\Tilde{R} \exp(-\Tilde{R}^2/2)(3+\Tilde{R}^2)\right] \nonumber \\ 
   & + & g(e_1,e_2)^{-4}\Tilde{R}^2 \frac{\pi}{2 (2\pi)^{3/2}}\left[ \frac{3 \sqrt{\pi}}{8 }2^{5/2}\text{Erf}\left(\frac{\Tilde{R}}{\sqrt{2}}\right)\right] \nonumber \\
    & - & \frac{g(e_1,e_2)^{-4}\Tilde{R}}{\sqrt{2\pi}} \frac{m^2}{m_{c,G}^2}\int_0^{\Tilde{R}} \text{e}^{-\frac{\Tilde{r}_s^2}{2}} \text{Erf}\left(\frac{\Tilde{r}_s}{\sqrt{2}}\right) \Tilde{r}_s \text{d}\Tilde{r}_s \nonumber \\
    &+& g(e_1,e_2)^{-4} \frac{\delta E_G'}{\Omega_{QM,G}L_{QM}}\left(\frac{m}{m_{c,G}},\Tilde{s},\Tilde{R},\frac{\Omega}{\Omega_{QM,G}}\right) \nonumber \\
    &-& \frac{k_1}{2(2\pi)^{1/4}} \frac{m}{m_{c,G}}g(e_1,e_2)^{-1}\Tilde{R}^{3/2}\nonumber \\
    &+& \left( \frac{m}{m_{c,G}}\right)^2 g(e_1,e_2)^{-2}f(e_1,e_2) \nonumber \\
   &- & \frac{\kappa_2}{10}\left(\frac{\Omega}{\Omega_{QM,R}}\right)^2 \frac{e_1^4}{(1-e_1^2)^{1/3}(1-e_2^2)^{1/3}(2-e_1^2)}\nonumber \\
   &+& \Tilde{\Omega}^2\frac{\kappa_2}{5} \left(\frac{m}{m_{c,G}}\right)^2\frac{3}{4}\frac{g(e_1,e_2)^{-2}e_1^4}{(1-e_1^2)^{1/3}(1-e_2^2)^{1/3}(2-e_1^2)}\ ,
\end{eqnarray}
with
\begin{eqnarray}
   \frac{\Omega}{\Omega_{QM,R}} &=& \Tilde{\Omega}\frac{m}{m_{c,G}}\frac{\sqrt{3}}{2}g(e_1,e_2)^{-1}\ , \\
    \frac{\Omega}{\Omega_{QM,G}} &=& \Tilde{\Omega}\frac{m}{m_{c,G}}\frac{\sqrt{3}}{2}g(e_1,e_2)^{3}\ .
\end{eqnarray}

\begin{figure*}
     \begin{minipage}[b]{0.5\linewidth}
      \centering\includegraphics[width=8cm]{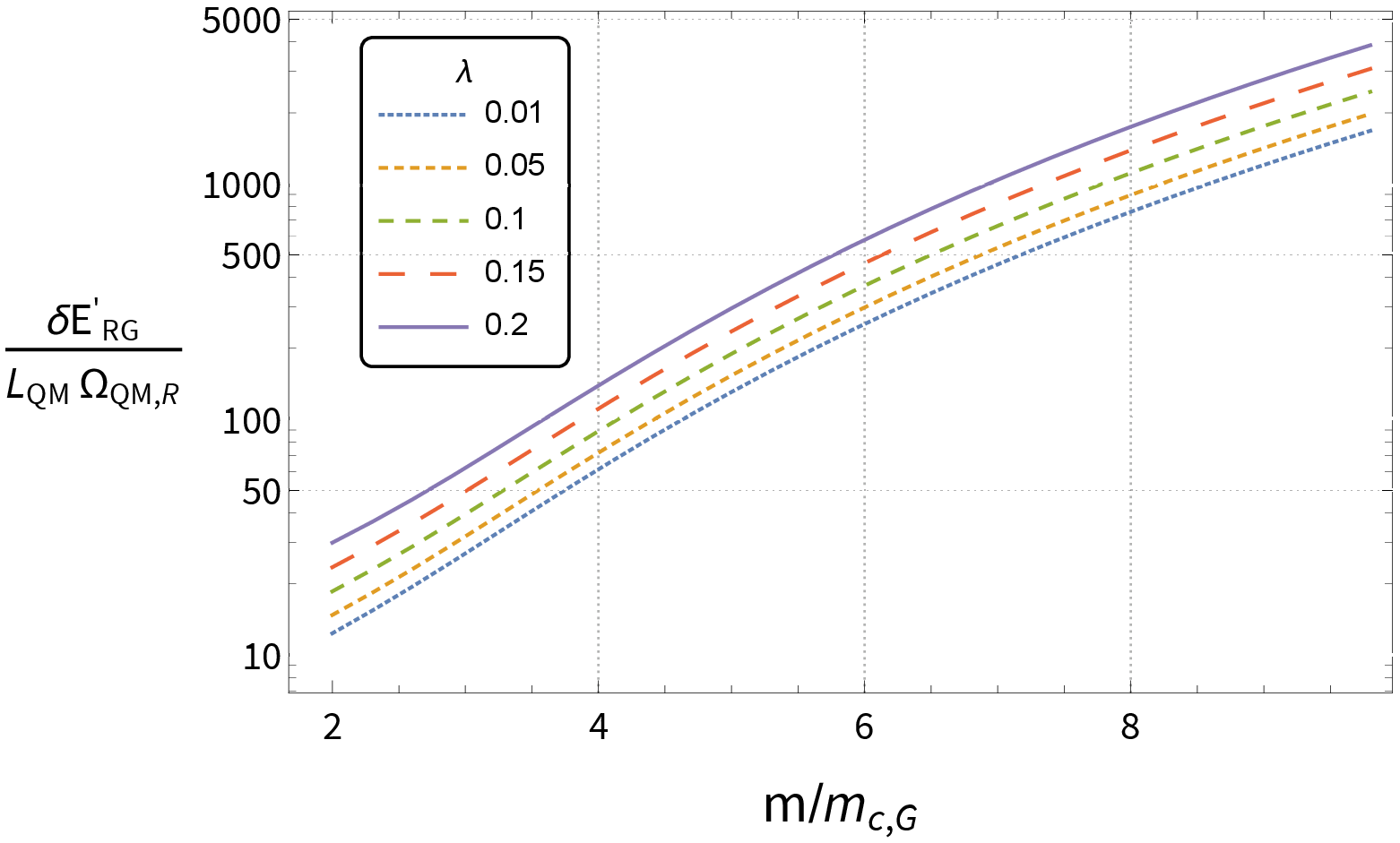}
     \hspace{0.1cm}
    \end{minipage}%
 \begin{minipage}[b]{0.5\linewidth}
      \centering\includegraphics[width=7cm]{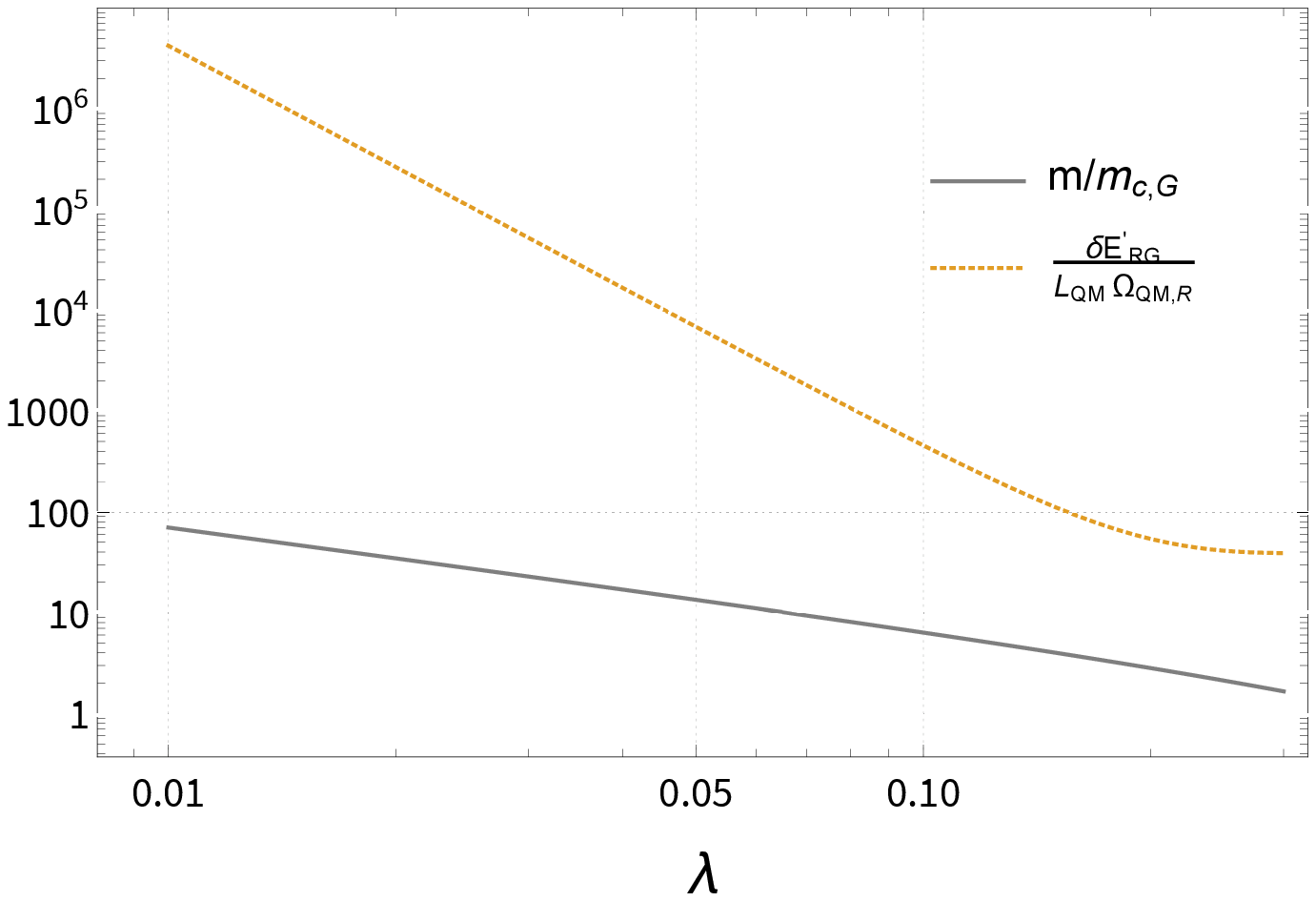}
     \hspace{0.1cm}
    \end{minipage}
 \caption{Left-hand-plot: Energy difference $\delta E_{RG}'/(\Omega_{QM,R}L_{QM})$ in the co-rotating frame, see (\ref{eq:total-riemann}), logarithmically plotted as a function of DM particle mass $m/ m_{c,G}$ according to (\ref{massratio}), for different spin-parameters $\lambda$. The angular velocity is determined according to Eq. (\ref{eq:omega-grav}) for given polytropic index $n=2$, $f_R = -2$ and the axis ratios are listed in \autoref{tab:lambda-table} for different $\lambda$. Right-hand-plot: Energy difference $\delta E_{RG}'/( \Omega_{QM,R}L_{QM})$ as a function of spin-parameter $\lambda$ for \textit{fixed} total angular momentum $L = L_{QM}$ (light, dotted curve). For comparison, $m/m_{c,G}$ as a function of $\lambda$ at \textit{fixed} $L = L_{QM}$ is shown again (dark, solid curve); see also \autoref{fig:m-lambda}, but note the difference between linear and logarithmic scaling.  }
 \label{fig:Riemann-plots}
\end{figure*}

It is important to stress that the energy difference $\delta E_{RG}'$ of halo model B is an entirely different quantity than $\delta E'$ of halo model A. The latter is the difference in energy between sphere-with-vortex and sphere-without-vortex, i.e. it is basically the vortex energy. However, $\delta E_{RG}'$ is the difference in energy between the ellipsoid-without-vortex (initial state) and the sphere-with-vortex (final state), i.e. it is more than just the vortex energy. In fact, the vortex energy $\delta E'$ is only a part of $\delta E_{RG}'$ and it is, up to a prefactor, the fourth term in equ.(\ref{eq:total-riemann}).

In order to understand the implications for $\delta E_{RG}'/ (\Omega_{QM,R}L_{QM})$, we are going to gradually incorporate the framework built by LRS93. First of all, the gravitational angular velocity $\Tilde{\Omega}$ cannot be independently chosen within this framework, but is directly coupled to the geometry of the ellipsoid, in other words to the axis ratios $a_2/a_1$ and $a_3/a_1$, or equally eccentricities $e_1, e_2$, for given polytropic index $n$ and ratio $f_R = -2$ via Eq. (\ref{eq:omega-grav}). A visualization of the dependence of the energy difference $\delta E_{RG}'$ on the vortex core radius $\Tilde{s}$ only requires a choice of axis ratios and setting $\Tilde{R} = 2.576$. However, this kind of plot would not provide any additional information, since it would just show the same functional shape, as the left panel of \autoref{fig:delta-E-s-lambda}, because it is only $\delta E_G' (=\delta E')$ which depends upon $\Tilde{s}$.

Instead, we plot $\delta E_{RG}'$ as a function of $m/m_{c,G}$, assuming again that the vortex core radius is of same order as the gravitational healing length  $\ell_{\rm{grav}}$ and thus a function of the particle mass for given $\Tilde{R} = 2.576$, using (\ref{eq: s-ersatz}) with the understanding that $m_c = m_{c,G}$.
In addition, the axis ratios and thus the energy difference are fixed by the spin-parameter $\lambda$, according to the system of equations (\ref{eq: axis-rel-2}) and (\ref{eq:lambda-final}-\ref{eq:t-final}); see also \autoref{tab:lambda-table}. The result is shown in \autoref{fig:Riemann-plots}.

In contrast to \autoref{fig:delta-E-s-lambda} (right panel), \autoref{fig:Riemann-plots} (left panel) shows a striking dependence on the spin-parameter $\lambda$. For given $m/m_{c,G}$, we see that for increasing $\lambda$-values the dimensionless energy difference, given in Eq. (\ref{eq:total-riemann}), yields higher (positive) values. 
It is important not to jump to the conclusion suggested by halo model A, where the energy difference $\delta E_G'/ (\Omega_{QM,G}L_{QM})$ (Eq. (\ref{eq:deltaELO})) was the direct consequence of an energy splitting procedure for one single object - the Gaussian sphere. Its shape is not a degree of freedom and hence not coupled to its rotational support. As a consequence, since a vortex "feeds" on rotation, the expectation that its impact on the system's total energy is such that it is increasingly lowered for higher $\lambda$-values is, indeed, confirmed in \autoref{fig:delta-E-s-lambda} (right panel). In contrast, halo model B here describes a transition from an ellipsoid to a sphere with vortex, where several features are subject to transition and  coupled to the rotational support of the object, as follows. The construction of the Riemann-S ellipsoid implies that its energy and relationship to the corresponding polytropic sphere of same mass are coupled to its shape (i.e. the eccentricity of the ellipsoid), which we, in turn, connect to its rotational support via Eq. (\ref{eq:lambda-final}): the higher $\lambda$, the higher the eccentricities of the Riemann-S ellipsoid, whose shape increasingly differs from a (Gaussian) sphere. Now, \autoref{tab:lambda-table} shows that increasing values for the spin-parameter $\lambda$ yield increasing values for $\Tilde{L}^2$ and $g(e_1,e_2)^{-2}$, but decreasing values for $f(e_1,e_2)$. All these factors combine to explain the dependence of $\delta E_{RG}'/(\Omega_{QM,R}L_{QM})$ on $\lambda$, not just the vortex which is part of the final object.  As a result, the energy difference between the initial and the final state, in the frame co-rotating with $\Tilde{\Omega}$ (or $\Bar{\Omega}$ respectively), \textit{increases} as $\lambda$ increases, thanks to the different ways in which the vortex and the global shape of the halo core each depend upon rotation.

Using the total angular momentum in units of $L_{QM}$ given in (\ref{eq:total-l-lqm}), we can constrain the transition from the Riemann-S ellipsoid into a Gaussian sphere with one central vortex further by imposing $L = L_{QM}$, i.e. the vortex shall take up the entire angular momentum of the system. This implies
\begin{equation}
\label{eq:m-e-cond}
    \frac{m}{m_{c,R}} \stackrel{!}{=} \left( \frac{\kappa_2}{5}  \frac{\sqrt{3}}{2} \Tilde{\Omega} \frac{e_1^4}{(1-e_1^2)^{1/3}(1-e_2^2)^{1/3}(2-e_1^2)}\right)^{-1} ,
\end{equation}
i.e. the DM particle mass is now a function of the halo core geometry. In other words, given a shape of the halo core and requiring the amount of angular momentum to be enough in order to sustain just one vortex yields a condition on the DM particle mass. Meeting this condition implies that the particle mass is fixed for given $\lambda$, or equally for given axis ratios of the halo core. Hence, this condition can only be met at one single point (when considering the energy difference as a function of particle mass; compare to the left panel of \autoref{fig:Riemann-plots}), i.e. fixing the halo core geometry by $\lambda$ at \textit{fixed} $L/L_{QM}$ immediately yields one value for $\delta E_{RG}'/( \Omega_{QM,R}L_{QM})$. This can be seen in the right panel of \autoref{fig:Riemann-plots}, where $\delta E_{RG}'/( \Omega_{QM,R}L_{QM})$ and $m/m_{c,G}$ are each plotted as a function of $\lambda$. Although the energy difference does \textit{decrease} with increasing spin-parameter (at \textit{fixed} $L/L_{QM}=1$), it remains much too high in order to make energetically favourable the formation of a vortex ever. Thus, we have shown that vortex formation is also not favoured in the context of halo model B.

\section{Conclusions and discussion}\label{sec:con}

We have studied gravitationally bound halo structures made of ultra-light DM bosons. These structures are referred to as SFDM haloes or BEC-DM haloes, whose bosons are described by a single scalar wavefunction. These haloes can be modelled, using the Gross-Pitaevskii-Poisson system of equations, which can be written in the form of quantum-mechanical fluid equations. The analysis in this work focused on the so-called fuzzy regime (fuzzy DM, or FDM), in which quantum pressure balances gravity.  In order for the wave nature of FDM to be potent on galactic scales, the de Broglie wavelength of bosons (evaluated with the virial velocity of the object) has to be of same order of magnitude than the global size of the system.  In fact, ground-state solutions of the Gross-Pitaevskii-Poisson system (also called Schr\"odinger-Poisson system for FDM) have a size of roughly the de Broglie wavelength and constitute attractor solutions, so-called "solitons", of these equations. They were the subject of our investigation in this paper. These ground-states describe the "solitonic cores" of large haloes, as well as the entire halo, if the latter only consists of the core (e.g. appropriate hosts for the smallest galaxies). 

The main objective of our analysis was the study of rotating FDM haloes and halo cores, and to study the question of vortex formation. The underlying equations of motion allow for vortex solutions; within vortices the density goes to zero and the velocity diverges. In general,
vortices are manifestations of the quantized vorticity in superfluids and they are the building blocks of quantum turbulence, phenomena which have been also studied for BEC-DM haloes.
As a quantum superfluid, BEC-DM is irrotational - vorticity-free, but, while starting from irrotational initial conditions,
it can become unstable to vortex formation.  Vortices can change the smooth halo background, with dynamical and structural consequences, e.g. affecting the density profiles of haloes, e.g. within halo cores, which could impact the dynamics of baryonic tracers, central star formation, or the formation of supermassive black holes. It is thus important to establish whether vortices can form in FDM, or BEC-DM more generally. 


This was demonstrated for the
case of haloes and halo cores that form in BEC-DM
with repulsive SI in the Thomas-Fermi (TF) regime,
when angular momentum is present, as expected during large-scale structure formation,
in \cite{2012MNRAS.422..135R}. In particular, the virialized halo cores
in that case, in which gravity was balanced
by SI pressure and rotation, were found
to be unstable to vortex formation
for a large range of boson parameters, mass $m$ and SI coupling strength $g$. However, since this instability required a minimum SI strength, we suggested there that the small SI regime (and especially FDM without SI) would
be found to be vortex-free inside halo cores. 
In \cite{2012MNRAS.422..135R}, we also introduced approximate, analytical equilibrium solutions for rotating BEC-DM halo cores, in particular we applied for the first time irrotational Riemann-S ellipsoids to BEC-DM haloes, for which an ($n=1$)-polytropic density profile was used, appropriate for the TF regime. These irrotational Riemann-S ellipsoids are useful models to describe halo cores which do not form vortices.

In \citet{2014MPLA...2930002R}, we further suggested that the polytropic
SI pressure support that set the
size of these halo cores in the TF regime
would be supplemented on larger scales by the
wave-motion-support generated by the
wave nature of BEC-DM and its quantum pressure,
during the virialization of haloes that assemble
from infall and mergers, making it possible for
haloes to be much larger than
their polytropic cores in which only SI dominates.
Later, BEC-DM without SI with boson masses around $m \sim 10^{-22}$ eV, i.e. FDM, has been studied with greater detail, including
simulations that report all haloes have
solitonic cores of the size of the de Broglie
wavelength (as evaluated inside haloes),
supported against gravity by quantum pressure.
BEC-DM haloes also show a wave-supported envelope
outside this solitonic cores, with a profile that resembles
that in CDM haloes, but in which wave motions provide the
random internal motions responsible for virial equilibrium to be dicussed shortly. 

In this paper, we extend the analysis of \cite{2012MNRAS.422..135R} and study the question of vortex formation in rotating FDM halo cores (without SI) in equilibrium, with the same analytical rigour than in this previous work. We stress that, in the fuzzy regime, all characteristic length scales are roughly comparable, including the size of perturbations of the system, like quantum vortices, which poses a hard problem for analytic studies.
As in RS12, we associate the typical amount of angular momentum, expected from large-scale structure, with the spin parameter.  
We studied the necessary and sufficient conditions for the formation of one centrally located, singly-quantized vortex in FDM halo cores. The sufficient condition is studied by employing a detailed energy analysis. 
We found that, unlike the TF regime, for solitonic
cores of FDM, vortex formation cannot be
triggered by angular momentum as it is in the TF regime, for neither of our models that we considered. 
For vortex formation to be triggered by angular momentum,
the specific angular momentum must first satisfy a necessary (minimum)
condition, that it exceed the minimum value that gives each
particle an angular momentum of $\hbar$. If this necessary condition
is satisfied, then it is further required that vortex formation be
energetically favoured, in order to establish that vortex formation
will take place.  In the TF regime, both conditions can be
met for the typical amounts of specific angular momentum for cosmological haloes, for a large range of $m$ and $g$.  However, for FDM, we have shown in this paper that the necessary condition is generally not
met for typical amounts of halo angular momentum.  
We have further shown that, even for angular momentum which is large enough to meet the
necessary condition,
vortex formation is nevertheless not energetically favoured.   
This is consistent with and can explain the fact that simulations
of structure formation in the FDM model do not find vortices
in the solitonic cores of FDM haloes.   

For our analysis, we considered two different analytic models for rotating, solitonic cores. First, a Gaussian profile cut off at a finite radius; we call it the Gaussian sphere. It served as a useful analytical model for the undisturbed halo core prior to vortex formation, but also thereafter, i.e. a disturbed Gaussian sphere with central vortex as a model for the halo core. This allowed us to perform our energy analysis in a very general way, allowing a clear exposition of the various parameter dependencies which enter to show that vortex formation is not favoured. However, the Gaussian sphere has two drawbacks, first that, by construction, it remains spherical, even for non-zero angular momentum and second, that it is not strictly irrotational in the laboratory frame.  Therefore, as a further model to overcome these drawbacks, we studied the irrotational Riemann-S ellipsoid with an $(n=2)$-polytropic density profile. The latter equation-of-state was shown to be a viable approximation for FDM solitons, and irrotationality prior to any vortex formation is guaranteed. We have thereby shown that irrotational Riemann-S ellipsoids can be used in the fuzzy regime, as well, not only in the TF regime. 

Now that we have described the implications of our results, 
let us finally compare them in more detail to some of the previous simulation works of FDM (i.e. no SI).
The cosmological simulations of \cite*{2014NatPh..10..496S} first showed convincingly that in the centre of gravitationally bound FDM haloes, one finds coherent standing waves, i.e. stable solitonic cores.     

\cite*{2016PhRvD..94d3513S} simulated the dynamics of these solitonic cores by investigating binary and multiple mergers of up to 13 such cores. Again, it is found that solitonic cores are embedded within bigger haloes, whose outer density profile declines like NFW density profiles. Furthermore, \cite*{2016PhRvD..94d3513S} find that their emerging solitonic cores are rotating ellipsoids, if the system is initialized with non-zero total angular momentum. In fact, the respective volume rendered images and velocity fields of the cores strongly indicate that they resemble irrotational Riemann-S ellipsoids, as the authors point out. The same conclusion was drawn by \cite{2018JCAP...10..027E} in their study of the dynamics of solitonic cores, and they also find ellipsoidal cores akin to irrotational Riemann-S ellipsoids. 
Therefore, our work established more generally that polytropic, irrotational Riemann-S ellipsoids provide useful analytical counterparts for the formed solitonic halo cores of BEC-DM halo formation simulations.

Moreover, vorticity is generated during structure formation
(from vorticity-free initital conditions), but only
\textit{outside} of the solitonic cores, as found in \cite*{2016PhRvD..94d3513S} and \cite{2017MNRAS.471.4559M}.
The origin of this vorticity has not been well-studied,
but its absence from solitonic cores is
consistent with the results of \cite{2012MNRAS.422..135R}, as well as with our new results of this paper.
More precisely, \cite{2017MNRAS.471.4559M} present a set of 100 numerical simulations in which a group consisting of 4 to 32 solitonic cores merge and form one final halo whose core is (like the final bound cores of \cite*{2016PhRvD..94d3513S}) well-fitted by the density profile introduced by \cite{2014PhRvL.113z1302S}, see Eq.(\ref{eq: schive}). However, in contrast to the simulations by \cite{2014PhRvL.113z1302S} and \cite*{2016PhRvD..94d3513S}, the work of \cite{2017MNRAS.471.4559M} specifically includes the study of quantum turbulence, found in the envelopes of their final haloes.  
Given the soliton fit (\ref{eq: schive}), they find that the break between the soliton profile and the outer NFW-like profile within the final virialized BEC-DM halo occurs universally at\footnote{Note that their notation of the core radius differs from ours in (\ref{eq: schive}); they are related as $r_c = 0.3c$.} $\approx 3.5r_c$, which approximately corresponds to the soliton radius. Regarding turbulence and vortices exhibited by the final halo, \cite{2017MNRAS.471.4559M} conclude by analysing the energy power spectra $E_k$, the radial energy density profiles and 2D slices of the wavefunction amplitude $|\psi|$ of their 100 simulations. Granules and turbulence appear everywhere in the domain, \textit{except} for the central solitonic core. The stable solitonic core remains free of substructure and turbulence. The radial energy density profiles show that the quantum-kinetic energy supports the structure up to $2.7r_c$. Beyond that radius, all three energy contributions become comparable, yielding a characteristic signature of turbulence, namely equipartition. The energy power spectra lack power for small $k$, show a mode which displays most of the turbulence, and finally follow a $k^{-1.1}$ power law for large $k$. This resembles the spectrum of thermally-driven and hence isotropic turbulence of superfluids (\cite{2016PhR...622....1T}).

Furthermore, \cite{2017MNRAS.471.4559M} show that the power spectra of their simulations peak at $2\pi / k_{peak} \approx 7.5r_c$ which corresponds to a scale of twice the soliton radius. This explains why the filamentary distribution of the $|\psi|$-field (outside the soliton) show preferentially soliton-sized granules. 

Now, \cite{2017MNRAS.471.4559M} explain the absence of vortices (and quantum turbulence, as a result) within solitonic cores by referring to their very equilibrium properties, as ground-state solutions of the fundamental equations.
However, this assessment is not accurate. For example, the analysis in \cite{2012MNRAS.422..135R} showed that vortices do arise in equilibrium halo cores in the TF regime, for a large parameter space, although these cores also correspond to ground-state solutions of the fundamental equations.  The analysis in this paper suggests that vortices add energetic "penalty" to the equilibrium of the cores, hence are strongly unfavoured, while the addition of a positive, large-enough particle self-interaction overcomes this penalty and vortices can be favoured, as shown in \cite{2012MNRAS.422..135R}. Moreover, BEC-DM in the fuzzy regime requires small $m/m_c$, see (\ref{massratio}) (corresponding to small $m/m_H$ in \cite{2012MNRAS.422..135R}), and for the lower end of that range, even the necessary condition for vortex formation is not met, not even for high spin parameters. In short, the FDM parameter space excludes vortices in halo cores, because either the necessary, or the sufficient condition fails to be fulfilled.

More recently, \cite{Hui_2021} followed up on the question of vortex formation in FDM haloes, with analytical arguments and numerical simulations. By simulating the merger of isolated solitons in the form of Gaussian "peaks", they formed structures similar to those seen in previous simulations of
virialized BEC-DM haloes, where a solitonic core is enshrouded by an envelope.  They were interested in studying vortices in this envelope region, where quantum turbulent dynamics causes vortex tangles. In fact, similar to the earlier simulation papers discussed above, \cite{Hui_2021} do also \textit{not} find vortices within the central solitonic cores of their simulated haloes, while vortices are identified only in the envelope.
We note that the "universal" properties of vortices featured in their abstract, such as the velocity or density profile close to their origin, as well as the fact that singly-quantized vortices are favoured over multiply-quantized vortices, were reported already 
in earlier papers (some of which we cited elsewhere
in this paper, incl. our own work). Defects were investigated analytically, but only in the absence of gravity, although gravitational effects were included in their numerical simulations. However, in the
latter case, they report that vortex gravity is not important. By contrast, for the objects of interest
to us here, all characteristic length scales in FDM - unlike the TF regime - are of similar order (e.g. perturbations characterized by the vortex core radius $s$ can be as large as the system size, in principle). Our analysis shows that the quantum-kinetic energy of such vortices need not be the leading-order term and the gravitational potential energy due to the vortex is not any less important than its quantum-kinetic energy.
Inasmuch as the findings of \cite{Hui_2021} imply that \emph{"envelope vortices"} are not
self-gravitating, as small parts of the much bigger envelope, they are quite different from the case described here, where vortices are studied within hydrostatic equilibrium objects - solitonic cores - which hold themselves up by balancing their own gravity by their own quantum pressure and rotation. 
Still, we expect some minimum condition for vortex formation to apply also within halo envelopes, as follows.
Our argument for solitonic cores was based upon the relationship between the size of the core, $\propto \lambda_\text{deB}$, set by the circular velocity of the core if supported by quantum pressure
with no rotational support, and the implied angular momentum per particle, expressed in units of $\hbar$, for different degrees of rotational support.
In order to provide $\hbar$ per particle to satisfy the minimum condition for vortex formation, we needed to provide a high
degree of rotational support, i.e. high spin-parameters $\lambda \gtrsim 0.1$, which are actually not expected for the cores. And even if we consider such and higher 
spin-parameters, the study of the sufficient condition revealed that vortices are not energetically favoured in that case.   
\\
However, if self-gravity is
unimportant for the envelope vortices, then the effective "angular momentum" of the vortex in terms of
$\hbar$ per particle summed over the particles there - the minimum condition -, can be untethered from the implied degree of
rotational support against the vortex' gravity. In that sense, there is no such "spin-parameter barrier" to vortex formation, if envelope vortices are not self-gravitating.   
As such, it is perhaps not surprising, in view of our finding that failing to overcome this barrier prevents solitonic cores from forming vortices, that the envelopes, which are not subject to the same barrier, can form them. 
Still, that leaves open the question of why those envelope vortices should be energetically favoured, or whether they are a very transient phenomenon, after all.
In any case, the subject of envelope vortices and
their distinct origin from the case studied here deserves further study in future work.  

Finally, we want to comment on another point by \cite{Hui_2021} in their discussion, that solutions exist with angular momentum that do not carry vortices, and that realistic haloes would be supported by velocity dispersion, rather than rotation. Of course, it is true that we do not expect rotationally-supported FDM haloes or halo cores, either, a point of view we also expressed in this paper and self-consistently quantified in terms of the spin-parameter and its statistically-likely values. That systems with angular momentum need not have vortices has long been known for laboratory BECs.  For the gravitationally-supported category relevant to haloes
and halo cores, 
this was one key result in \cite{2012MNRAS.422..135R}, where irrotational (vortex-free) Riemann-S ellipsoids were shown to be viable approximate solutions for TF cores with spin parameters similar to CDM, for model parameters which would not favour vortex formation.  
And as we have shown in this paper, using analytical methods, angular momentum by itself will never be sufficient to create vortices within the cores of FDM haloes, in perfect agreement with our earlier work \cite{2012MNRAS.422..135R}, and with simulations of FDM halo formation.

\section*{Acknowledgements}

Schobesberger and Rindler-Daller acknowledge the financial support by the Austrian Science Fund FWF through an Elise Richter fellowship (grant nr. V 656-N28) to Rindler-Daller. The computations were done by means of the technical computing system \citet{Mathematica}.

\section*{Data availability statement}

All data are incorporated into the article and its online supplementary material.




\bibliographystyle{mnras}
\bibliography{example} 




\appendix

\section{Polytropic spheres}\label{appendix-polytrope}

In order to clarify some notation, we include this appendix on polytropes.
Classical literature on stellar structure like \cite{1939isss.book.....C} or \cite*{kippenhahn2012stellar} elaborate on a special class of equilibrium configurations of gas spheres, so-called polytropic spheres.

For spheres in hydrostatic equilibrium, we require
\begin{equation}
\label{eq:hy}
    \frac{d P }{dr} = - \frac{d \Phi}{dr} \rho\ ,
\end{equation}
considering static and spherically symmetric solutions only, and combine with the Poisson equation
\begin{equation}
\label{eq:poiss}
    \frac{1}{r^2}\frac{d}{dr}\left( r^2 \frac{d \Phi }{dr}\right) = 4 \pi G \rho \ ,
\end{equation}
where $P$ denotes the pressure, $\rho$ the density and $\Phi$ the gravitational potential of the system.

This class of configurations can be provided with a simple so-called polytropic relation between the pressure and the density of the form
\begin{equation}
\label{eq:polyP}
    P = K_p \rho^{1 + \frac{1}{n}},
\end{equation}
where the polytropic "constant" $K_p$ and the polytropic index $n$ are fixed. \cite*{kippenhahn2012stellar} elaborate on two reasons for a polytropic relation specifically in stars. On the one hand, the equation of state of the gas can have the form (\ref{eq:polyP}), in which case $K_p$ is fixed by natural constants. On the other hand, the equation of state may contain the temperature $T$ and in addition we have a relation between $T$ and $P$. These two relations then yield a polytropic relation, where $K_p$ is a free parameter that can vary from star to star.

We are interested in cases where (\ref{eq:hy}-\ref{eq:polyP}) describe gravitationally bound DM structures. Then, $K_p$ is fixed by the DM particle parameters. If the DM is SFDM, we have $n=1$ for the TF regime, see (\ref{eq:PSI}), and $n=2$ for the fuzzy regime, see (\ref{eq:kfix}).

By introducing the dimensionless density and radius,
\begin{equation}\label{poly}
    \theta = \left(\frac{\rho}{\rho_c} \right)^{1/n}, \ \ \ \ \xi = \frac{r}{b},
\end{equation}
where $\rho_c$ denotes the central density and 
\begin{equation}
    b = \rho_c^{\frac{1}{2}(\frac{1}{n}-1)}\sqrt{\frac{K_p (n+1)}{4 \pi G}},
\end{equation}
after combining equations (\ref{eq:hy}-\ref{eq:polyP}), one obtains the famous Lane-Emden differential equation
\begin{equation}
\label{eq: lane}
    \frac{1}{\xi^2}\frac{d}{d\xi}\left( \xi^2 \frac{d\theta}{d\xi}\right) = - \theta^n \ ,
\end{equation}
 with boundary conditions
\begin{eqnarray}
\theta(0) &=& 1, \\
\theta'(0) &=& 0\ .
\end{eqnarray}
Here, the prime denotes differentiation with respect to $\xi$. Solutions corresponding to $0 \leq n < 5$ have a compact support. Thus, they become zero at a finite radius $\xi_1$:
\begin{equation}
    \theta |_{\xi = \xi_1} = 0\ .
\end{equation}
In that case, one can define a so-called "complete polytrope" with surface at $\xi = \xi_1$ and subsequently its radius, mass and mass-radius relationship, depending upon the polytropic index $0 \leq n < 5$, can be explicitly given as
\begin{eqnarray}\label{A12}
R &=& \xi_1 \sqrt{\frac{K_p (n+1)}{4 \pi G}} \rho_c^{-\frac{n-1}{2n}}, \\
M &=& -4\pi \frac{\theta_1'}{\xi_1}\rho_c R^3, \\
M^{(n-1)/n}R^{(3-n)/n} &=&  -\xi_1^{\frac{n+1}{n}} \theta_1'^{\frac{n-1}{n}}\frac{K_p (1+n)}{G (4 \pi)^{1/n}}\ ,
\end{eqnarray}
where $\theta_1' \equiv \theta'(\xi_1)$.\\
Closed-form expressions for the density profile $\theta = \theta(\xi)$ exist only for $n \in \{ 0,1,5 \}$. \cite{1939isss.book.....C} presents several tables of numerical values for polytropic models with index $n$. Some of the basic values are listed in \autoref{tab:polya}. We derived numerical solutions for polytropic density profiles governed by the Lane-Emden equation for any index $n$ as follows. An expansion of Eq. (\ref{eq: lane}) shows that the point $\xi =0$ represents a regular singularity of the ordinary differential equation. Thus, any numerical method will have difficulty starting at $\xi =0$. Therefore, we started integrating at some value $\xi = \xi^*$ near zero. However, accurate solutions then require initial conditions at $\xi^*$. 
We employed a standard approach where we calculate a Taylor series expansion of $\theta$ about $\xi =0$ and use this series in order to determine the initial conditions at $\xi^*$.

\begin{table}
\centering
\caption{Numerical values of polytropes with index $n$ according to \\ \protect\cite{1939isss.book.....C}. $\protect\Bar{\rho}$ denotes the mean density and $\protect\rho_c$ the central density.}
\label{tab:polya}
\begin{tabular}{llll}
$n$    &    $\xi_1$     &    $\left( - \xi^2 \frac{d \theta}{d \xi}\right)|_{\xi = \xi_1}$     &    $ \rho_c/ \Bar{\rho}$     \\ \hline
0   & 2.4494  & 4.8988  & 1.0000   \\
1   & 3.1416 & 3.1416 & 3.2899  \\
1.5 & 3.6538 & 2.7141 & 5.9907  \\
2   & 4.3529 & 2.4111 & 11.4025
\end{tabular}
\end{table}

The global energy terms given by LRS93 include the constants $\kappa_n$ and $q_n$, depending on the polytropic index $n$. Through numerical integration we get
\begin{equation}
\label{eq:kappan}
   \kappa_n \equiv \frac{5}{3 \xi_1^4 |\theta_1'|}\int_0^{\xi_1} \theta^n \xi^4 d\xi \begin{cases}
= \frac{5}{3}\left(1- \frac{6}{\pi^2} \right) \approx 0.653 & \text{for } n = 1 \\
\approx 1.448 & \text{for } n = 2 \\
\end{cases}
\end{equation}
and

\begin{equation}
\label{eq:qnvalues}
   q_n \equiv \kappa_n \left( 1- \frac{n}{5}\right) \begin{cases}
= \frac{4}{3}\left( 1- \frac{6}{\pi^2} \right) \approx 0.523 & \text{for } n = 1 \\
\approx 0.869 & \text{for } n = 2\ . \\
\end{cases}
\end{equation}

\section{Approximate equilibrium rotating ellipsoidal figures }\label{appendix-riemann}

The ellipsoidal approximation of LRS93 includes two crucial assumptions:
\begin{itemize}
    \item The isodensity surfaces are assumed to be self-similar ellipsoids. Thereby, the three principal axes $a_1,\ a_2$ and $a_3$ of the outer surface, where $\rho = P = 0$, or equivalently the eccentricities given in (\ref{eq:eccen-def}) solely specify the geometry. 
    \item The density profile $\rho(m)$ and specific internal energy profile $u(m)$, where $m$ denotes the mass inside an isodensity surface, are set identical to those of a spherical polytrope of same $n$, $K_p$ and volume, i.e. whose radius is the mean radius (\ref{eq:Rmean}).
\end{itemize}
The ellipticity of these equilibrium figures is a pure result of rotation, reflected by the dimensionless factor $f$ given by
\begin{equation}
\label{eq: f-ratio-riemann}
    f = \frac{1}{2}\frac{A_1a_1^2+A_2a_2^2+A_3a_3^2}{(a_1a_2a_3)^{2/3}}\ .
\end{equation}
Its dimensionless coefficients $A_i$ are given in \cite{1969efe..book.....C} and can be written in terms of the axis ratios:
\begin{eqnarray*}
A_1 &=& 2 \frac{a_2}{a_1} \frac{a_3}{a_1}  \frac{F(\theta,\phi)-E(\theta,\phi)}{\sin^3\phi \sin^2\theta} \\
A_2 &=& 2 \frac{a_2}{a_1} \frac{a_3}{a_1}  \frac{E(\theta,\phi)-F(\theta,\phi)\cos^2\theta-\frac{a_3}{a_2}\sin^2\theta\sin\phi}{\sin^3\phi \sin^2\theta \cos^2\theta} \\
A_3 &=& 2 \frac{a_2}{a_1} \frac{a_3}{a_1}  \frac{(a_2/a_3)\sin \phi-E(\theta,\phi)}{\sin^3\phi \sin^2\theta}\ ,
\end{eqnarray*}
where $\cos \phi = a_3/a_1$, $\sin \theta = \sqrt{\frac{1-(a_2/a_1)^2}{1-(a_3/a_1)^2}}$ and the standard incomplete elliptic integrals are given by
\begin{eqnarray}
E(\theta, \phi) &=& \int_0^{\phi}(1- \sin^2\theta \sin^2\phi')^{1/2}\text{d}\phi'\ , \\
F(\theta, \phi) &=& \int_0^{\phi}(1- \sin^2\theta \sin^2\phi')^{-1/2}\text{d}\phi'\ .
\end{eqnarray}
Note that in all these expressions $\theta$ and $\phi$ denote here the standard spherical angular coordinates.

In addition, LRS93 introduce the quantities
\begin{equation}
\label{eq:B12}
    A_{12} = \frac{A_1-A_2}{a_2^2-a_1^2} \ \ \text{and} \ \ B_{12} = A_2 - a_1^2A_{12}\ .
\end{equation}
The underlying polytropic density profile enters the energy expressions of LRS93 for the Riemann-S ellipsoid, also via the following constants given by
\begin{eqnarray}
\label{eq:k1}
k_1 & \equiv & \frac{n(n+1)}{5-n} \xi_1 |\theta'_1|\ , \\
\label{eq:k2}
k_2 & \equiv & \frac{3}{5-n} \left( \frac{4 \pi |\theta'_1|}{\xi_1} \right)^{1/3} \ .
\end{eqnarray}
Solving the system of equations (\ref{eq: axis-rel-2}) and (\ref{eq:lambda-final}), given the typical range of values for the spin-parameter, $\lambda \in [0.01,0.1]$, in addition to values beyond that range, $\lambda \in [0.1,0.3]$, yields the corresponding axis ratios and values for several dimensionless global quantities and properties of the Riemann-S ellipsoid which only depend on the ratios $a_2/a_1$ and $a_3/a_1$ shown in \autoref{tab:lambda-table}. The numerical values of the dimensionless quantities\ ,
\begin{eqnarray}
\label{eq:tild-lam}
   \Tilde{\Lambda} &\equiv& \frac{\Lambda}{\Omega_{\text{grav},R}} = \Tilde{\Omega} \frac{2\sqrt{1-e_1^2}}{2-e_1^2}\ , \\  
   \Tilde{L}^2 &\equiv& \frac{L^2}{GM^3R_R} =  \frac{\left(\kappa_2/5\right)^2 \frac{3}{4}\Tilde{\Omega}^2e_1^8}{(2-e_1^2)^2(1-e_1^2)^{\frac{2}{3}}(1-e_2^2)^{\frac{2}{3}}},\\
   |\Tilde{W}| &\equiv& \frac{|W|}{GM^2/ R_R} = f(e_1,e_2)\ , \\
   \label{eq:tild-L}
   \frac{R_R}{R_0} &=& g(e_1,e_2)^{-2}\ ,
\end{eqnarray}
which are functions of the eccentricities or equivalently axis ratios, as well as $\Tilde{\Omega}$ and $\kappa_2$ were computed once the system of equations (\ref{eq: axis-rel-2}) and (\ref{eq:lambda-final}) was solved for $a_2/a_1$ and $a_3/a_1$. From \autoref{tab:lambda-table} it is evident that $(n=2)$-polytropic, irrotational Riemann-S ellipsoids are prolate figures, see also \autoref{fig:ellipsoids}. They share this feature with the $(n=1)$-polytropic version of RS12 (in fact, this feature is shared by all irrotational Riemann-S ellipsoids).

\begin{figure*}
     \begin{minipage}[b]{0.5\linewidth}
      \centering\includegraphics[width=7cm]{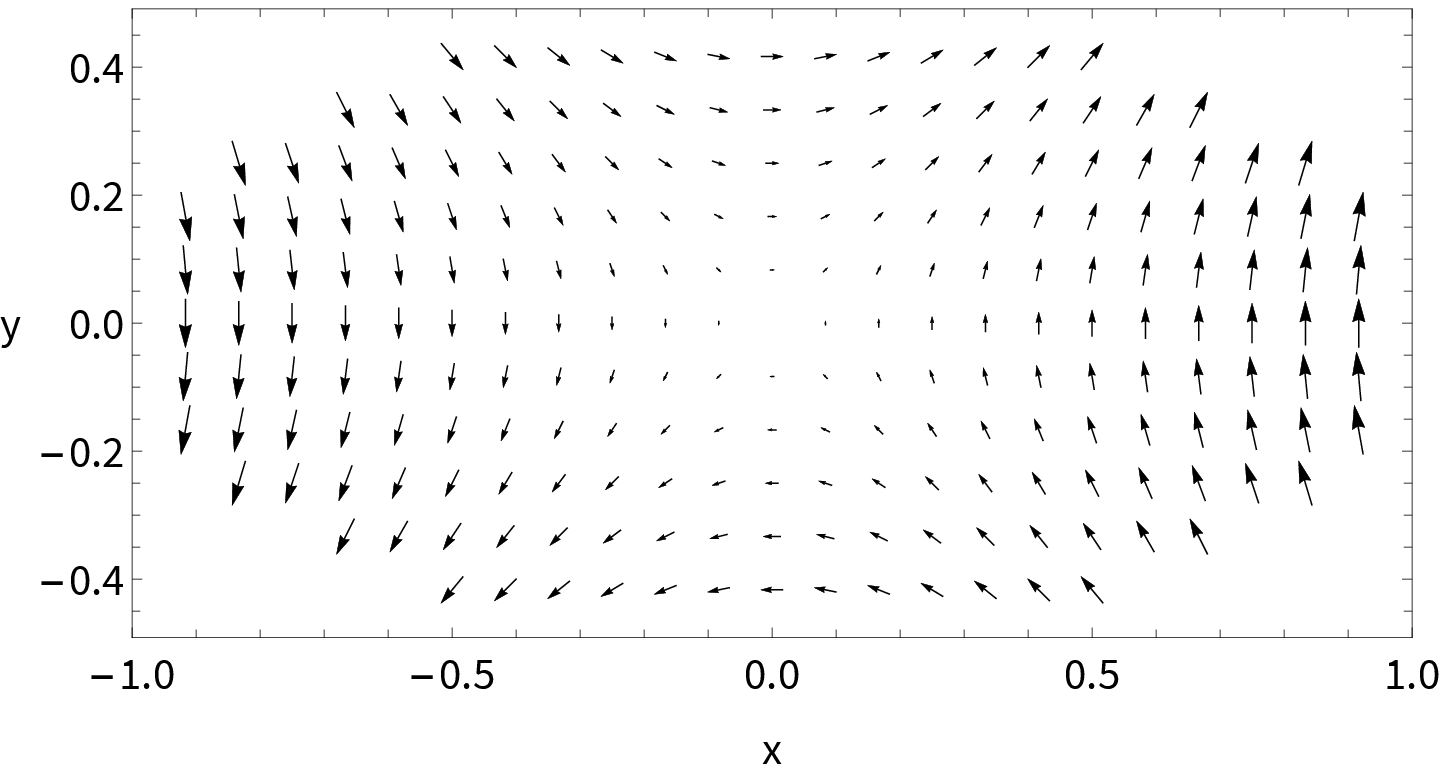}
     \hspace{0.1cm}
    \end{minipage}%
 \begin{minipage}[b]{0.5\linewidth}
      \centering\includegraphics[width=7cm]{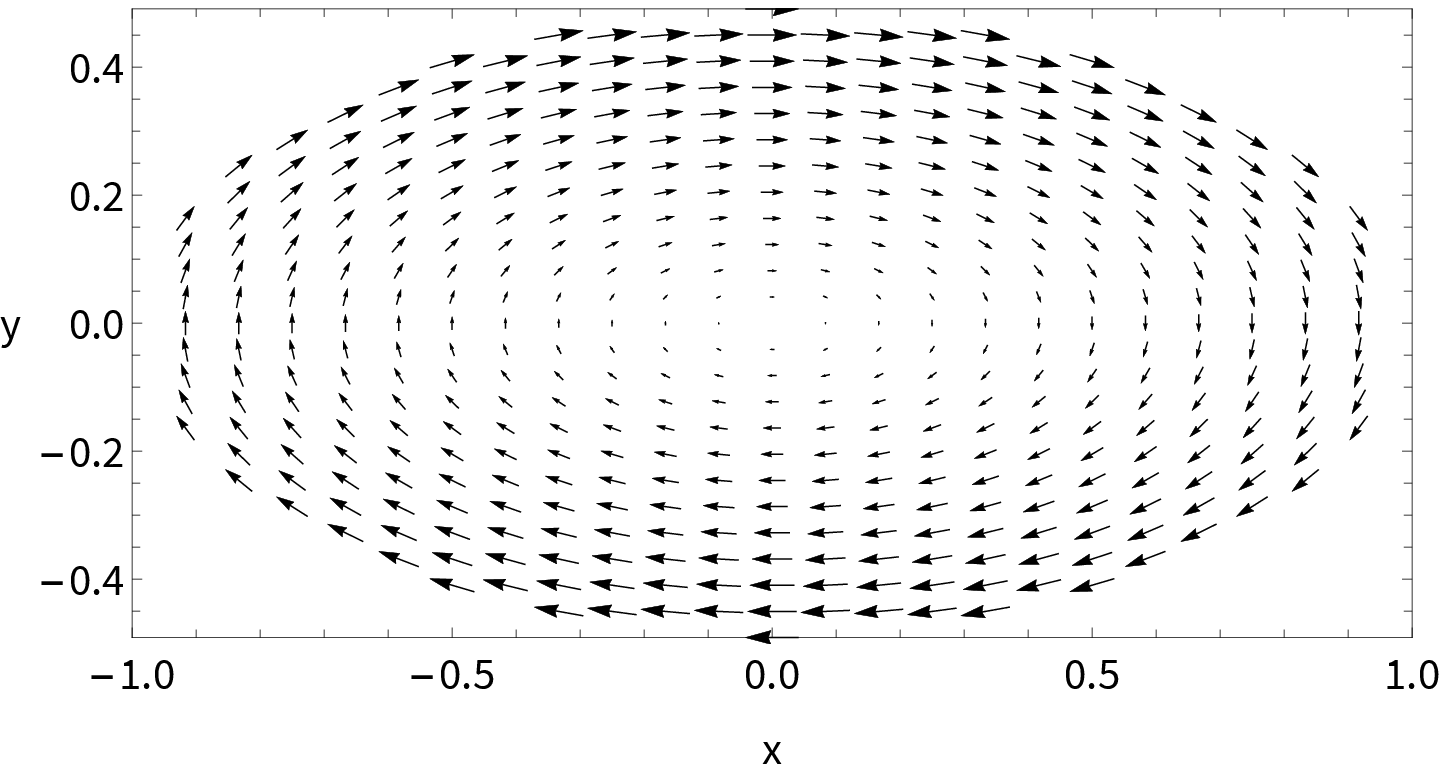}
     \hspace{0.1cm}
    \end{minipage}
 \caption{Illustrative velocity fields of an $(n=2)$-polytropic, irrotational Riemann-S ellipsoid in the rest frame (left-hand plot) according to Eq. (\ref{eq:vel-restframe}) and in the co-rotating frame (right-hand plot) according to Eq. (\ref{eq:vel-rotation}) with eccentricities $e_1 = 0.87098$ and $e_2 = 0.78203$ (or $\lambda = 0.1$, see \autoref{tab:lambda-table}).}
 \label{fig:velocity-fields}
\end{figure*}

\begin{figure*}
     \begin{minipage}[b]{0.5\linewidth}
      \centering\includegraphics[width=7cm]{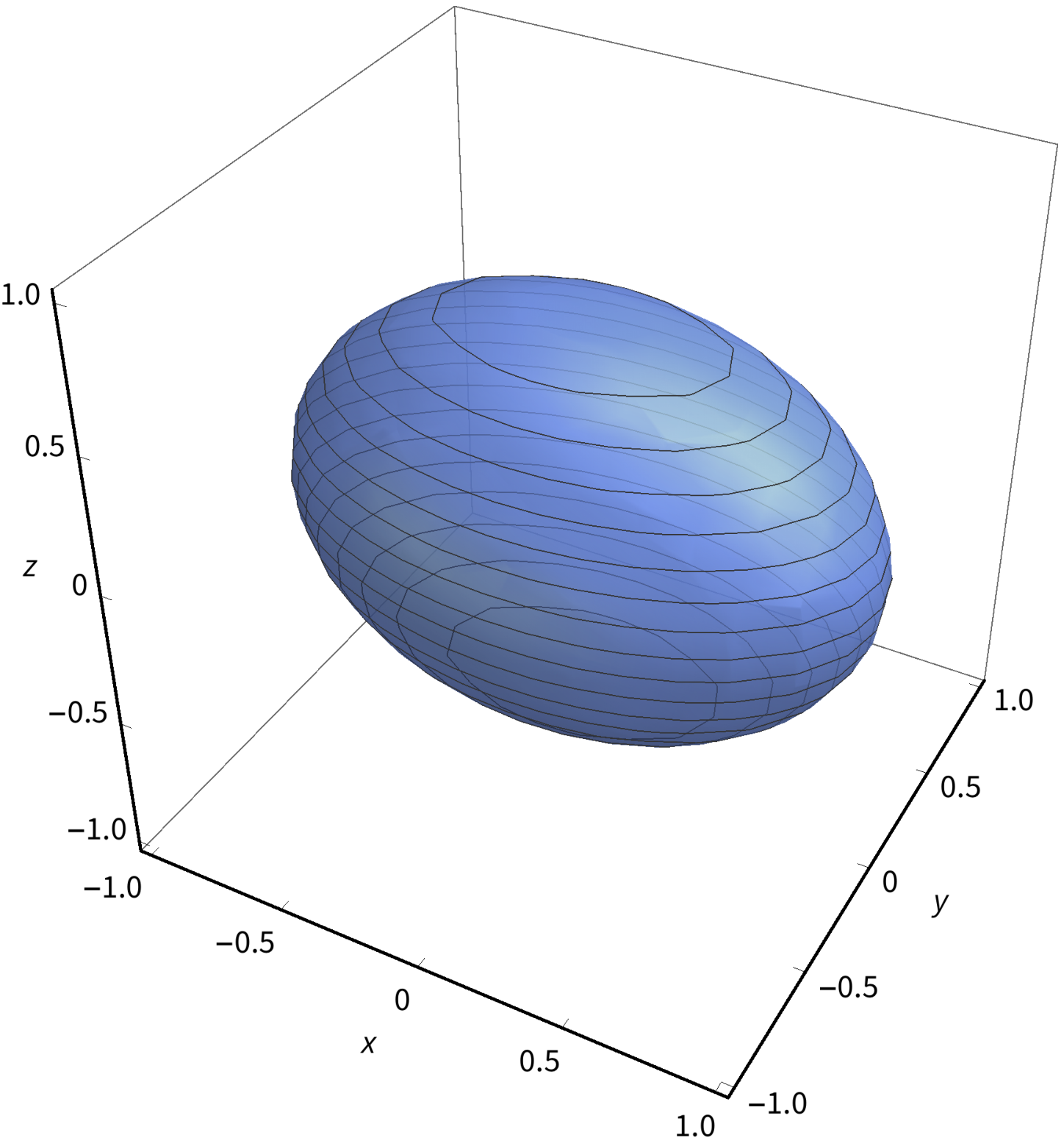}
     \hspace{0.1cm}
    \end{minipage}%
 \begin{minipage}[b]{0.5\linewidth}
      \centering\includegraphics[width=7cm]{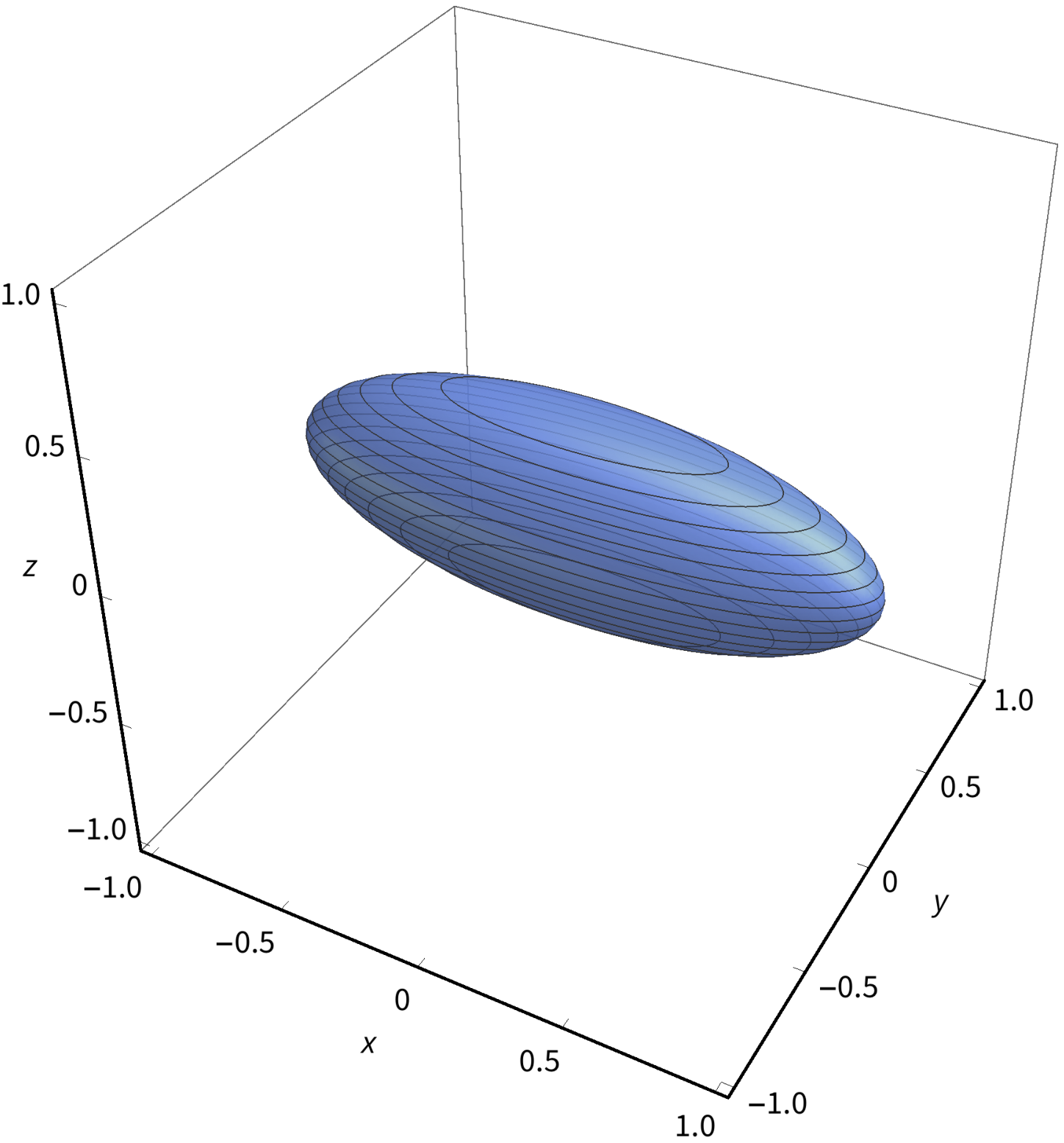}
     \hspace{0.1cm}
    \end{minipage}
 \caption{Irrotational Riemann-S ellipsoids rotating about the z-axis with $a_1=1$ and $\lambda= 0.05$ (left-hand plot) and $\lambda= 0.3$ (right-hand plot).}
 \label{fig:ellipsoids}
\end{figure*}

RS12 derive the velocity field of the irrotational (i.e. $f_R = -2$) Riemann-S ellipsoid for a given polytropic index $n$ in the rotating frame,
\begin{equation}
\label{eq:vel-rotation}
    \Vec{v}^{\ '}  = 2 \Omega_{\text{grav},R} \left(\frac{2 B_{12}}{q_n}\right)^{\frac{1}{2}} (8(1-e_1^2)+e_1^4)^{-1/2}\ (y,-(1-e_1^2)x,0)\ ,
\end{equation}
and in the rest frame,
\begin{equation}
\label{eq:vel-restframe}
    \Vec{v}  = \Omega_{\text{grav},R} \left(\frac{2 B_{12}}{q_n}\right)^{1/2} (1+8(1-e_1^2)/e_1^4)^{-1/2}\ (y,x,0)\ ,
\end{equation}
respectively, by using (\ref{eq:vel-1}-\ref{eq:xi-lambda}), (\ref{eq:omega-grav}) and (\ref{eq:B12}). $\Vec{v}^{\ '}/ (\Omega_{\text{grav},R} l)$ and $\Vec{v}/ (\Omega_{\text{grav},R} l)$, with $l$ denoting a quantity with dimension of length, are plotted in \autoref{fig:velocity-fields} for polytropic index $(n=2)$. Obviously, the vorticity vanishes in the rest frame in the case of the $f_R = -2$ irrotational sequence. In addition, two illustrative examples of halo core shapes in the form of Riemann-S ellipsoids can be found in \autoref{fig:ellipsoids}.

\begin{table}
\centering
\caption{Parameters (defined in (\ref{eq:tild-lam})-(\ref{eq:tild-L})) of the irrotational, $(n=2)$-polytropic Riemann-S ellipsoid as a function of $\lambda$.}
\label{tab:lambda-table}
\begin{tabular}{c|ccccc}
\hline
$\lambda$ & $e_1$            & $e_2$             & $a_2/a_1$              & $a_3/a_1$     & $t$                    \\ \hline
0.01      & $0.60246 $       & $0.46823 $        & 0.79815                & 0.88361       & $3.4110 \cdot 10^{-3}$  \\
0.03      & $0.73601$        & $0.60587$         & 0.67697                & 0.79556       & $1.0328 \cdot 10^{-2}$ \\
0.05      & $0.79654$        & $0.67861 $        & 0.60458                & 0.73450       & $1.7337 \cdot 10^{-2}$ \\
0.1       & $0.87098 $       & $0.78203$         & 0.49132                & 0.62324       & $3.5048 \cdot 10^{-2}$ \\
0.15      &  0.90803         &  0.84132          & 0.41891                & 0.54054       & $5.2718 \cdot 10^{-2}$ \\
0.2       & $0.93067 $       & $0.88056$         & 0.36585                & 0.47394       & $7.0077 \cdot 10^{-2}$ \\
0.3       & $0.95701 $       & $0.92861$         & 0.29005                & 0.37107       & $1.0326 \cdot 10^{-1}$ \\ \hline
$\lambda$ & $\Tilde{\Omega}$ & $\Tilde{\Lambda}$ & $\Tilde{L}^2$          & $|\Tilde{W}|$ & $R_R/R_0$              \\ \hline
0.01      & 0.55513          & 0.54131           & $1.9999 \cdot 10^{-4}$    & 0.99661       & 1.0207                 \\
0.03      & 0.55644          & 0.51663           & $1.7999 \cdot 10^{-3}$ & 0.98981       & 1.0642                 \\
0.05      & 0.55659           & 0.49287           & $4.9998 \cdot 10^{-3}$ & 0.98299         & 1.1106                 \\
0.1       & 0.55266          & 0.43746           & $2.0007 \cdot 10^{-2}$ & 0.96580        & 1.2398                 \\
0.15      & 0.54376          & 0.38756           & $4.5078 \cdot 10^{-2}$ & 0.94827        & 1.3897                 \\
0.2       & 0.53113          & 0.34275           & $8.0359 \cdot 10^{-2}$ & 0.93034       & 1.5627                 \\
0.3       & 0.49831          & 0.26663           & $1.8267 \cdot 10^{-1}$ & 0.89313       & 1.9911                
\end{tabular}
\end{table}

\bsp	
\label{lastpage}
\end{document}